\documentclass[a4paper,11pt]{article}
\pdfoutput=1
\usepackage{jheppub}
\usepackage[utf8]{inputenc}
\usepackage{appendix}

\usepackage{pdflscape}
\usepackage{booktabs}
\usepackage{afterpage}
\usepackage{hyperref}
\usepackage[colorinlistoftodos]{todonotes}

\definecolor{goodgreen}{RGB}{55,169,49}

\newcommand{\verticalcenter}[1]{\raisebox{-0.5\height}{\begin{tikzpicture}
    \node at (0,0) {#1};
\end{tikzpicture}}}

\usepackage{pifont}
\newcommand{\cmark}{\text{\ding{51}}}
\newcommand{\xmark}{\text{\ding{55}}}

\colorlet{colors01}{red!10}
\colorlet{colors02}{yellow!10}
\colorlet{colors03}{green!10}
\colorlet{colors04}{blue!10}
\colorlet{colors05}{purple!10}

\usepackage{enumerate}
\usepackage{oubraces}
\usepackage{tikz}
\usetikzlibrary{shapes.geometric}
\tikzset{gauge1/.style={draw=none,minimum size=0.5cm,fill=white,circle, draw}}
\tikzset{dotsize/.style={draw=none,minimum size=0.6pt,fill=black,circle,inner sep=1pt, draw}}
\tikzset{mini/.style={draw=none,minimum size=1pt,fill=white,circle,inner sep=3pt, draw}}
\tikzset{miniG/.style={draw=none,minimum size=1pt,fill=black,circle,inner sep=3pt, draw}}
\tikzset{cyane/.style={draw=none,minimum size=0.5cm,fill=cyan,circle, draw}}
\tikzset{pinklinet/.style={draw=none,minimum size=0.5cm,fill=magenta,circle, draw}}
\tikzset{greenlinet/.style={draw=none,minimum size=0.5cm,fill=green,circle, draw}}
\tikzset{brownlinet/.style={draw=none,minimum size=0.5cm,fill=olive,circle, draw}}
\tikzset{magicmintlinet/.style={draw=none,minimum size=0.5cm,fill=red,circle, draw}}
\tikzset{orangeet/.style={draw=none,minimum size=0.5cm,fill=orange,circle, draw}}
\tikzset{grayet/.style={draw=none,minimum size=0.5cm,fill=gray,circle, draw}}
\tikzset{blueet/.style={draw=none,minimum size=0.5cm,fill=blue,circle, draw}}
\tikzset{flavour1/.style={draw=none,minimum size=0.8cm,fill=white, regular polygon,regular polygon sides=4,draw}}
\tikzset{none/.style={draw=none}}
\tikzset{new edge style 1/.style={dashed}}
\tikzset{brace1/.style={decorate,decoration={brace,amplitude=5pt,mirror}}}
\tikzset{bluee/.style={line width=0.5mm,blue}}
\tikzset{orangee/.style={line width=0.5mm,orange}}
\tikzset{magentae/.style={line width=0.5mm,magenta}}
\tikzset{rede/.style={line width=0.5mm,red}}
\tikzset{greene/.style={line width=0.5mm,green}}
\tikzset{darke/.style={line width=0.5mm,black}}
\tikzset{cyaneX/.style={line width=0.5mm,cyan}}
\tikzset{new edge style 3/.style={dashed,red}}
\tikzset{magicmintline/.style={line width=0.5mm,gray}}
\tikzset{brownline/.style={line width=0.5mm,brown}}
\tikzset{greenline/.style={line width=0.5mm,green}}
\tikzset{oliveline/.style={line width=0.5mm,green}}
\tikzset{darkgreenline/.style={line width=0.5mm,olive}}
\tikzset{pinkline/.style={line width=0.5mm,magenta}}
\tikzset{brace2/.style={decorate,decoration={brace,amplitude=5pt}}}
\pgfdeclarelayer{edgelayer}
\pgfdeclarelayer{nodelayer}
\usetikzlibrary{decorations.pathreplacing}
\pgfsetlayers{edgelayer,nodelayer,main}

\usepackage{xcolor,colortbl}

\usepackage{ytableau}
\tikzstyle{brane}=[draw]
\tikzset{D7/.style={circle, draw=black, inner sep=0pt, fill=white, minimum size=3mm}}
\tikzset{hasse/.style={circle, fill,inner sep=2pt}}
\tikzset{flavor/.style={regular polygon,regular polygon sides=4,inner sep=2.5pt, draw}}
\tikzset{gauge/.style={circle, draw,inner sep=2.5pt}}
\tikzset{gaugeb/.style={circle, draw,fill=black,inner sep=2.5pt}}
\tikzset{gaugered/.style={circle, draw,fill=red,inner sep=2.5pt}}
\tikzset{gaugeblue/.style={circle, draw,fill=blue,inner sep=2.5pt}}
\tikzset{gaugegreen/.style={circle, draw,fill=green,inner sep=2.5pt}}
\tikzset{bd/.style={circle, draw=black, inner sep=0pt, fill=black, minimum size=2mm}}
\tikzset{wd/.style={circle, draw=black, inner sep=0pt, fill=white, minimum size=2mm}}
\tikzset{Dynkin/.style={circle, draw=black, inner sep=0pt, fill=white, minimum size=2mm}}
\tikzstyle{ligne}=[draw, thick] 
\tikzset{doublearrow/.style={ draw=black!75, color=black!75, thick, double distance=3pt, }} 
\newcommand{\midarrow}{\tikz 
\draw[-triangle 90] (0,0) -- +(.1,0);} 
\newcommand{\midarrowrev}{\tikz 
\draw[-triangle 90] +(.1,0) -- (0,0);}

\usetikzlibrary{calc}
\tikzset{gaugeJ/.style={inner sep=1mm,draw=none,fill=white,minimum size=2mm,circle, draw}}
\tikzset{flavourJ/.style={draw=none,minimum size=0.3mm,fill=white, regular polygon,regular polygon sides=4,draw}}
\tikzset{hasseJ/.style={circle, fill,inner sep=2pt}}

\usepackage{multirow}

\newcommand{\convexpath}[2]{
  [   
  create hullcoords/.code={
    \global\edef\namelist{#1}
    \foreach [count=\counter] \nodename in \namelist {
      \global\edef\numberofnodes{\counter}
      \coordinate (hullcoord\counter) at (\nodename);
    }
    \coordinate (hullcoord0) at (hullcoord\numberofnodes);
    \pgfmathtruncatemacro\lastnumber{\numberofnodes+1}
    \coordinate (hullcoord\lastnumber) at (hullcoord1);
  },
  create hullcoords
  ]
  ($(hullcoord1)!#2!-90:(hullcoord0)$)
  \foreach [
  evaluate=\currentnode as \previousnode using \currentnode-1,
  evaluate=\currentnode as \nextnode using \currentnode+1
  ] \currentnode in {1,...,\numberofnodes} {
    let \p1 = ($(hullcoord\currentnode) - (hullcoord\previousnode)$),
    \n1 = {atan2(\y1,\x1) + 90}, 
    \p2 = ($(hullcoord\nextnode) - (hullcoord\currentnode)$),
    \n2 = {atan2(\y2,\x2) + 90},
    \n{delta} = {Mod(\n2-\n1,360) - 360}
    in 
    {arc [start angle=\n1, delta angle=\n{delta}, radius=#2]}
    -- ($(hullcoord\nextnode)!#2!-90:(hullcoord\currentnode)$) 
  }
}

\title{Higgs Branches of U/SU Quivers via Brane Locking}
\preprint{Imperial/AH/21/7}

\author[1,2]{Antoine Bourget, }
\author[3]{Julius F. Grimminger, }
\author[3]{Amihay Hanany, }
\author[3]{Rudolph Kalveks, }
\author[3]{and Zhenghao Zhong}
\affiliation[1]{Université Paris-Saclay, CNRS, CEA, Institut de physique théorique, 91191, Gif-sur-Yvette, France}
\affiliation[2]{Laboratoire de Physique de l’Ecole normale supérieure, ENS, Université PSL, CNRS, Sorbonne
Université, Université de Paris, F-75005 Paris, France}
\affiliation[3]{Theoretical Physics Group, The Blackett Laboratory, Imperial College London, Prince Consort Road
London, SW7 2AZ, UK}
\emailAdd{antoine.bourget@polytechnique.org}
\emailAdd{julius.grimminger17@imperial.ac.uk}
\emailAdd{a.hanany@imperial.ac.uk}
\emailAdd{rudolph.kalveks09@imperial.ac.uk}
\emailAdd{zhenghao.zhong14@imperial.ac.uk}

\abstract{We solve a long standing problem on the computation of the Higgs branch $\mathcal{H}$ of linear quivers with 8 supercharges and with both unitary and special unitary gauge nodes. The solution uses the concept of magnetic quivers, where components of $\mathcal{H}$ are described as 3d $\mathcal{N}=4$ Coulomb branches. When the starting quiver is good, there is a single component in $\mathcal{H}$ and the magnetic quiver is a 3d mirror. The magnetic quivers are obtained from studying the brane web for an auxiliary 5d theory (with only special unitary gauge groups), constrained by a new notion called \emph{brane locking}, where some branes are required to move together. We view this as a computational tool rather than an operation in 5d. A detailed algorithm is given, and implemented in a code available for download.}

\begin{document} 

\maketitle

\section{Introduction}
Duality has long played a crucial role in string theory and supersymmetric gauge theories. 
Finding two equivalent descriptions of the same object sheds light on its intrinsic properties, and often offers alternative ways of computing physical quantities. 
In this note, we provide a dual description of the Higgs branch of theories with 8 supercharges, using the concept of magnetic quivers. Concretely, such a Higgs branch is described as a union of the 3d $\mathcal{N}=4$ Coulomb branches of a collection of quiver gauge theories. 
In favorable cases, which will be discussed below, this provides a large class of 3d mirror pairs \cite{Intriligator:1996ex}.

Our focus will be on the Higgs branch of \emph{linear quivers}, which we treat as electric, and their magnetic quivers. Linear quivers with only unitary gauge groups are well studied in the context of 3d $\mathcal{N}=4$ gauge theories. Linear quivers with only special unitary gauge theories make their appearance in 4d $\mathcal{N}=2$, 5d $\mathcal{N}=1$ and $6d$ $\mathcal{N}=(1,0)$ theories. However, studies of linear quivers with \emph{mixed} unitary and special unitary gauge groups are rare.

The lack of investigation of quivers with mixed U \& SU gauge groups can be attributed to the lack of a brane description. For example, unitary quivers can arise as 3d $\mathcal{N}=4$ effective field theories living on the worldvolume of D3 branes in D3-D5-NS5 brane systems \cite{Hanany:1996ie}. On the other hand, the description of special unitary quivers requires a 
different brane system. In particular, they can be described as 5d $\mathcal{N}=1$ effective field theories living on the worldvolume of 5-branes in certain brane web systems with 5-branes and 7-branes \cite{Aharony:1997ju,Aharony:1997bh,DeWolfe:1999hj}. The gauge groups appearing naturally in theories described by brane webs is special unitary, rather than unitary, due to the linear bending of 5-branes. 

Recently, it was found that Higgs branches of certain classes of Argyres-Douglas theories can be described as Higgs branches of linear quivers with mixed U \& SU gauge groups, which allows the construction of magnetic quivers for these theories \cite{Closset:2020afy,Giacomelli:2020ryy,Xie:2021ewm,Dey:2021rxw}. The present paper confirms these results and extends them to U / SU quivers not coming from Argyres-Douglas theories.

In this paper, we use brane webs to describe changes in the Higgs branch when special unitary gauge groups are turned to unitary gauge groups. The main idea can be summarized in the following embedding, on which we comment in Appendix \ref{app:quotient}: 
\begin{equation}
\label{inclusion1}
\mathcal{H}_{\mathrm{SU}} \supset \mathcal{H}_{\mathrm{U}/\mathrm{SU}}  \, .
\end{equation}
The left-hand side denotes the Higgs branch of a linear quiver with SU gauge nodes; the right-hand side is the Higgs branch of the same quiver with some SU nodes replaced by U nodes. The first theory can be constructed on a brane web configuration, and its Higgs branch is realized by movements of irreducible subwebs; the inclusion corresponds to \emph{restricting certain irreducible subwebs to move together}, which we call \emph{web locking}\footnote{Combining irreducible subwebs into one is also needed e.g.\ when computing the intersection of several cones \cite{Bourget:2019rtl}. The new feature here is that the subwebs we lock together don't need to intersect each other.}.  The 3d $\mathcal{N}=4$ magnetic quivers can then be readily read off the brane web following the algorithm proposed in \cite{Cabrera:2018jxt}. 
In other words, the inclusion (\ref{inclusion1}) is realized on the magnetic side as 
\begin{equation}
\label{inclusion2}
\mathcal{H}_{\mathrm{SU}} = \bigcup\limits_{\textrm{Cones}} \mathcal{C} \left( \mathrm{MQ} \left(  \begin{array}{c}
  \textrm{Brane web} \\
 \textrm{without locking}
\end{array} \right) \right) \supset  \bigcup\limits_{\textrm{Cones}} \mathcal{C} \left( \mathrm{MQ} \left( \begin{array}{c}
  \textrm{Brane web} \\
 \textrm{with locking}
\end{array} \right) \right)  = \mathcal{H}_{\mathrm{U}/\mathrm{SU}}  \, .
\end{equation}
Here $\mathcal{C}$ denotes the 3d $\mathcal{N}=4$ Coulomb branch, and MQ denotes the magnetic quivers read from brane web intersections. In general, there are several maximal decompositions of a given brane web into irreducible subwebs. When this happens, the Higgs branch is a union of cones (which can intersect in a non-trivial way), one for each decomposition, and each cone is described by a magnetic quiver. Equation (\ref{inclusion2}) can be seen as the main message of this paper: 
\begin{itemize}
    \item The embedding (\ref{inclusion1}) is a hyper-K\"ahler quotient by $\mathrm{U}(1)^r$, where $r$ is the number of SU nodes that are turned into U nodes. Geometrically, seeing $\mathcal{H}_{\mathrm{SU}}$ as a complex algebraic variety, it corresponds to imposing a quadratic algebraic equation (given by the complex moment map), followed by a projection onto the $\left(\mathrm{U}(1)_\mathbb{C}\right)^r$ invariant locus. 
    \item This embedding has the \emph{dual} realization (\ref{inclusion2}). Here it suffices to impose a \emph{linear} constraint -- the locking -- on the brane web degrees of freedom, after which one can read the associated magnetic quiver(s). 
\end{itemize}
We present an algorithm that provides the magnetic quivers in (\ref{inclusion2}) in full generality; this algorithm is implemented in a Mathematica code available at the \emph{arXiv} or at
\begin{center}
\url{http://www.antoinebourget.org/attachments/files/SUquivers.nb} 
\end{center}
 
Note that we provide magnetic quivers for electric theories which are good, bad or ugly.\footnote{See conventions at the end of the introduction.} If the electric theory is good, there is one magnetic quiver which is a 3d mirror. If the electric theory is ugly, there is one magnetic quiver, which is a 3d mirror if free hypers are added. If the electric theory is bad then in general no 3d mirror exists, see e.g.\ Section \ref{sec:incomplete3d} and Appendix \ref{bad}, and there may be multiple magnetic quivers.

Before dealing with the details of brane lockings, a word of warning is in order. When the number of flavors is small enough (see fourth column in Table \ref{tab:effects}), the classical Higgs branch, as defined by generators and relations, can be a non-reduced scheme \cite{Bourget:2019rtl}. The corresponding ring has nilpotent operators, to which brane webs are not known to be sensitive. Therefore when this happens, the magnetic quivers produced by our algorithm correspond to the underlying Higgs variety, i.e.\ the reduced part of the Higgs branch \cite{Bourget:2019rtl}. It should be noted that in these circumstances, it is challenging to cross-check our results, as direct computation of Higgs branches require algorithms involving Gr\"obner bases, which have prohibitive computational cost. 

\paragraph{Conventions.}
Following \cite{Gaiotto:2008ak} a 3d $\mathcal{N}=4$ gauge theory is called \textbf{good} if the R-charge $q$ (highest weight under SU$(2)_R$) of any monopole operator satisfies $q \geq 1$; we say that it is \textbf{ugly} if this bound is replaced by $q \geq \frac{1}{2}$ and is saturated by at least one monopole operator; and finally it is \textbf{bad} if at least one monopole operator has $q \leq 0$.

A useful notion when dealing with a quiver gauge theory is the \textbf{balance} of a gauge node. For a gauge node U$(N_c)$ or SU$(N_c)$ with $N_f$ hypermultiplets in its fundamental representation, we define the balance $b$ as
\begin{equation}
    b=N_f-2N_c\;.
\end{equation}
We call a node balanced, if $b=0$, overbalanced if $b>0$ and underbalanced if $b<0$. If no node has negative balance we say the quiver has non-negative balance, if any node has negative balance we say the quiver has negative balance.

For U$(N_c)$ SQCD: the theory is good if $b\geq0$ (for $b=0$ the Coulomb branch global symmetry enhances from U$(1)_{\mathrm{topological}}$ to SU$(2)$), the quiver is ugly if $b=-1$ and the quiver is bad if $b<-1$.

For SU$(N_c)$ SQCD: the theory is good if $b\geq-1$ (a U$(1)$ Coulomb branch global symmetry emerges for $b=-1$) and it is bad if $b<-1$.

Note, however, that it is in general not enough to inspect the balance of the nodes in a quiver to tell whether it is good. Counterexamples are discussed in Section \ref{sec:incomplete3d}.

\paragraph{Organization of the paper.}

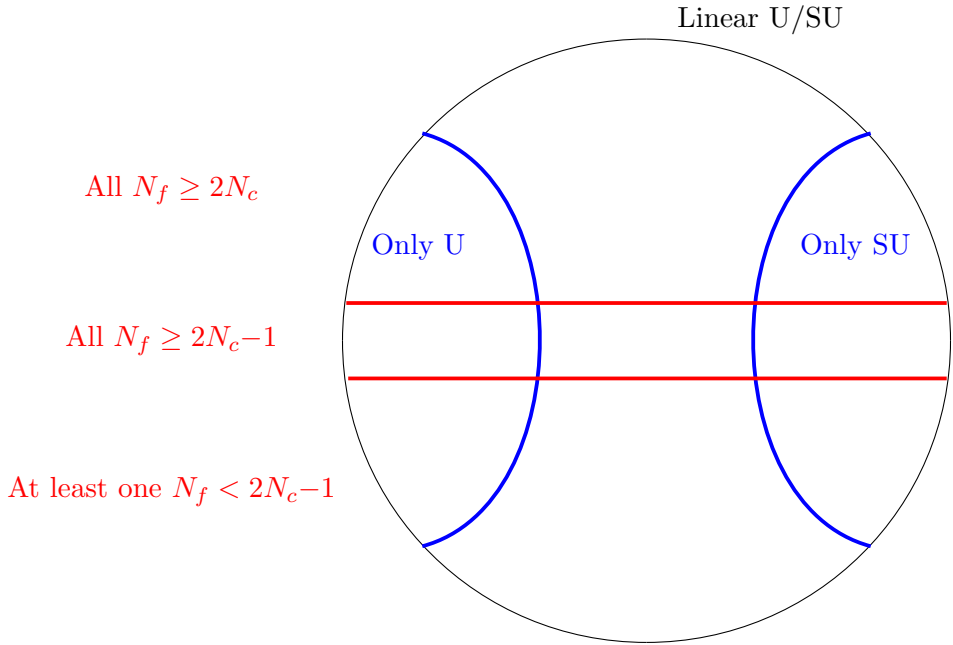
\begin{figure}
    \centering
\begin{tikzpicture}
	\begin{pgfonlayer}{nodelayer}
		\node [style=none] (0) at (0, 4) {};
		\node [style=none] (1) at (-4, 0) {};
		\node [style=none] (2) at (0, -4) {};
		\node [style=none] (3) at (4, 0) {};
		\node [style=none] (6) at (2.95, 2.75) {};
		\node [style=none] (7) at (2.95, -2.725) {};
		\node [style=none] (8) at (-2.95, 2.75) {};
		\node [style=none] (9) at (-2.95, -2.725) {};
		\node [style=none] (10) at (-3.925, -0.5) {};
		\node [style=none] (11) at (3.95, -0.5) {};
		\node [style=none] (12) at (-3.95, 0.5) {};
		\node [style=none] (13) at (3.95, 0.5) {};
		\node [style=none] (14) at (1.5, 4.25) {Linear U/SU};
		\node [style=none] (15) at (-6.25, 2) {\color{red}{All $N_f \geq 2N_c$}};
		\node [style=none] (16) at (-6.25, 0) {\color{red}{All $N_f \geq 2N_c{-}1$}};
		\node [style=none] (17) at (-6.25, -2) {\color{red}{At least one $N_f < 2N_c{-}1$}};
		\node [style=none] (18) at (-3, 1.25) {\color{blue}{Only U}};
		\node [style=none] (19) at (2.75, 1.25) {\color{blue}{Only SU}};
	\end{pgfonlayer}
	\begin{pgfonlayer}{edgelayer}
		\draw [bend right=45] (0.center) to (1.center);
		\draw [bend right=45] (1.center) to (2.center);
		\draw [bend right=45] (2.center) to (3.center);
		\draw [bend right=45] (3.center) to (0.center);
		\draw [style=bluee, in=165, out=-165] (6.center) to (7.center);
		\draw [style=bluee, in=15, out=-15] (8.center) to (9.center);
		\draw [style=rede] (10.center) to (11.center);
		\draw [style=rede] (12.center) to (13.center);
	\end{pgfonlayer}
\end{tikzpicture}
    \caption{Venn diagram of the different types of quivers discussed in this paper. $N_c$ stands for the gauge rank of a node in the quiver, while $N_f$ stands for the number of hypermultiplets connected to it. The circle represents all linear quivers with unitary and special unitary gauge groups, and arbitrary numbers of flavors. In the left blue region, where all gauge groups are unitary, magnetic quivers can be computed using D3-D5-NS5 systems (One can also use (fully) locked brane webs), while in the right blue region, where all the gauge groups are special unitary, one can use 5-brane webs. In the generic (middle) region, brane webs with lockings are needed. In the top region, all gauge groups have enough matter to ensure that the magnetic quiver is a 3d mirror theory. Below this, in the middle stripe, the same applies up to free hypermultiplets (see discussion in Section \ref{section4}) if only one gauge node has $N_f = 2N_c -1$; if two gauge nodes or more satisfy this equality, the situation is more complex, see Section \ref{section4}. In the last region, a collection of effects can happen: the Higgs branch can contain one cone or more, and the Higgs ring can possibly contain nilpotent elements, see Table \ref{tab:effects}.}
    \label{fig:venn}
\end{figure}

The rest of the paper is organized as follows. In Section \ref{branes} we start by looking at specific examples with quivers containing only special unitary gauge groups. We then replace some of the gauge groups with unitary gauge groups and see how to interpret this in a brane web configuration and how to arrive at the corresponding mirror quiver. Section \ref{prescription} generalizes this prescription by offering a simple and systematic way for obtaining the mirror quiver of any linear quiver with mixed U/SU gauge groups. Section \ref{section4} discusses more examples and general lessons that can be learned from the general algorithm concerning the various quiver types according to Figure \ref{fig:venn}.

\section{Web locking: first examples}\label{branes}

In this section, we introduce the concept of brane locking on a family of basic examples, which are all good 3d $\mathcal{N}=4$ quiver theories (in the sense described in the introduction), which means $N_f \geq 2N_c$ for every gauge node. For this kind of theory, brane locking provides 3d $\mathcal{N}=4$ mirror pairs. We confirm these findings using Hilbert series computations for the Higgs and Coulomb branches of both quivers in the pair. More general quivers will be dealt with later on, using the same principles. 

The traditional way of computing a 3d $\mathcal{N}=4$ mirror of a linear quiver with unitary gauge groups makes use of brane set ups with D3, D5 and NS5 branes and S-duality.\footnote{One can also include orientifold planes, but these will not be needed in this article. }
Let us start with a $T(\mathrm{SU}(4))$ quiver. Utilizing brane set ups, one finds that this theory is 3d self mirror:
\begin{equation}
\raisebox{-.5\height}{\scalebox{.8}{\begin{tikzpicture}
	\begin{pgfonlayer}{nodelayer}
		\node [style=gauge1] (0) at (-2, 0) {};
		\node [style=gauge1] (1) at (-0.5, 0) {};
		\node [style=gauge1] (2) at (1, 0) {};
		\node [style=none] (4) at (-2, -0.575) {U(3)};
		\node [style=none] (5) at (-0.5, -0.575) {U(2)};
		\node [style=none] (6) at (1, -0.575) {U(1)};
		\node [style=none] (7) at (-3.5, -0.575) {4};
		\node [style=flavour1] (8) at (-3.5, 0) {};
		\node [style=gauge1] (9) at (8, 0) {};
		\node [style=gauge1] (10) at (6.5, 0) {};
		\node [style=gauge1] (11) at (5, 0) {};
		\node [style=none] (12) at (8, -0.575) {U(3)};
		\node [style=none] (13) at (6.5, -0.575) {U(2)};
		\node [style=none] (14) at (5, -0.575) {U(1)};
		\node [style=none] (15) at (9.5, -0.575) {4};
		\node [style=flavour1] (16) at (9.5, 0) {};
		\node [style=none] (17) at (2, 0) {};
		\node [style=none] (18) at (4, 0) {};
		\node [style=none] (19) at (3, 0.5) {3d mirror};
	\end{pgfonlayer}
	\begin{pgfonlayer}{edgelayer}
		\draw (0) to (1);
		\draw (1) to (2);
		\draw (0) to (8);
		\draw (9) to (10);
		\draw (10) to (11);
		\draw (9) to (16);
		\draw [style=->] (17.center) to (18.center);
		\draw [style=->] (18.center) to (17.center);
	\end{pgfonlayer}
\end{tikzpicture}}}
\label{tsu4}
\end{equation}
This can be checked through explicit Coulomb branch and Higgs branch Hilbert series computations as shown in Table \ref{tsu4table}.

We now want to know what happens if we replace \textit{all} the unitary gauge nodes with special unitary gauge nodes:
\begin{equation}
   \raisebox{-.5\height}{\scalebox{.8}{ \begin{tikzpicture}
	\begin{pgfonlayer}{nodelayer}
		\node [style=gauge1] (0) at (-1, 0) {};
		\node [style=gauge1] (1) at (0.5, 0) {};
		\node [style=gauge1] (2) at (2, 0) {};
		\node [style=none] (4) at (-1, -0.575) {SU(3)};
		\node [style=none] (5) at (0.5, -0.575) {SU(2)};
		\node [style=none] (6) at (2, -0.575) {SU(1)};
		\node [style=none] (7) at (-2.5, -0.575) {4};
		\node [style=flavour1] (8) at (-2.5, 0) {};
	\end{pgfonlayer}
	\begin{pgfonlayer}{edgelayer}
		\draw (0) to (1);
		\draw (1) to (2);
		\draw (0) to (8);
	\end{pgfonlayer}
\end{tikzpicture}}}
\label{tsu4SU}
\end{equation}
In 3d, we do not have a brane system for such a quiver, as stacks of D3 branes stretched between 5-branes only give rise to unitary gauge groups. However we can construct a brane configuration in 5d using brane webs \cite{Aharony:1997ju,Aharony:1997bh,DeWolfe:1999hj}. Our quiver theories are now effective gauge theories living on $(p,q)$ 5-branes suspended between $[p,q]$ 7-branes. 
Quantum effects cause the 5-branes to bend when they end on each other, forming a web. This process freezes a $\mathrm{U}(1)$ factor, turning unitary gauge groups into special unitary gauge groups. For the 5d set up, the occupations of space-time directions by the different branes are given in Table \ref{spacetime}.

\begin{table}[t]
\begin{center}
\begin{tabular}{|c|c|c|c|c|c|c|c|c|c|c|}
\hline
Type IIB & $x^0$ & $x^1$ & $x^2$ & $x^3$ & $x^4$ & $x^5$ & $x^6$ & $x^7$ & $x^8$ & $x^9$\\
\hline
$(p,q)5$-brane & $\times$ & $\times$ & $\times$ & $\times$ & $\times$ & \multicolumn{2}{c|}{angle $\alpha$} & & & \\
\hline
$[p,q]7$-Brane & $\times$ & $\times$ & $\times$ & $\times$ & $\times$ & & & $\times$ & $\times$ & $\times$ \\ \hline
\end{tabular}
\caption{Occupation of space-time directions of the $(p,q)$5-branes and $[p,q]$7-branes in Type IIB are denoted by $\times$. The angle $\alpha$ depends on the $(p,q)$ charges and the axio-dilaton $\tau$; $\alpha=\arg(p+\tau q)$. We set $\tau=i$ in the rest of the paper, s.t.\ $\tan(\alpha)=q/p$. Our 5d $\mathcal{N}=1$ theories exist as effective field theories living on 5-branes suspended between 7-branes.}
\label{spacetime}
\end{center}
\end{table}

The 5d brane configuration of (\ref{tsu4SU}) with bare masses turned on takes the form:
\begin{equation}
   \raisebox{-.5\height}{\scalebox{.789}{
\begin{tikzpicture}
	\begin{pgfonlayer}{nodelayer}
		\node [style=none] (200) at (-4, -2) {};
		\node [style=none] (210) at (-4, -4) {};
		\node [style=none] (220) at (-2, -4) {};
		\node [style=gauge1] (230) at (-4, -4) {};
		\node [style=none] (250) at (-1.575, -4) {$x^6$};
		\node [style=none] (260) at (-4, -4.6) {$x^7,x^8,x^9$};
		\node [style=none] (280) at (-4, -4) {{\Huge $\times$}};
		\node [style=gauge1] (0) at (-1.75, 2) {};
		\node [style=gauge1] (1) at (0, 3.25) {};
		\node [style=gauge1] (2) at (-1.75, 0.25) {};
		\node [style=gauge1] (3) at (-1.75, -1) {};
		\node [style=gauge1] (4) at (-1.75, -2) {};
		\node [style=none] (5) at (0, 2) {};
		\node [style=none] (6) at (0.75, 1.25) {};
		\node [style=none] (7) at (3.25, 1.25) {};
		\node [style=none] (8) at (0.75, 0.25) {};
		\node [style=none] (9) at (1.25, -0.25) {};
		\node [style=gauge1] (10) at (3.25, 3.25) {};
		\node [style=none] (11) at (4, 0.5) {};
		\node [style=none] (12) at (4, -0.25) {};
		\node [style=none] (13) at (5.5, 0.5) {};
		\node [style=gauge1] (14) at (5.5, 3.25) {};
		\node [style=none] (15) at (4.75, -1) {};
		\node [style=none] (16) at (6.25, -0.25) {};
		\node [style=none] (17) at (6.25, -1) {};
		\node [style=none] (18) at (1.25, -1) {};
		\node [style=none] (19) at (1.75, -1.5) {};
		\node [style=none] (20) at (1.75, -2) {};
		\node [style=none] (21) at (4.75, -1.5) {};
		\node [style=gauge1] (22) at (8.25, -3) {};
		\node [style=gauge1] (23) at (6.25, -3) {};
		\node [style=gauge1] (24) at (2.75, -3) {};
		\node [style=gauge1] (25) at (8, 3.25) {};
		\node [style=gauge1] (26) at (10.75, -3) {};
		\node [style=none] (27) at (8, -0.25) {};
		\node [style=none] (281) at (-4, -1.5) {$x^5$};
	\end{pgfonlayer}
	\begin{pgfonlayer}{edgelayer}
		\draw (210.center) to (230);
		\draw [style=->] (210.center) to (200.center);
		\draw [style=->] (210.center) to (220.center);
		\draw (0) to (5.center);
		\draw (1) to (5.center);
		\draw (5.center) to (6.center);
		\draw (6.center) to (8.center);
		\draw (6.center) to (7.center);
		\draw (8.center) to (9.center);
		\draw (2) to (8.center);
		\draw (10) to (7.center);
		\draw (7.center) to (11.center);
		\draw (11.center) to (12.center);
		\draw (9.center) to (12.center);
		\draw (11.center) to (13.center);
		\draw (14) to (13.center);
		\draw (12.center) to (15.center);
		\draw (13.center) to (16.center);
		\draw (15.center) to (17.center);
		\draw (16.center) to (17.center);
		\draw (3) to (18.center);
		\draw (9.center) to (18.center);
		\draw (18.center) to (19.center);
		\draw (15.center) to (21.center);
		\draw (19.center) to (21.center);
		\draw (20.center) to (19.center);
		\draw (17.center) to (22);
		\draw (21.center) to (23);
		\draw (20.center) to (24);
		\draw (20.center) to (4);
		\draw (16.center) to (27.center);
		\draw (25) to (27.center);
		\draw (27.center) to (26);
	\end{pgfonlayer}
\end{tikzpicture}}}
\label{braneweb1}
\end{equation}
where nodes represent 7-branes and lines represent 5-branes (Specifically, horizontal lines are D5 branes, vertical lines are NS5 branes, and lines at an angle here are $(1,-1)$ 5-branes.) The coordinate system we adopted here will be the same throughout the paper. The areas of the polygons represent moduli of the Coulomb branch. For the case above we see three polygons and hence the Coulomb branch has real dimension 3.

\subsubsection*{Going to the Higgs branch}
As outlined in \cite{Cabrera:2018jxt, Bourget:2019rtl}, we can go to the Higgs branch by first setting all the masses (given by the vertical distance between the D7 branes) to zero. The resulting configuration is:
\begin{equation}
   \raisebox{-.5\height}{\scalebox{.789}{
\begin{tikzpicture}
	\begin{pgfonlayer}{nodelayer}
		\node [style=gauge1] (0) at (-1, -0.05) {};
		\node [style=gauge1] (1) at (0.5, -0.05) {};
		\node [style=gauge1] (2) at (2, -0.05) {};
		\node [style=gauge1] (3) at (3.5, -0.05) {};
		\node [style=gauge1] (4) at (5, 1.25) {};
		\node [style=gauge1] (5) at (6, -1) {};
		\node [style=gauge1] (6) at (7.5, -1) {};
		\node [style=gauge1] (7) at (9.175, -1) {};
		\node [style=gauge1] (8) at (10.75, -1) {};
		\node [style=gauge1] (9) at (6.5, 1.25) {};
		\node [style=gauge1] (10) at (8, 1.25) {};
		\node [style=gauge1] (11) at (9.5, 1.25) {};
		\node [style=none] (13) at (5, -0.25) {};
		\node [style=none] (14) at (6.5, -0.2) {};
		\node [style=none] (15) at (8, -0.075) {};
		\node [style=none] (16) at (9.5, 0) {};
		\node [style=none] (17) at (0.5, 0.075) {};
		\node [style=none] (18) at (0.5, -0.15) {};
		\node [style=none] (19) at (2, 0.075) {};
		\node [style=none] (20) at (2, -0.15) {};
		\node [style=none] (21) at (2, 0.075) {};
		\node [style=none] (22) at (2, -0.05) {};
		\node [style=none] (23) at (2, -0.175) {};
		\node [style=none] (24) at (3.5, 0.075) {};
		\node [style=none] (25) at (3.5, -0.05) {};
		\node [style=none] (26) at (3.5, -0.175) {};
		\node [style=none] (27) at (9.5, 0.025) {};
		\node [style=none] (28) at (3.5, 0.25) {};
		\node [style=none] (29) at (3.5, 0.125) {};
		\node [style=none] (30) at (3.5, 0) {};
		\node [style=none] (31) at (3.5, -0.125) {};
		\node [style=none] (32) at (3.5, -0.25) {};
		\node [style=none] (33) at (8, -0.075) {};
		\node [style=none] (34) at (6.5, -0.175) {};
		\node [style=none] (35) at (5, -0.275) {};
	\end{pgfonlayer}
	\begin{pgfonlayer}{edgelayer}
		\draw (4) to (13.center);
		\draw (13.center) to (5);
		\draw (9) to (14.center);
		\draw (14.center) to (6);
		\draw (10) to (15.center);
		\draw (15.center) to (7);
		\draw (11) to (16.center);
		\draw (16.center) to (8);
		\draw (0) to (1);
		\draw (17.center) to (19.center);
		\draw (18.center) to (20.center);
		\draw (21.center) to (24.center);
		\draw (22.center) to (25.center);
		\draw (23.center) to (26.center);
		\draw (29.center) to (27.center);
		\draw (31.center) to (34.center);
		\draw (30.center) to (33.center);
		\draw (32.center) to (35.center);
	\end{pgfonlayer}
\end{tikzpicture}}}
\label{tsu4brane}
\end{equation}
This diagram can be made clearer by pulling the fourth 7-brane from the left all the way to the right. The process involves several brane creations and annihilations \cite{Hanany:1996ie}. As a result, $(1,-1)$ branes become NS5 branes after passing through the monodromy cuts originating from the 7-branes:
\begin{equation}
   \raisebox{-.5\height}{\scalebox{.789}{
    \begin{tikzpicture}
	\begin{pgfonlayer}{nodelayer}
		\node [style=gauge1] (0) at (-1, -0.05) {};
		\node [style=gauge1] (1) at (0.5, -0.05) {};
		\node [style=gauge1] (2) at (2, -0.05) {};
		\node [style=gauge1] (3) at (8, -0.075) {};
		\node [style=gauge1] (4) at (2.75, 1) {};
		\node [style=gauge1] (5) at (2.75, -1) {};
		\node [style=gauge1] (6) at (4.25, -1) {};
		\node [style=gauge1] (7) at (5.75, -1) {};
		\node [style=gauge1] (8) at (7.25, -1) {};
		\node [style=gauge1] (9) at (4.25, 1) {};
		\node [style=gauge1] (10) at (5.75, 1) {};
		\node [style=gauge1] (11) at (7.25, 1) {};
		\node [style=none] (13) at (2.75, -0.25) {};
		\node [style=none] (14) at (4.25, -0.25) {};
		\node [style=none] (15) at (5.75, -0.25) {};
		\node [style=none] (16) at (7.25, 0) {};
		\node [style=none] (17) at (0.5, 0.075) {};
		\node [style=none] (18) at (0.5, -0.15) {};
		\node [style=none] (19) at (2, 0.075) {};
		\node [style=none] (20) at (2, -0.15) {};
		\node [style=none] (21) at (2, 0.075) {};
		\node [style=none] (22) at (2, -0.05) {};
		\node [style=none] (23) at (2, -0.175) {};
		\node [style=none] (24) at (8, 0.05) {};
		\node [style=none] (25) at (8, -0.075) {};
		\node [style=none] (26) at (8, -0.2) {};
		\node [style=none] (27) at (7.25, 0.025) {};
		\node [style=none] (28) at (8, 0.225) {};
		\node [style=none] (29) at (8, 0.1) {};
		\node [style=none] (30) at (8, -0.025) {};
		\node [style=none] (31) at (8, -0.15) {};
		\node [style=none] (32) at (8, -0.275) {};
		\node [style=none] (33) at (5.75, -0.075) {};
		\node [style=none] (34) at (4.25, -0.175) {};
		\node [style=none] (35) at (2.75, -0.275) {};
	\end{pgfonlayer}
	\begin{pgfonlayer}{edgelayer}
		\draw (4) to (13.center);
		\draw (13.center) to (5);
		\draw (9) to (14.center);
		\draw (14.center) to (6);
		\draw (10) to (15.center);
		\draw (15.center) to (7);
		\draw (11) to (16.center);
		\draw (16.center) to (8);
		\draw (0) to (1);
		\draw (17.center) to (19.center);
		\draw (18.center) to (20.center);
		\draw (21.center) to (24.center);
		\draw (22.center) to (25.center);
		\draw (23.center) to (26.center);
	\end{pgfonlayer}
\end{tikzpicture}}}
\label{tsu4braneX}
\end{equation}
The Higgs branch moduli correspond to brane segments moving in the $x^7,x^8,x^9$ directions. 

\subsubsection*{Getting the magnetic quiver}
Now, we employ the method pioneered in \cite{Cabrera:2018jxt} to arrive at the \emph{magnetic quiver}. The magnetic quiver is a quiver whose 3d $\mathcal{N}=4$ Coulomb branch is equal to the Higgs branch of theory (\ref{tsu4SU}), the original 5d $\mathcal{N}=1$ theory\footnote{In general, the Higgs branch of the 5d $\mathcal{N}=1$ theory can be made of several hyper-K\"ahler cones. In such cases, there will be a magnetic quiver for each of the cones. However, for the examples in this section, the Higgs branch of the 5d theory is only a single cone and therefore there is only one magnetic quiver. We will discuss Higgs branch of several cones in later sections for \emph{bad} quivers.}. The process involves doing a maximal decomposition of the web  into subwebs that move freely apart from each other. By looking at the intersection numbers, we can construct the magnetic quiver.

To demonstrate maximal decomposition, we color the different subwebs that are free to move with respect to each together:
\begin{equation}
   \raisebox{-.5\height}{\scalebox{.789}{
      \begin{tikzpicture}
	\begin{pgfonlayer}{nodelayer}
		\node [style=gauge1] (0) at (-1, -0.05) {};
		\node [style=gauge1] (1) at (0.5, -0.05) {};
		\node [style=gauge1] (2) at (2, -0.05) {};
		\node [style=gauge1] (3) at (8, -0.075) {};
		\node [style=gauge1] (4) at (2.75, 1) {};
		\node [style=gauge1] (5) at (2.75, -1) {};
		\node [style=gauge1] (6) at (4.25, -1) {};
		\node [style=gauge1] (7) at (5.75, -1) {};
		\node [style=gauge1] (8) at (7.25, -1) {};
		\node [style=gauge1] (9) at (4.25, 1) {};
		\node [style=gauge1] (10) at (5.75, 1) {};
		\node [style=gauge1] (11) at (7.25, 1) {};
		\node [style=none] (17) at (0.5, 0.075) {};
		\node [style=none] (18) at (0.5, -0.15) {};
		\node [style=none] (19) at (2, 0.075) {};
		\node [style=none] (20) at (2, -0.15) {};
		\node [style=none] (21) at (2, 0.075) {};
		\node [style=none] (22) at (2, -0.05) {};
		\node [style=none] (23) at (2, -0.175) {};
		\node [style=none] (24) at (8, 0.05) {};
		\node [style=none] (25) at (8, -0.075) {};
		\node [style=none] (26) at (8, -0.2) {};
		\node [style=none] (28) at (8, 0.225) {};
		\node [style=none] (29) at (8, 0.1) {};
		\node [style=none] (30) at (8, -0.025) {};
		\node [style=none] (31) at (8, -0.15) {};
		\node [style=none] (32) at (8, -0.275) {};
	\end{pgfonlayer}
	\begin{pgfonlayer}{edgelayer}
		\draw [style=gray,line width=0.5mm](0) to (1);
		\draw [style=magentae] (17.center) to (19.center);
		\draw [style=magentae] (18.center) to (20.center);
		\draw [style=orange,line width=0.3mm] (21.center) to (24.center);
		\draw [style=orange,line width=0.3mm](22.center) to (25.center);
		\draw [style=orange,line width=0.3mm](23.center) to (26.center);
		\draw [style=bluee] (4) to (5);
		\draw [style=cyaneX] (9) to (6);
		\draw [style=rede] (10) to (7);
		\draw  [style=olive,line width=0.5mm](11) to (8);
	\end{pgfonlayer}
\end{tikzpicture}}}
\label{firstweb}
\end{equation}
We now read off the magnetic quiver where each subweb represents a unitary gauge node with rank given by the number of coincident branes. The multiplicity of hypermultiplets between the gauge nodes are then given by the intersection number between each pair of subwebs. For the current example, each pair of subwebs intersect at most once so the edges in the quiver have multiplicity at most one. The resulting magnetic quiver is:
\begin{equation}
\raisebox{-.5\height}{\scalebox{.8}{\begin{tikzpicture}
	\begin{pgfonlayer}{nodelayer}
		\node [style=gauge1] (0) at (3.5, 0) {};
		\node [style=gauge1] (1) at (2, 0) {};
		\node [style=gauge1] (2) at (0.5, 0) {};
		\node [style=none] (3) at (3.5, -0.575) {3};
		\node [style=none] (4) at (2, -0.575) {2};
		\node [style=none] (5) at (0.5, -0.575) {1};
		\node [style=gauge1] (6) at (2.75, 1.25) {};
		\node [style=gauge1] (7) at (4.25, 1.5) {};
		\node [style=gauge1] (8) at (5.25, 0.5) {};
		\node [style=gauge1] (9) at (5, -0.75) {};
		\node [style=none] (10) at (2.5, 1.75) {1};
		\node [style=none] (11) at (4.5, 2) {1};
		\node [style=none] (12) at (6, 0.75) {1};
		\node [style=none] (13) at (5.6, -1.35) {1};
		\node [style=grayet] (14) at (0.5, 0) {};
		\node [style=pinklinet] (15) at (2, 0) {};
		\node [style=orangeet] (16) at (3.5, 0) {};
		\node [style=blueet] (17) at (2.75, 1.25) {};
		\node [style=cyane] (18) at (4.25, 1.5) {};
		\node [style=magicmintlinet] (19) at (5.25, 0.5) {};
		\node [style=brownlinet] (20) at (5, -0.75) {};
	\end{pgfonlayer}
	\begin{pgfonlayer}{edgelayer}
		\draw (0) to (1);
		\draw (1) to (2);
		\draw (6) to (0);
		\draw (7) to (0);
		\draw (8) to (0);
		\draw (9) to (0);
	\end{pgfonlayer}
\end{tikzpicture}}} 
\label{mirrorTSU4F} 
\end{equation}
where the colored nodes correspond to the different subwebs in \eqref{firstweb}. Here and throughout the paper, magnetic quivers are denoted with only the ranks of the gauge groups on each node, following the convention of \cite{Cabrera:2018jxt}; it is understood that every node is unitary, and that a global $\mathrm{U}(1)$ should be ungauged -- the choice of this $\mathrm{U}(1)$ does not affect the Coulomb branch, nor the Higgs branch. 
\eqref{mirrorTSU4F} reproduces results computed in \cite{Dancer:2020wll,Collinucci:2020kdm}. 

Now, \eqref{mirrorTSU4F} is only the magnetic quiver for \eqref{tsu4SU} when the gauge couplings of the special unitary gauge groups are all finite. Crucially, at finite gauge coupling, the Higgs branch of the 5d quiver is classical and thus the same in $3-6$ dimensions. This allows us to establish the following relationship:
\begin{equation}
      \mathcal{H}_{\mathrm{classical}}^{5d}\eqref{tsu4SU} =  \mathcal{H}^{3d}\eqref{tsu4SU}= \mathcal{C}^{3d}\eqref{mirrorTSU4F}  
      \label{5d3drelation}
\end{equation}
The right equality of \eqref{5d3drelation} is now an equality amongst 3d $\mathcal{N}=4$ theories. This motivates us to conjecture that the two quivers form a 3d mirror pair:
\begin{equation}
 \mathcal{C}^{3d}\eqref{tsu4SU}= \mathcal{H}^{3d}\eqref{mirrorTSU4F}  
      \label{5d3drelation1}
\end{equation}
This is checked explicitly through Hilbert series computations in Table \ref{tsu4table}.

\subsubsection*{Changing $\mathrm{SU}(3)$ to $\mathrm{U}(3)$}
Now, consider gauging the $\mathrm{SU}(3)$ node to a $\mathrm{U}(3)$:
\begin{equation}
  \raisebox{-.5\height}{\scalebox{.8}{  \begin{tikzpicture}
	\begin{pgfonlayer}{nodelayer}
		\node [style=gauge1] (0) at (-1, 0) {};
		\node [style=gauge1] (1) at (0.5, 0) {};
		\node [style=gauge1] (2) at (2, 0) {};
		\node [style=none] (4) at (-1, -0.575) {U(3)};
		\node [style=none] (5) at (0.5, -0.575) {SU(2)};
		\node [style=none] (6) at (2, -0.575) {SU(1)};
		\node [style=none] (7) at (-2.5, -0.575) {4};
		\node [style=flavour1] (8) at (-2.5, 0) {};
	\end{pgfonlayer}
	\begin{pgfonlayer}{edgelayer}
		\draw (0) to (1);
		\draw (1) to (2);
		\draw (0) to (8);
	\end{pgfonlayer}
\end{tikzpicture}}}
\label{tsu4USS}
\end{equation}
The Higgs branch of this quiver is now a subspace of the Higgs branch of (\ref{tsu4SU}). To demonstrate this gauging process in the brane set up, we introduce the notion of \emph{locking}. Two subwebs are locked if they are forced to move together. The gauging of $\mathrm{SU}(3)$ to $\mathrm{U}(3)$ then translates to locking the two left most NS5 branes (colored in blue). On the other hand, the two remaining NS5 branes (in red and olive) are still free to move independently. 
\begin{equation}
   \raisebox{-.5\height}{\scalebox{.789}{
\begin{tikzpicture}
	\begin{pgfonlayer}{nodelayer}
		\node [style=gauge1] (0) at (-1, -0.05) {};
		\node [style=gauge1] (1) at (0.5, -0.05) {};
		\node [style=gauge1] (2) at (2, -0.05) {};
		\node [style=gauge1] (3) at (8, -0.075) {};
		\node [style=gauge1] (4) at (2.75, 1) {};
		\node [style=gauge1] (5) at (2.75, -1) {};
		\node [style=gauge1] (6) at (4.25, -1) {};
		\node [style=gauge1] (7) at (5.75, -1) {};
		\node [style=gauge1] (8) at (7.25, -1) {};
		\node [style=gauge1] (9) at (4.25, 1) {};
		\node [style=gauge1] (10) at (5.75, 1) {};
		\node [style=gauge1] (11) at (7.25, 1) {};
		\node [style=none] (17) at (0.5, 0.075) {};
		\node [style=none] (18) at (0.5, -0.15) {};
		\node [style=none] (19) at (2, 0.075) {};
		\node [style=none] (20) at (2, -0.15) {};
		\node [style=none] (21) at (2, 0.075) {};
		\node [style=none] (22) at (2, -0.05) {};
		\node [style=none] (23) at (2, -0.175) {};
		\node [style=none] (24) at (8, 0.05) {};
		\node [style=none] (25) at (8, -0.075) {};
		\node [style=none] (26) at (8, -0.2) {};
		\node [style=none] (28) at (8, 0.225) {};
		\node [style=none] (29) at (8, 0.1) {};
		\node [style=none] (30) at (8, -0.025) {};
		\node [style=none] (31) at (8, -0.15) {};
		\node [style=none] (32) at (8, -0.275) {};
	\end{pgfonlayer}
	\begin{pgfonlayer}{edgelayer}
		\draw [style=gray,line width=0.5mm](0) to (1);
		\draw [style=magentae] (17.center) to (19.center);
		\draw [style=magentae] (18.center) to (20.center);
		\draw [style=orange,line width=0.3mm] (21.center) to (24.center);
		\draw [style=orange,line width=0.3mm](22.center) to (25.center);
		\draw [style=orange,line width=0.3mm](23.center) to (26.center);
		\draw [style=bluee] (4) to (5);
		\draw [style=bluee] (6) to (9);
		\draw [style=rede] (10) to (7);
		\draw  [style=olive,line width=0.5mm](11) to (8);
	\end{pgfonlayer}
\end{tikzpicture}}}
\label{tsu4three}
\end{equation}
Physically, the explanation for locking is that separating the NS5 branes in the $x^7,x^8,x^9$ direction corresponds to moving on the baryonic branches of the 5d theory; as a consequence, preventing the NS5 from moving apart is equivalent to removing one of the baryonic branches, removing a baryonic U$(1)$ global symmetry by gauging it, therefore producing the Higgs branch of a theory with a unitary gauge group. Indeed, the Higgs branches of an $\mathrm{SU}(3)$ gauge theory and the Higgs branch of a $\mathrm{U}(3)$ gauge theory share the same mesonic branch, and the only difference is that the $\mathrm{SU}(3)$ theory has a baryonic branch in addition. We can read off the magnetic quiver from \eqref{tsu4three}:
\begin{equation}
\raisebox{-.5\height}{\scalebox{.8}{\begin{tikzpicture}
	\begin{pgfonlayer}{nodelayer}
		\node [style=gauge1] (0) at (-0.1, 0) {};
		\node [style=gauge1] (1) at (-1.6, 0) {};
		\node [style=gauge1] (2) at (-3.1, 0) {};
		\node [style=none] (4) at (-0.1, -0.575) {3};
		\node [style=none] (5) at (-1.6, -0.575) {2};
		\node [style=none] (6) at (-3.1, -0.575) {1};
		\node [style=gauge1] (9) at (0.65, 1.25) {};
		\node [style=gauge1] (10) at (1.4, 0.5) {};
		\node [style=none] (13) at (0.9, 1.75) {1};
		\node [style=none] (14) at (2, 0.9) {1};
		\node [style=none] (16) at (-0.85, 1.75) {1};
		\node [style=gauge1] (17) at (-0.85, 1.25) {};
		\node [style=none] (18) at (-0.725, 1.3) {};
		\node [style=none] (19) at (-0.95, 1.225) {};
		\node [style=none] (20) at (-0.05, 0.05) {};
		\node [style=none] (21) at (-0.275, -0.025) {};
		\node [style=grayet] (22) at (-3.1, 0) {};
		\node [style=pinklinet] (23) at (-1.6, 0) {};
		\node [style=blueet] (24) at (-0.85, 1.25) {};
		\node [style=magicmintlinet] (25) at (0.65, 1.25) {};
		\node [style=brownlinet] (26) at (1.4, 0.5) {};
		\node [style=orangeet] (27) at (-0.1, 0) {};
	\end{pgfonlayer}
	\begin{pgfonlayer}{edgelayer}
		\draw (0) to (1);
		\draw (1) to (2);
		\draw (9) to (0);
		\draw (10) to (0);
		\draw (19.center) to (21.center);
		\draw (20.center) to (18.center);
	\end{pgfonlayer}
\end{tikzpicture}}}
\label{tsu4ussmirror}
\end{equation}
One of the $\mathrm{U}(1)$ nodes is now connected to the $\mathrm{U}(3)$ node with an edge of multiplicity two, which simply means there are twice as many hypermultiplets. This is because the intersection number between the blue subweb and orange subweb is two. Edge multiplicity naturally arises when studying magnetic quivers \cite{Cabrera:2018jxt, Bourget:2019rtl}. Once again, the Coulomb branch and Higgs branch computation shows that, at least on the level of Hilbert series, \eqref{tsu4three} and \eqref{tsu4ussmirror} are indeed mirror pairs.  

\subsubsection*{Changing $\mathrm{SU}(3)$ to $\mathrm{U}(3)$ and $\mathrm{SU}(1)$ to $\mathrm{U}(1)$}
Let's see what happens if we gauge the baryonic U$(1)$s associated to both the $\mathrm{SU}(3)$ and $\mathrm{SU}(1)$ nodes:
\begin{equation}
   \raisebox{-.5\height}{\scalebox{.8}{ \begin{tikzpicture}
	\begin{pgfonlayer}{nodelayer}
		\node [style=gauge1] (0) at (-1, 0) {};
		\node [style=gauge1] (1) at (0.5, 0) {};
		\node [style=gauge1] (2) at (2, 0) {};
		\node [style=none] (4) at (-1, -0.575) {U(3)};
		\node [style=none] (5) at (0.5, -0.575) {SU(2)};
		\node [style=none] (6) at (2, -0.575) {U(1)};
		\node [style=none] (7) at (-2.5, -0.575) {4};
		\node [style=flavour1] (8) at (-2.5, 0) {};
	\end{pgfonlayer}
	\begin{pgfonlayer}{edgelayer}
		\draw (0) to (1);
		\draw (1) to (2);
		\draw (0) to (8);
	\end{pgfonlayer}
\end{tikzpicture}}}
\label{tsu4USU}
\end{equation}

In the brane set up, this is equivalent to locking both the pair of branes on the left (blue) and on the right (green):
\begin{equation}
   \raisebox{-.5\height}{\scalebox{.789}{
\begin{tikzpicture}
	\begin{pgfonlayer}{nodelayer}
		\node [style=gauge1] (0) at (-1, -0.05) {};
		\node [style=gauge1] (1) at (0.5, -0.05) {};
		\node [style=gauge1] (2) at (2, -0.05) {};
		\node [style=gauge1] (3) at (8, -0.075) {};
		\node [style=gauge1] (4) at (2.75, 1) {};
		\node [style=gauge1] (5) at (2.75, -1) {};
		\node [style=gauge1] (6) at (4.25, -1) {};
		\node [style=gauge1] (7) at (5.75, -1) {};
		\node [style=gauge1] (8) at (7.25, -1) {};
		\node [style=gauge1] (9) at (4.25, 1) {};
		\node [style=gauge1] (10) at (5.75, 1) {};
		\node [style=gauge1] (11) at (7.25, 1) {};
		\node [style=none] (17) at (0.5, 0.075) {};
		\node [style=none] (18) at (0.5, -0.15) {};
		\node [style=none] (19) at (2, 0.075) {};
		\node [style=none] (20) at (2, -0.15) {};
		\node [style=none] (21) at (2, 0.075) {};
		\node [style=none] (22) at (2, -0.05) {};
		\node [style=none] (23) at (2, -0.175) {};
		\node [style=none] (24) at (8, 0.05) {};
		\node [style=none] (25) at (8, -0.075) {};
		\node [style=none] (26) at (8, -0.2) {};
		\node [style=none] (28) at (8, 0.225) {};
		\node [style=none] (29) at (8, 0.1) {};
		\node [style=none] (30) at (8, -0.025) {};
		\node [style=none] (31) at (8, -0.15) {};
		\node [style=none] (32) at (8, -0.275) {};
	\end{pgfonlayer}
	\begin{pgfonlayer}{edgelayer}
		\draw [style=gray,line width=0.5mm](0) to (1);
		\draw [style=magentae] (17.center) to (19.center);
		\draw [style=magentae] (18.center) to (20.center);
		\draw [style=orange,line width=0.3mm] (21.center) to (24.center);
		\draw [style=orange,line width=0.3mm](22.center) to (25.center);
		\draw [style=orange,line width=0.3mm](23.center) to (26.center);
		\draw [style=bluee] (4) to (5);
		\draw [style=bluee] (6) to (9);
		\draw [style=olive,line width=0.5mm] (10) to (7);
		\draw  [style=olive,line width=0.5mm](11) to (8);
	\end{pgfonlayer}
\end{tikzpicture}}}
\end{equation}
 Now, amongst the NS5 branes, there are only two pieces (rather than three in (\ref{tsu4three})) that move independently. Looking at the intersection numbers, one quickly obtains the magnetic quiver:
\begin{equation}
\raisebox{-.5\height}{\scalebox{.8}{
\begin{tikzpicture}
	\begin{pgfonlayer}{nodelayer}
		\node [style=gauge1] (0) at (0.875, 0) {};
		\node [style=gauge1] (1) at (-0.625, 0) {};
		\node [style=gauge1] (2) at (-2.125, 0) {};
		\node [style=none] (4) at (0.875, -0.575) {3};
		\node [style=none] (5) at (-0.625, -0.575) {2};
		\node [style=none] (6) at (-2.125, -0.575) {1};
		\node [style=gauge1] (10) at (1.9, 1.225) {};
		\node [style=none] (14) at (1.875, 1.775) {1};
		\node [style=none] (17) at (2, 1.125) {};
		\node [style=none] (18) at (1.8, 1.3) {};
		\node [style=none] (19) at (1, -0.125) {};
		\node [style=none] (20) at (0.8, 0.05) {};
		\node [style=gauge1] (23) at (0.05, 1.3) {};
		\node [style=none] (24) at (0, 1.8) {1};
		\node [style=none] (25) at (0.175, 1.3) {};
		\node [style=none] (26) at (-0.1, 1.275) {};
		\node [style=none] (27) at (0.95, 0.075) {};
		\node [style=none] (28) at (0.7, -0.05) {};
		\node [style=grayet] (29) at (-2.125, 0) {};
		\node [style=pinklinet] (30) at (-0.625, 0) {};
		\node [style=orangeet] (31) at (0.875, 0) {};
		\node [style=blueet] (32) at (0.05, 1.3) {};
		\node [style=brownlinet] (33) at (1.9, 1.225) {};
	\end{pgfonlayer}
	\begin{pgfonlayer}{edgelayer}
		\draw (0) to (1);
		\draw (1) to (2);
		\draw (18.center) to (20.center);
		\draw (17.center) to (19.center);
		\draw (26.center) to (28.center);
		\draw (25.center) to (27.center);
	\end{pgfonlayer}
\end{tikzpicture}}}
\end{equation}
where the edges of both $\mathrm{U}(1)$s connecting the $\mathrm{U}(3)$  have multiplicity two. The unrefined Coulomb and Higgs branch Hilbert series are presented in Table \ref{tsu4table} and is consistent with the conjecture that they are 3d mirror pairs. 

\subsubsection*{Summary}

We summarize all the different combinations of U / SU nodes for $T(SU(4))$ theories in Table \ref{tsu4table1}, along with their unrefined Coulomb and Higgs branch Hilbert series in Table \ref{tsu4table} and global symmetries in Table \ref{tsu4tableGlobal}. As is clear from the magnetic quivers in Table \ref{tsu4table1} and 3d mirror symmetry, the Higgs and Coulomb branches of these quivers only depend on the partition of $4$ which defines the locking (see the colored cells in the second column). For each partition, we compute the Higgs branch Hasse diagram using the quiver subtraction algorithm \cite{Bourget:2019aer}. The results are shown in Table \ref{tab:tsu4HasseDiagrams}. 
This illustrates a general lesson: when unitary gauge groups are changed to special unitary gauge groups, the Higgs branch Hasse diagram gets more complicated.

\begin{table}[p]
    \centering
  \hspace*{-.5cm}\scalebox{0.77}{
  \begin{tabular}{|c|c|c|}
    \hline
    Electric Quiver & Magnetic Quiver & Brane web \\ \hline
\raisebox{-.5\height}{\begin{tikzpicture}
	\begin{pgfonlayer}{nodelayer}
		\node [style=gauge1] (0) at (-1, 0) {};
		\node [style=gauge1] (1) at (0.5, 0) {};
		\node [style=gauge1] (2) at (2, 0) {};
		\node [style=none] (4) at (-1, -0.575) {U(3)};
		\node [style=none] (5) at (0.5, -0.575) {U(2)};
		\node [style=none] (6) at (2, -0.575) {U(1)};
		\node [style=flavour1] (12) at (-2.5, 0) {};
		\node [style=none] (13) at (-2.5, -0.575) {4};
	\end{pgfonlayer}
	\begin{pgfonlayer}{edgelayer}
		\draw (0) to (1);
		\draw (1) to (2);
		\draw (0) to (12);
	\end{pgfonlayer}
\end{tikzpicture}}
 & \cellcolor{colors01}\raisebox{-.5\height}{\begin{tikzpicture}
	\begin{pgfonlayer}{nodelayer}
		\node [style=gauge1] (0) at (-1, 0) {};
		\node [style=none] (4) at (-1, -0.575) {1};
		\node [style=none] (7) at (-2.5, -0.575) {3};
		\node [style=none] (8) at (-1, 0.175) {};
		\node [style=none] (9) at (-1, 0.05) {};
		\node [style=none] (10) at (-1, -0.075) {};
		\node [style=none] (11) at (-1, -0.2) {};
		\node [style=none] (12) at (-2.5, 0.175) {};
		\node [style=none] (13) at (-2.5, 0.05) {};
		\node [style=none] (14) at (-2.5, -0.075) {};
		\node [style=none] (15) at (-2.5, -0.2) {};
		\node [style=gauge1] (16) at (-2.5, 0) {};
		\node [style=none] (19) at (-4, -0.575) {2};
		\node [style=none] (20) at (-5.5, -0.55) {1};
		\node [style=gauge1] (21) at (-4, 0) {};
		\node [style=gauge1] (22) at (-5.5, 0) {};
	\end{pgfonlayer}
	\begin{pgfonlayer}{edgelayer}
		\draw (12.center) to (8.center);
		\draw (9.center) to (13.center);
		\draw (14.center) to (10.center);
		\draw (11.center) to (15.center);
		\draw (21) to (16);
		\draw (21) to (22);
	\end{pgfonlayer}
\end{tikzpicture}}
&   \raisebox{-.5\height}{ \scalebox{.789}{   \begin{tikzpicture}
	\begin{pgfonlayer}{nodelayer}
		\node [style=gauge1] (0) at (-1, -0.05) {};
		\node [style=gauge1] (1) at (0.5, -0.05) {};
		\node [style=gauge1] (2) at (2, -0.05) {};
		\node [style=gauge1] (3) at (8, -0.075) {};
		\node [style=gauge1] (4) at (2.75, 1) {};
		\node [style=gauge1] (5) at (2.75, -1) {};
		\node [style=gauge1] (6) at (4.25, -1) {};
		\node [style=gauge1] (7) at (5.75, -1) {};
		\node [style=gauge1] (8) at (7.25, -1) {};
		\node [style=gauge1] (9) at (4.25, 1) {};
		\node [style=gauge1] (10) at (5.75, 1) {};
		\node [style=gauge1] (11) at (7.25, 1) {};
		\node [style=none] (17) at (0.5, 0.075) {};
		\node [style=none] (18) at (0.5, -0.15) {};
		\node [style=none] (19) at (2, 0.075) {};
		\node [style=none] (20) at (2, -0.15) {};
		\node [style=none] (21) at (2, 0.075) {};
		\node [style=none] (22) at (2, -0.05) {};
		\node [style=none] (23) at (2, -0.175) {};
		\node [style=none] (24) at (8, 0.05) {};
		\node [style=none] (25) at (8, -0.075) {};
		\node [style=none] (26) at (8, -0.2) {};
		\node [style=none] (28) at (8, 0.225) {};
		\node [style=none] (29) at (8, 0.1) {};
		\node [style=none] (30) at (8, -0.025) {};
		\node [style=none] (31) at (8, -0.15) {};
		\node [style=none] (32) at (8, -0.275) {};
		\node [style=none] (33) at (8, -1.5) {};
		\node [style=none] (34) at (8, 1.5) {};
	\end{pgfonlayer}
	\begin{pgfonlayer}{edgelayer}
		\draw [style=gray,line width=0.5mm](0) to (1);
		\draw [style=magentae] (17.center) to (19.center);
		\draw [style=magentae] (18.center) to (20.center);
		\draw [style=orange,line width=0.3mm] (21.center) to (24.center);
		\draw [style=orange,line width=0.3mm](22.center) to (25.center);
		\draw [style=orange,line width=0.3mm](23.center) to (26.center);
		\draw [style=bluee] (4) to (5);
		\draw [style=bluee] (9) to (6);
		\draw [style=bluee] (10) to (7);
		\draw  [style=bluee](11) to (8);
	\end{pgfonlayer}
\end{tikzpicture}}}
 \\ \hline
\raisebox{-.5\height}{\begin{tikzpicture}
	\begin{pgfonlayer}{nodelayer}
		\node [style=gauge1] (0) at (-1, 0) {};
		\node [style=gauge1] (1) at (0.5, 0) {};
		\node [style=gauge1] (2) at (2, 0) {};
		\node [style=none] (4) at (-1, -0.575) {SU(3)};
		\node [style=none] (5) at (0.5, -0.575) {U(2)};
		\node [style=none] (6) at (2, -0.575) {U(1)};
		\node [style=flavour1] (12) at (-2.5, 0) {};
		\node [style=none] (13) at (-2.5, -0.575) {4};
	\end{pgfonlayer}
	\begin{pgfonlayer}{edgelayer}
		\draw (0) to (1);
		\draw (1) to (2);
		\draw (0) to (12);
	\end{pgfonlayer}
\end{tikzpicture}}
& \cellcolor{colors02} \raisebox{-.5\height}{\begin{tikzpicture}
	\begin{pgfonlayer}{nodelayer}
		\node [style=gauge1] (0) at (-1, 0) {};
		\node [style=none] (4) at (-1, -0.575) {1};
		\node [style=none] (7) at (-2.5, -0.575) {3};
		\node [style=none] (8) at (-1, 0.25) {};
		\node [style=none] (9) at (-1, 0.125) {};
		\node [style=none] (10) at (-1, 0) {};
		\node [style=none] (11) at (-1, -0.125) {};
		\node [style=none] (12) at (-2.5, 0.25) {};
		\node [style=none] (13) at (-2.5, 0.125) {};
		\node [style=none] (14) at (-2.6, 0) {};
		\node [style=none] (15) at (-2.6, -0.125) {};
		\node [style=gauge1] (16) at (-2.5, 0) {};
		\node [style=none] (19) at (-4, -0.575) {2};
		\node [style=none] (20) at (-5.5, -0.55) {1};
		\node [style=gauge1] (21) at (-4, 0) {};
		\node [style=gauge1] (22) at (-5.5, 0) {};
		\node [style=gauge1] (23) at (-1, 1.25) {};
		\node [style=none] (24) at (-1, 1.75) {1};
	\end{pgfonlayer}
	\begin{pgfonlayer}{edgelayer}
		\draw (9.center) to (13.center);
		\draw (14.center) to (10.center);
		\draw (11.center) to (15.center);
		\draw (21) to (16);
		\draw (21) to (22);
		\draw (23) to (16);
	\end{pgfonlayer}
\end{tikzpicture}}
&    \raisebox{-.5\height}{\scalebox{.789}{   \begin{tikzpicture}
	\begin{pgfonlayer}{nodelayer}
		\node [style=gauge1] (0) at (-1, -0.05) {};
		\node [style=gauge1] (1) at (0.5, -0.05) {};
		\node [style=gauge1] (2) at (2, -0.05) {};
		\node [style=gauge1] (3) at (8, -0.075) {};
		\node [style=gauge1] (4) at (2.75, 1) {};
		\node [style=gauge1] (5) at (2.75, -1) {};
		\node [style=gauge1] (6) at (4.25, -1) {};
		\node [style=gauge1] (7) at (5.75, -1) {};
		\node [style=gauge1] (8) at (7.25, -1) {};
		\node [style=gauge1] (9) at (4.25, 1) {};
		\node [style=gauge1] (10) at (5.75, 1) {};
		\node [style=gauge1] (11) at (7.25, 1) {};
		\node [style=none] (17) at (0.5, 0.075) {};
		\node [style=none] (18) at (0.5, -0.15) {};
		\node [style=none] (19) at (2, 0.075) {};
		\node [style=none] (20) at (2, -0.15) {};
		\node [style=none] (21) at (2, 0.075) {};
		\node [style=none] (22) at (2, -0.05) {};
		\node [style=none] (23) at (2, -0.175) {};
		\node [style=none] (24) at (8, 0.05) {};
		\node [style=none] (25) at (8, -0.075) {};
		\node [style=none] (26) at (8, -0.2) {};
		\node [style=none] (28) at (8, 0.225) {};
		\node [style=none] (29) at (8, 0.1) {};
		\node [style=none] (30) at (8, -0.025) {};
		\node [style=none] (31) at (8, -0.15) {};
		\node [style=none] (32) at (8, -0.275) {};
	\end{pgfonlayer}
	\begin{pgfonlayer}{edgelayer}
		\draw [style=gray,line width=0.5mm](0) to (1);
		\draw [style=magentae] (17.center) to (19.center);
		\draw [style=magentae] (18.center) to (20.center);
		\draw [style=orange,line width=0.3mm] (21.center) to (24.center);
		\draw [style=orange,line width=0.3mm](22.center) to (25.center);
		\draw [style=orange,line width=0.3mm](23.center) to (26.center);
		\draw [style=bluee] (4) to (5);
		\draw [style=olive,line width=0.5mm] (9) to (6);
		\draw[style=olive,line width=0.5mm] (10) to (7);
		\draw  [style=olive,line width=0.5mm](11) to (8);
	\end{pgfonlayer}
\end{tikzpicture}}}
\\ \hline
\raisebox{-.5\height}{\begin{tikzpicture}
	\begin{pgfonlayer}{nodelayer}
		\node [style=gauge1] (0) at (-1, 0) {};
		\node [style=gauge1] (1) at (0.5, 0) {};
		\node [style=gauge1] (2) at (2, 0) {};
		\node [style=none] (4) at (-1, -0.575) {U(3)};
		\node [style=none] (5) at (0.5, -0.575) {U(2)};
		\node [style=none] (6) at (2, -0.575) {SU(1)};
		\node [style=flavour1] (12) at (-2.5, 0) {};
		\node [style=none] (13) at (-2.5, -0.575) {4};
	\end{pgfonlayer}
	\begin{pgfonlayer}{edgelayer}
		\draw (0) to (1);
		\draw (1) to (2);
		\draw (0) to (12);
	\end{pgfonlayer}
\end{tikzpicture}}

 & \cellcolor{colors02} \raisebox{-.5\height}{ \begin{tikzpicture}
	\begin{pgfonlayer}{nodelayer}
		\node [style=gauge1] (0) at (-1, 0) {};
		\node [style=none] (4) at (-1, -0.575) {1};
		\node [style=none] (7) at (-2.5, -0.575) {3};
		\node [style=gauge1] (16) at (-2.5, 0) {};
		\node [style=none] (19) at (-4, -0.575) {2};
		\node [style=none] (20) at (-5.5, -0.55) {1};
		\node [style=gauge1] (21) at (-4, 0) {};
		\node [style=gauge1] (22) at (-5.5, 0) {};
		\node [style=gauge1] (23) at (-1, 1.25) {};
		\node [style=none] (24) at (-1, 1.75) {1};
		\node [style=none] (25) at (-2.575, 0.15) {};
		\node [style=none] (26) at (-2.475, 0.025) {};
		\node [style=none] (27) at (-1.125, 1.35) {};
		\node [style=none] (28) at (-1.025, 1.225) {};
		\node [style=none] (29) at (-0.875, 1.125) {};
		\node [style=none] (30) at (-2.425, -0.125) {};
	\end{pgfonlayer}
	\begin{pgfonlayer}{edgelayer}
		\draw (21) to (16);
		\draw (21) to (22);
		\draw (25.center) to (27.center);
		\draw (26.center) to (28.center);
		\draw (29.center) to (30.center);
		\draw (0) to (16);
	\end{pgfonlayer}
\end{tikzpicture}}

& \raisebox{-.5\height}{ \scalebox{.789}{     \begin{tikzpicture}
	\begin{pgfonlayer}{nodelayer}
		\node [style=gauge1] (0) at (-1, -0.05) {};
		\node [style=gauge1] (1) at (0.5, -0.05) {};
		\node [style=gauge1] (2) at (2, -0.05) {};
		\node [style=gauge1] (3) at (8, -0.075) {};
		\node [style=gauge1] (4) at (2.75, 1) {};
		\node [style=gauge1] (5) at (2.75, -1) {};
		\node [style=gauge1] (6) at (4.25, -1) {};
		\node [style=gauge1] (7) at (5.75, -1) {};
		\node [style=gauge1] (8) at (7.25, -1) {};
		\node [style=gauge1] (9) at (4.25, 1) {};
		\node [style=gauge1] (10) at (5.75, 1) {};
		\node [style=gauge1] (11) at (7.25, 1) {};
		\node [style=none] (17) at (0.5, 0.075) {};
		\node [style=none] (18) at (0.5, -0.15) {};
		\node [style=none] (19) at (2, 0.075) {};
		\node [style=none] (20) at (2, -0.15) {};
		\node [style=none] (21) at (2, 0.075) {};
		\node [style=none] (22) at (2, -0.05) {};
		\node [style=none] (23) at (2, -0.175) {};
		\node [style=none] (24) at (8, 0.05) {};
		\node [style=none] (25) at (8, -0.075) {};
		\node [style=none] (26) at (8, -0.2) {};
		\node [style=none] (28) at (8, 0.225) {};
		\node [style=none] (29) at (8, 0.1) {};
		\node [style=none] (30) at (8, -0.025) {};
		\node [style=none] (31) at (8, -0.15) {};
		\node [style=none] (32) at (8, -0.275) {};
	\end{pgfonlayer}
	\begin{pgfonlayer}{edgelayer}
		\draw [style=gray,line width=0.5mm](0) to (1);
		\draw [style=magentae] (17.center) to (19.center);
		\draw [style=magentae] (18.center) to (20.center);
		\draw [style=orange,line width=0.3mm] (21.center) to (24.center);
		\draw [style=orange,line width=0.3mm](22.center) to (25.center);
		\draw [style=orange,line width=0.3mm](23.center) to (26.center);
		\draw [style=bluee] (4) to (5);
		\draw [style=bluee] (9) to (6);
		\draw [style=bluee] (10) to (7);
		\draw  [style=olive,line width=0.5mm](11) to (8);
	\end{pgfonlayer}
\end{tikzpicture} }}
     \\  \hline
\raisebox{-.5\height}{\begin{tikzpicture}
	\begin{pgfonlayer}{nodelayer}
		\node [style=gauge1] (0) at (-1, 0) {};
		\node [style=gauge1] (1) at (0.5, 0) {};
		\node [style=gauge1] (2) at (2, 0) {};
		\node [style=none] (4) at (-1, -0.575) {U(3)};
		\node [style=none] (5) at (0.5, -0.575) {SU(2)};
		\node [style=none] (6) at (2, -0.575) {U(1)};
		\node [style=flavour1] (12) at (-2.5, 0) {};
		\node [style=none] (13) at (-2.5, -0.575) {4};
	\end{pgfonlayer}
	\begin{pgfonlayer}{edgelayer}
		\draw (0) to (1);
		\draw (1) to (2);
		\draw (0) to (12);
	\end{pgfonlayer}
\end{tikzpicture}}
& \cellcolor{colors03} \raisebox{-.5\height}{\begin{tikzpicture}
	\begin{pgfonlayer}{nodelayer}
		\node [style=gauge1] (0) at (-1, 0) {};
		\node [style=none] (4) at (-1, -0.575) {1};
		\node [style=none] (7) at (-2.5, -0.575) {3};
		\node [style=none] (10) at (-1, 0.075) {};
		\node [style=none] (11) at (-1, -0.125) {};
		\node [style=none] (14) at (-2.6, 0.075) {};
		\node [style=none] (15) at (-2.6, -0.125) {};
		\node [style=gauge1] (16) at (-2.5, 0) {};
		\node [style=none] (19) at (-4, -0.575) {2};
		\node [style=none] (20) at (-5.5, -0.55) {1};
		\node [style=gauge1] (21) at (-4, 0) {};
		\node [style=gauge1] (22) at (-5.5, 0) {};
		\node [style=gauge1] (23) at (-1, 1.25) {};
		\node [style=none] (24) at (-1, 1.75) {1};
		\node [style=none] (25) at (-2.525, 0.175) {};
		\node [style=none] (26) at (-2.375, 0) {};
		\node [style=none] (27) at (-1.125, 1.275) {};
		\node [style=none] (28) at (-0.975, 1.1) {};
	\end{pgfonlayer}
	\begin{pgfonlayer}{edgelayer}
		\draw (14.center) to (10.center);
		\draw (11.center) to (15.center);
		\draw (21) to (16);
		\draw (21) to (22);
		\draw (25.center) to (27.center);
		\draw (26.center) to (28.center);
	\end{pgfonlayer}
\end{tikzpicture}}

&    \raisebox{-.5\height}{\scalebox{.789}{   \begin{tikzpicture}
	\begin{pgfonlayer}{nodelayer}
		\node [style=gauge1] (0) at (-1, -0.05) {};
		\node [style=gauge1] (1) at (0.5, -0.05) {};
		\node [style=gauge1] (2) at (2, -0.05) {};
		\node [style=gauge1] (3) at (8, -0.075) {};
		\node [style=gauge1] (4) at (2.75, 1) {};
		\node [style=gauge1] (5) at (2.75, -1) {};
		\node [style=gauge1] (6) at (4.25, -1) {};
		\node [style=gauge1] (7) at (5.75, -1) {};
		\node [style=gauge1] (8) at (7.25, -1) {};
		\node [style=gauge1] (9) at (4.25, 1) {};
		\node [style=gauge1] (10) at (5.75, 1) {};
		\node [style=gauge1] (11) at (7.25, 1) {};
		\node [style=none] (17) at (0.5, 0.075) {};
		\node [style=none] (18) at (0.5, -0.15) {};
		\node [style=none] (19) at (2, 0.075) {};
		\node [style=none] (20) at (2, -0.15) {};
		\node [style=none] (21) at (2, 0.075) {};
		\node [style=none] (22) at (2, -0.05) {};
		\node [style=none] (23) at (2, -0.175) {};
		\node [style=none] (24) at (8, 0.05) {};
		\node [style=none] (25) at (8, -0.075) {};
		\node [style=none] (26) at (8, -0.2) {};
		\node [style=none] (28) at (8, 0.225) {};
		\node [style=none] (29) at (8, 0.1) {};
		\node [style=none] (30) at (8, -0.025) {};
		\node [style=none] (31) at (8, -0.15) {};
		\node [style=none] (32) at (8, -0.275) {};
	\end{pgfonlayer}
	\begin{pgfonlayer}{edgelayer}
		\draw [style=gray,line width=0.5mm](0) to (1);
		\draw [style=magentae] (17.center) to (19.center);
		\draw [style=magentae] (18.center) to (20.center);
		\draw [style=orange,line width=0.3mm] (21.center) to (24.center);
		\draw [style=orange,line width=0.3mm](22.center) to (25.center);
		\draw [style=orange,line width=0.3mm](23.center) to (26.center);
		\draw [style=bluee] (4) to (5);
		\draw [style=bluee] (9) to (6);
		\draw [style=rede] (10) to (7);
		\draw  [style=rede](11) to (8);
	\end{pgfonlayer}
\end{tikzpicture} }}
\\ \hline 
\raisebox{-.5\height}{\begin{tikzpicture}
	\begin{pgfonlayer}{nodelayer}
		\node [style=gauge1] (0) at (-1, 0) {};
		\node [style=gauge1] (1) at (0.5, 0) {};
		\node [style=gauge1] (2) at (2, 0) {};
		\node [style=none] (4) at (-1, -0.575) {SU(3)};
		\node [style=none] (5) at (0.5, -0.575) {SU(2)};
		\node [style=none] (6) at (2, -0.575) {U(1)};
		\node [style=flavour1] (12) at (-2.5, 0) {};
		\node [style=none] (13) at (-2.5, -0.575) {4};
	\end{pgfonlayer}
	\begin{pgfonlayer}{edgelayer}
		\draw (0) to (1);
		\draw (1) to (2);
		\draw (0) to (12);
	\end{pgfonlayer}
\end{tikzpicture}}

 & \cellcolor{colors04} \raisebox{-.5\height}{\begin{tikzpicture}
	\begin{pgfonlayer}{nodelayer}
		\node [style=gauge1] (0) at (-1, 0) {};
		\node [style=none] (4) at (-1, -0.575) {1};
		\node [style=none] (7) at (-2.5, -0.575) {3};
		\node [style=none] (10) at (-1, 0.075) {};
		\node [style=none] (11) at (-1, -0.125) {};
		\node [style=none] (14) at (-2.6, 0.075) {};
		\node [style=none] (15) at (-2.6, -0.125) {};
		\node [style=gauge1] (16) at (-2.5, 0) {};
		\node [style=none] (19) at (-4, -0.575) {2};
		\node [style=none] (20) at (-5.5, -0.55) {1};
		\node [style=gauge1] (21) at (-4, 0) {};
		\node [style=gauge1] (22) at (-5.5, 0) {};
		\node [style=gauge1] (23) at (-1.25, 1) {};
		\node [style=none] (24) at (-0.85, 1.5) {1};
		\node [style=gauge1] (25) at (-2.25, 1.5) {};
		\node [style=none] (26) at (-2.25, 2) {1};
	\end{pgfonlayer}
	\begin{pgfonlayer}{edgelayer}
		\draw (14.center) to (10.center);
		\draw (11.center) to (15.center);
		\draw (21) to (16);
		\draw (21) to (22);
		\draw (25) to (16);
		\draw (23) to (16);
	\end{pgfonlayer}
\end{tikzpicture}}
&    \raisebox{-.5\height}{ \scalebox{.789}{  \begin{tikzpicture}
	\begin{pgfonlayer}{nodelayer}
		\node [style=gauge1] (0) at (-1, -0.05) {};
		\node [style=gauge1] (1) at (0.5, -0.05) {};
		\node [style=gauge1] (2) at (2, -0.05) {};
		\node [style=gauge1] (3) at (8, -0.075) {};
		\node [style=gauge1] (4) at (2.75, 1) {};
		\node [style=gauge1] (5) at (2.75, -1) {};
		\node [style=gauge1] (6) at (4.25, -1) {};
		\node [style=gauge1] (7) at (5.75, -1) {};
		\node [style=gauge1] (8) at (7.25, -1) {};
		\node [style=gauge1] (9) at (4.25, 1) {};
		\node [style=gauge1] (10) at (5.75, 1) {};
		\node [style=gauge1] (11) at (7.25, 1) {};
		\node [style=none] (17) at (0.5, 0.075) {};
		\node [style=none] (18) at (0.5, -0.15) {};
		\node [style=none] (19) at (2, 0.075) {};
		\node [style=none] (20) at (2, -0.15) {};
		\node [style=none] (21) at (2, 0.075) {};
		\node [style=none] (22) at (2, -0.05) {};
		\node [style=none] (23) at (2, -0.175) {};
		\node [style=none] (24) at (8, 0.05) {};
		\node [style=none] (25) at (8, -0.075) {};
		\node [style=none] (26) at (8, -0.2) {};
		\node [style=none] (28) at (8, 0.225) {};
		\node [style=none] (29) at (8, 0.1) {};
		\node [style=none] (30) at (8, -0.025) {};
		\node [style=none] (31) at (8, -0.15) {};
		\node [style=none] (32) at (8, -0.275) {};
	\end{pgfonlayer}
	\begin{pgfonlayer}{edgelayer}
		\draw [style=gray,line width=0.5mm](0) to (1);
		\draw [style=magentae] (17.center) to (19.center);
		\draw [style=magentae] (18.center) to (20.center);
		\draw [style=orange,line width=0.3mm] (21.center) to (24.center);
		\draw [style=orange,line width=0.3mm](22.center) to (25.center);
		\draw [style=orange,line width=0.3mm](23.center) to (26.center);
		\draw [style=bluee] (4) to (5);
		\draw [style=cyaneX] (9) to (6);
		\draw [style=olive,line width=0.5mm] (10) to (7);
		\draw  [style=olive,line width=0.5mm](11) to (8);
	\end{pgfonlayer}
\end{tikzpicture} }}
 \\ \hline
\raisebox{-.5\height}{\begin{tikzpicture}
	\begin{pgfonlayer}{nodelayer}
		\node [style=gauge1] (0) at (-1, 0) {};
		\node [style=gauge1] (1) at (0.5, 0) {};
		\node [style=gauge1] (2) at (2, 0) {};
		\node [style=none] (4) at (-1, -0.575) {SU(3)};
		\node [style=none] (5) at (0.5, -0.575) {U(2)};
		\node [style=none] (6) at (2, -0.575) {SU(1)};
		\node [style=flavour1] (12) at (-2.5, 0) {};
		\node [style=none] (13) at (-2.5, -0.575) {4};
	\end{pgfonlayer}
	\begin{pgfonlayer}{edgelayer}
		\draw (0) to (1);
		\draw (1) to (2);
		\draw (0) to (12);
	\end{pgfonlayer}
\end{tikzpicture}}

 & \cellcolor{colors04} \raisebox{-.5\height}{\begin{tikzpicture}
	\begin{pgfonlayer}{nodelayer}
		\node [style=gauge1] (0) at (-1.25, 1.25) {};
		\node [style=none] (4) at (-1, -0.575) {1};
		\node [style=none] (7) at (-2.5, -0.575) {3};
		\node [style=none] (10) at (-1.275, 1.35) {};
		\node [style=none] (11) at (-1.2, 1.125) {};
		\node [style=none] (14) at (-2.575, 0.025) {};
		\node [style=none] (15) at (-2.475, -0.125) {};
		\node [style=gauge1] (16) at (-2.5, 0) {};
		\node [style=none] (19) at (-4, -0.575) {2};
		\node [style=none] (20) at (-5.5, -0.55) {1};
		\node [style=gauge1] (21) at (-4, 0) {};
		\node [style=gauge1] (22) at (-5.5, 0) {};
		\node [style=gauge1] (23) at (-1, 0) {};
		\node [style=none] (24) at (-0.6, 1.5) {1};
		\node [style=gauge1] (25) at (-2.25, 1.5) {};
		\node [style=none] (26) at (-2.25, 2) {1};
	\end{pgfonlayer}
	\begin{pgfonlayer}{edgelayer}
		\draw (14.center) to (10.center);
		\draw (11.center) to (15.center);
		\draw (21) to (16);
		\draw (21) to (22);
		\draw (25) to (16);
		\draw (23) to (16);
	\end{pgfonlayer}
\end{tikzpicture}}

&    \raisebox{-.5\height}{ \scalebox{.789}{  \begin{tikzpicture}
	\begin{pgfonlayer}{nodelayer}
		\node [style=gauge1] (0) at (-1, -0.05) {};
		\node [style=gauge1] (1) at (0.5, -0.05) {};
		\node [style=gauge1] (2) at (2, -0.05) {};
		\node [style=gauge1] (3) at (8, -0.075) {};
		\node [style=gauge1] (4) at (2.75, 1) {};
		\node [style=gauge1] (5) at (2.75, -1) {};
		\node [style=gauge1] (6) at (4.25, -1) {};
		\node [style=gauge1] (7) at (5.75, -1) {};
		\node [style=gauge1] (8) at (7.25, -1) {};
		\node [style=gauge1] (9) at (4.25, 1) {};
		\node [style=gauge1] (10) at (5.75, 1) {};
		\node [style=gauge1] (11) at (7.25, 1) {};
		\node [style=none] (17) at (0.5, 0.075) {};
		\node [style=none] (18) at (0.5, -0.15) {};
		\node [style=none] (19) at (2, 0.075) {};
		\node [style=none] (20) at (2, -0.15) {};
		\node [style=none] (21) at (2, 0.075) {};
		\node [style=none] (22) at (2, -0.05) {};
		\node [style=none] (23) at (2, -0.175) {};
		\node [style=none] (24) at (8, 0.05) {};
		\node [style=none] (25) at (8, -0.075) {};
		\node [style=none] (26) at (8, -0.2) {};
		\node [style=none] (28) at (8, 0.225) {};
		\node [style=none] (29) at (8, 0.1) {};
		\node [style=none] (30) at (8, -0.025) {};
		\node [style=none] (31) at (8, -0.15) {};
		\node [style=none] (32) at (8, -0.275) {};
	\end{pgfonlayer}
	\begin{pgfonlayer}{edgelayer}
		\draw [style=gray,line width=0.5mm](0) to (1);
		\draw [style=magentae] (17.center) to (19.center);
		\draw [style=magentae] (18.center) to (20.center);
		\draw [style=orange,line width=0.3mm] (21.center) to (24.center);
		\draw [style=orange,line width=0.3mm](22.center) to (25.center);
		\draw [style=orange,line width=0.3mm](23.center) to (26.center);
		\draw [style=bluee] (4) to (5);
		\draw [style=cyaneX] (9) to (6);
		\draw [style=cyaneX] (10) to (7);
		\draw  [style=olive,line width=0.5mm](11) to (8);
	\end{pgfonlayer}
\end{tikzpicture}}}
 \\ \hline
\raisebox{-.5\height}{\begin{tikzpicture}
	\begin{pgfonlayer}{nodelayer}
		\node [style=gauge1] (0) at (-1, 0) {};
		\node [style=gauge1] (1) at (0.5, 0) {};
		\node [style=gauge1] (2) at (2, 0) {};
		\node [style=none] (4) at (-1, -0.575) {U(3)};
		\node [style=none] (5) at (0.5, -0.575) {SU(2)};
		\node [style=none] (6) at (2, -0.575) {SU(1)};
		\node [style=flavour1] (12) at (-2.5, 0) {};
		\node [style=none] (13) at (-2.5, -0.575) {4};
	\end{pgfonlayer}
	\begin{pgfonlayer}{edgelayer}
		\draw (0) to (1);
		\draw (1) to (2);
		\draw (0) to (12);
	\end{pgfonlayer}
\end{tikzpicture}}
 & \cellcolor{colors04} \raisebox{-.5\height}{\begin{tikzpicture}
	\begin{pgfonlayer}{nodelayer}
		\node [style=gauge1] (0) at (-2.725, 1.6) {};
		\node [style=none] (4) at (-1, -0.575) {1};
		\node [style=none] (7) at (-2.5, -0.575) {3};
		\node [style=none] (10) at (-2.8, 1.425) {};
		\node [style=none] (11) at (-2.6, 1.475) {};
		\node [style=none] (14) at (-2.625, 0) {};
		\node [style=none] (15) at (-2.4, -0.125) {};
		\node [style=gauge1] (16) at (-2.5, 0) {};
		\node [style=none] (19) at (-4, -0.575) {2};
		\node [style=none] (20) at (-5.5, -0.55) {1};
		\node [style=gauge1] (21) at (-4, 0) {};
		\node [style=gauge1] (22) at (-5.5, 0) {};
		\node [style=gauge1] (23) at (-1, 0) {};
		\node [style=none] (24) at (-2.825, 2.1) {1};
		\node [style=gauge1] (25) at (-1.5, 1.25) {};
		\node [style=none] (26) at (-1.25, 1.75) {1};
	\end{pgfonlayer}
	\begin{pgfonlayer}{edgelayer}
		\draw (14.center) to (10.center);
		\draw (11.center) to (15.center);
		\draw (21) to (16);
		\draw (21) to (22);
		\draw (25) to (16);
		\draw (23) to (16);
	\end{pgfonlayer}
\end{tikzpicture}}
&    \raisebox{-.5\height}{\scalebox{.789}{   \begin{tikzpicture}
	\begin{pgfonlayer}{nodelayer}
		\node [style=gauge1] (0) at (-1, -0.05) {};
		\node [style=gauge1] (1) at (0.5, -0.05) {};
		\node [style=gauge1] (2) at (2, -0.05) {};
		\node [style=gauge1] (3) at (8, -0.075) {};
		\node [style=gauge1] (4) at (2.75, 1) {};
		\node [style=gauge1] (5) at (2.75, -1) {};
		\node [style=gauge1] (6) at (4.25, -1) {};
		\node [style=gauge1] (7) at (5.75, -1) {};
		\node [style=gauge1] (8) at (7.25, -1) {};
		\node [style=gauge1] (9) at (4.25, 1) {};
		\node [style=gauge1] (10) at (5.75, 1) {};
		\node [style=gauge1] (11) at (7.25, 1) {};
		\node [style=none] (17) at (0.5, 0.075) {};
		\node [style=none] (18) at (0.5, -0.15) {};
		\node [style=none] (19) at (2, 0.075) {};
		\node [style=none] (20) at (2, -0.15) {};
		\node [style=none] (21) at (2, 0.075) {};
		\node [style=none] (22) at (2, -0.05) {};
		\node [style=none] (23) at (2, -0.175) {};
		\node [style=none] (24) at (8, 0.05) {};
		\node [style=none] (25) at (8, -0.075) {};
		\node [style=none] (26) at (8, -0.2) {};
		\node [style=none] (28) at (8, 0.225) {};
		\node [style=none] (29) at (8, 0.1) {};
		\node [style=none] (30) at (8, -0.025) {};
		\node [style=none] (31) at (8, -0.15) {};
		\node [style=none] (32) at (8, -0.275) {};
	\end{pgfonlayer}
	\begin{pgfonlayer}{edgelayer}
		\draw [style=gray,line width=0.5mm](0) to (1);
		\draw [style=magentae] (17.center) to (19.center);
		\draw [style=magentae] (18.center) to (20.center);
		\draw [style=orange,line width=0.3mm] (21.center) to (24.center);
		\draw [style=orange,line width=0.3mm](22.center) to (25.center);
		\draw [style=orange,line width=0.3mm](23.center) to (26.center);
		\draw [style=bluee] (4) to (5);
		\draw [style=bluee] (9) to (6);
		\draw [style=rede] (10) to (7);
		\draw  [style=olive,line width=0.5mm](11) to (8);
	\end{pgfonlayer}
\end{tikzpicture}}}
 \\ \hline
\raisebox{-.5\height}{ \begin{tikzpicture}
	\begin{pgfonlayer}{nodelayer}
		\node [style=gauge1] (0) at (-1, 0) {};
		\node [style=gauge1] (1) at (0.5, 0) {};
		\node [style=gauge1] (2) at (2, 0) {};
		\node [style=none] (4) at (-1, -0.575) {SU(3)};
		\node [style=none] (5) at (0.5, -0.575) {SU(2)};
		\node [style=none] (6) at (2, -0.575) {SU(1)};
		\node [style=flavour1] (12) at (-2.5, 0) {};
		\node [style=none] (13) at (-2.5, -0.575) {4};
	\end{pgfonlayer}
	\begin{pgfonlayer}{edgelayer}
		\draw (0) to (1);
		\draw (1) to (2);
		\draw (0) to (12);
	\end{pgfonlayer}
\end{tikzpicture}}
& \cellcolor{colors05}\raisebox{-.5\height}{\begin{tikzpicture}
	\begin{pgfonlayer}{nodelayer}
		\node [style=gauge1] (0) at (-2.725, 1.6) {};
		\node [style=none] (4) at (-1, -0.575) {1};
		\node [style=none] (7) at (-2.5, -0.575) {3};
		\node [style=gauge1] (16) at (-2.5, 0) {};
		\node [style=none] (19) at (-4, -0.575) {2};
		\node [style=none] (20) at (-5.5, -0.55) {1};
		\node [style=gauge1] (21) at (-4, 0) {};
		\node [style=gauge1] (22) at (-5.5, 0) {};
		\node [style=gauge1] (23) at (-1, 0) {};
		\node [style=none] (24) at (-2.825, 2.1) {1};
		\node [style=gauge1] (25) at (-1.75, 1.5) {};
		\node [style=none] (26) at (-1.35, 2) {1};
		\node [style=gauge1] (27) at (-1, 1) {};
		\node [style=none] (28) at (-0.575, 1.5) {1};
	\end{pgfonlayer}
	\begin{pgfonlayer}{edgelayer}
		\draw (21) to (16);
		\draw (21) to (22);
		\draw (25) to (16);
		\draw (23) to (16);
		\draw (0) to (16);
		\draw (27) to (16);
	\end{pgfonlayer}
\end{tikzpicture}}
&     
\raisebox{-.5\height}{\scalebox{.789}{ \begin{tikzpicture}
	\begin{pgfonlayer}{nodelayer}
		\node [style=gauge1] (0) at (-1, -0.05) {};
		\node [style=gauge1] (1) at (0.5, -0.05) {};
		\node [style=gauge1] (2) at (2, -0.05) {};
		\node [style=gauge1] (3) at (8, -0.075) {};
		\node [style=gauge1] (4) at (2.75, 1) {};
		\node [style=gauge1] (5) at (2.75, -1) {};
		\node [style=gauge1] (6) at (4.25, -1) {};
		\node [style=gauge1] (7) at (5.75, -1) {};
		\node [style=gauge1] (8) at (7.25, -1) {};
		\node [style=gauge1] (9) at (4.25, 1) {};
		\node [style=gauge1] (10) at (5.75, 1) {};
		\node [style=gauge1] (11) at (7.25, 1) {};
		\node [style=none] (17) at (0.5, 0.075) {};
		\node [style=none] (18) at (0.5, -0.15) {};
		\node [style=none] (19) at (2, 0.075) {};
		\node [style=none] (20) at (2, -0.15) {};
		\node [style=none] (21) at (2, 0.075) {};
		\node [style=none] (22) at (2, -0.05) {};
		\node [style=none] (23) at (2, -0.175) {};
		\node [style=none] (24) at (8, 0.05) {};
		\node [style=none] (25) at (8, -0.075) {};
		\node [style=none] (26) at (8, -0.2) {};
		\node [style=none] (28) at (8, 0.225) {};
		\node [style=none] (29) at (8, 0.1) {};
		\node [style=none] (30) at (8, -0.025) {};
		\node [style=none] (31) at (8, -0.15) {};
		\node [style=none] (32) at (8, -0.275) {};
	\end{pgfonlayer}
	\begin{pgfonlayer}{edgelayer}
		\draw [style=gray,line width=0.5mm](0) to (1);
		\draw [style=magentae] (17.center) to (19.center);
		\draw [style=magentae] (18.center) to (20.center);
		\draw [style=orange,line width=0.3mm] (21.center) to (24.center);
		\draw [style=orange,line width=0.3mm](22.center) to (25.center);
		\draw [style=orange,line width=0.3mm](23.center) to (26.center);
		\draw [style=bluee] (4) to (5);
		\draw [style=cyaneX] (9) to (6);
		\draw [style=rede] (10) to (7);
		\draw  [style=olive,line width=0.5mm](11) to (8);
	\end{pgfonlayer}
\end{tikzpicture}}}
\\ \hline
    \end{tabular}}
    \caption{The left column shows extensions of the $T(SU(4))$ quiver with different combinations of $U/SU$ nodes. The middle column shows their respective magnetic quivers, which in this particular case are in fact 3d mirrors. These magnetic quivers are derived from 5d brane webs, which yield quivers with all unitary gauge nodes. The right column shows the maximal decompositions of the brane webs into subwebs, with the necessary locking imposed. Note that the two magnetic quivers in blue cells and the three magnetic quivers in yellow cells are identical: this shows that Higgs branches for the family of theories considered in this table depends only on partitions of $4$. This is reflected in the next three tables by merging the corresponding cells.}
    \label{tsu4table1}
\end{table}

\afterpage{
\begin{landscape}
\begin{table}[p]
    \centering
  \scalebox{0.65}{
  \begin{tabular}{|c|c|c|}
    \hline
    \Large Electric Quiver & \Large Higgs branch unrefined Hilbert Series & \Large Coulomb branch unrefined Hilbert Series \\ \hline
\raisebox{-.5\height}{\begin{tikzpicture}
	\begin{pgfonlayer}{nodelayer}
		\node [style=gauge1] (0) at (-1, 0) {};
		\node [style=gauge1] (1) at (0.5, 0) {};
		\node [style=gauge1] (2) at (2, 0) {};
		\node [style=none] (4) at (-1, -0.575) {U(3)};
		\node [style=none] (5) at (0.5, -0.575) {U(2)};
		\node [style=none] (6) at (2, -0.575) {U(1)};
		\node [style=flavour1] (12) at (-2.5, 0) {};
		\node [style=none] (13) at (-2.5, -0.575) {4};
		\node at (0,.5) {};
	\end{pgfonlayer}
	\begin{pgfonlayer}{edgelayer}
		\draw (0) to (1);
		\draw (1) to (2);
		\draw (0) to (12);
	\end{pgfonlayer}
\end{tikzpicture}}
 & \verticalcenter{$\dfrac{(1 - t^4) (1 - t^6) (1 - t^8)}{(1 - t^2)^{15}}$}
&  \verticalcenter{$\dfrac{(1 - t^4) (1 - t^6) (1 - t^8)}{(1 - t^2)^{15}}$}
 \\ \hline
\raisebox{-.5\height}{\begin{tikzpicture}
	\begin{pgfonlayer}{nodelayer}
		\node [style=gauge1] (0) at (-1, 0) {};
		\node [style=gauge1] (1) at (0.5, 0) {};
		\node [style=gauge1] (2) at (2, 0) {};
		\node [style=none] (4) at (-1, -0.575) {SU(3)};
		\node [style=none] (5) at (0.5, -0.575) {U(2)};
		\node [style=none] (6) at (2, -0.575) {U(1)};
		\node [style=flavour1] (12) at (-2.5, 0) {};
		\node [style=none] (13) at (-2.5, -0.575) {4};
		\node at (0,.5) {};
	\end{pgfonlayer}
	\begin{pgfonlayer}{edgelayer}
		\draw (0) to (1);
		\draw (1) to (2);
		\draw (0) to (12);
	\end{pgfonlayer}
\end{tikzpicture}}
& \multirow{2}{*}{ \rule{0pt}{8ex} \verticalcenter{$\dfrac{1 + 3 t^2 + 4 t^3 + 7 t^4 + 4 t^5 + 7 t^6 + 4 t^7 + 3 t^8 + t^{10}}
   {(1 - t^2)^{13} (1 - t^3)^4 (1 - t^4)^{-2} (1 - t^6)^{-1}}$}}
&  \multirow{2}{*}{  \rule{0pt}{8ex}  \verticalcenter{$\dfrac{(1 - t^6) (1 - t^8) (1 + 4 t^4 + t^8)}
{(1 - t^2)^8 (1 - t^4)^4}$}}
     \\ \cline{1-1}
\raisebox{-.5\height}{
\begin{tikzpicture}
	\begin{pgfonlayer}{nodelayer}
		\node [style=gauge1] (0) at (-1, 0) {};
		\node [style=gauge1] (1) at (0.5, 0) {};
		\node [style=gauge1] (2) at (2, 0) {};
		\node [style=none] (4) at (-1, -0.575) {U(3)};
		\node [style=none] (5) at (0.5, -0.575) {U(2)};
		\node [style=none] (6) at (2, -0.575) {SU(1)};
		\node [style=flavour1] (12) at (-2.5, 0) {};
		\node [style=none] (13) at (-2.5, -0.575) {4};
		\node at (0,.5) {};
	\end{pgfonlayer}
	\begin{pgfonlayer}{edgelayer}
		\draw (0) to (1);
		\draw (1) to (2);
		\draw (0) to (12);
	\end{pgfonlayer}
\end{tikzpicture}}
 & 
& 
\\ \hline 
\raisebox{-.5\height}{\begin{tikzpicture}
	\begin{pgfonlayer}{nodelayer}
		\node [style=gauge1] (0) at (-1, 0) {};
		\node [style=gauge1] (1) at (0.5, 0) {};
		\node [style=gauge1] (2) at (2, 0) {};
		\node [style=none] (4) at (-1, -0.575) {U(3)};
		\node [style=none] (5) at (0.5, -0.575) {SU(2)};
		\node [style=none] (6) at (2, -0.575) {U(1)};
		\node [style=flavour1] (12) at (-2.5, 0) {};
		\node [style=none] (13) at (-2.5, -0.575) {4};
		\node at (0,.5) {};
	\end{pgfonlayer}
	\begin{pgfonlayer}{edgelayer}
		\draw (0) to (1);
		\draw (1) to (2);
		\draw (0) to (12);
	\end{pgfonlayer}
\end{tikzpicture}}
 & \verticalcenter{$\dfrac{1 + 5 t^2 + 23 t^4 + 62 t^6 + 110 t^8 + 130 t^{10} + 110 t^{12} + 
  62 t^{14} + 23 t^{16} + 5 t^{18} + t^{20}}
  {(1 - t^2)^{11} (1 - t^4)^3}$ }
  &
  \verticalcenter{$\dfrac{(1 - t^6) (1 - t^8)^2 (1 + 8 t^4 + t^8)}
  {(1 - t^2)^6 (1 - t^4)^7}$}
\\ \hline
\raisebox{-.5\height}{\begin{tikzpicture}
	\begin{pgfonlayer}{nodelayer}
		\node [style=gauge1] (0) at (-1, 0) {};
		\node [style=gauge1] (1) at (0.5, 0) {};
		\node [style=gauge1] (2) at (2, 0) {};
		\node [style=none] (4) at (-1, -0.575) {SU(3)};
		\node [style=none] (5) at (0.5, -0.575) {SU(2)};
		\node [style=none] (6) at (2, -0.575) {U(1)};
		\node [style=flavour1] (12) at (-2.5, 0) {};
		\node [style=none] (13) at (-2.5, -0.575) {4};
		\node at (0,.5) {};
	\end{pgfonlayer}
	\begin{pgfonlayer}{edgelayer}
		\draw (0) to (1);
		\draw (1) to (2);
		\draw (0) to (12);
	\end{pgfonlayer}
\end{tikzpicture}}
 &  \multirow{3}{*}{   \rule{0pt}{9ex}  \verticalcenter{$\dfrac{\left(\begin{array}{c}(1 + 3 t + 15 t^2 + 46 t^3 + 148 t^4 + 386 t^5 + 954 t^6 + 2064 t^7 +  4183 t^8 + 7649 t^9 + 13081 t^{10} + 20490 t^{11}\\ 
 + 30060 t^{12} +  40738 t^{13} + 51804 t^{14} + 61138 t^{15} + 67790 t^{16} + 69920 t^{17} + 
 ...\text{palindrome}+... + t^{34})\end{array}\right)}
 {(1 - t)^{-3} (1 - t^2)^8 (1 - t^3)^7 (1 - t^4)^4}$}}
&   \multirow{3}{*}{    \rule{0pt}{9ex}  \verticalcenter{$\dfrac{ 1 + t^2 + 4 t^4 + 9 t^6 + 13 t^8 + 12 t^{10} + 13 t^{12} + 
   9 t^{14} + 4 t^{16} + t^{18} + t^{20}}
   {(1 - t^2)^2 (1 - t^4)^5 (1 - t^6)^2 (1 - t^8)^{-1}}$}}
 \\ \cline{1-1}
\raisebox{-.5\height}{\begin{tikzpicture}
	\begin{pgfonlayer}{nodelayer}
		\node [style=gauge1] (0) at (-1, 0) {};
		\node [style=gauge1] (1) at (0.5, 0) {};
		\node [style=gauge1] (2) at (2, 0) {};
		\node [style=none] (4) at (-1, -0.575) {SU(3)};
		\node [style=none] (5) at (0.5, -0.575) {U(2)};
		\node [style=none] (6) at (2, -0.575) {SU(1)};
		\node [style=flavour1] (12) at (-2.5, 0) {};
		\node [style=none] (13) at (-2.5, -0.575) {4};
		\node at (0,.5) {};
	\end{pgfonlayer}
	\begin{pgfonlayer}{edgelayer}
		\draw (0) to (1);
		\draw (1) to (2);
		\draw (0) to (12);
	\end{pgfonlayer}
\end{tikzpicture}}
& 
 &  
 \\ \cline{1-1}
\raisebox{-.5\height}{\begin{tikzpicture}
	\begin{pgfonlayer}{nodelayer}
		\node [style=gauge1] (0) at (-1, 0) {};
		\node [style=gauge1] (1) at (0.5, 0) {};
		\node [style=gauge1] (2) at (2, 0) {};
		\node [style=none] (4) at (-1, -0.575) {U(3)};
		\node [style=none] (5) at (0.5, -0.575) {SU(2)};
		\node [style=none] (6) at (2, -0.575) {SU(1)};
		\node [style=flavour1] (12) at (-2.5, 0) {};
		\node [style=none] (13) at (-2.5, -0.575) {4};
		\node at (0,.5) {};
	\end{pgfonlayer}
	\begin{pgfonlayer}{edgelayer}
		\draw (0) to (1);
		\draw (1) to (2);
		\draw (0) to (12);
	\end{pgfonlayer}
\end{tikzpicture}}
& &  
 \\ \hline
 \raisebox{-.5\height}{\begin{tikzpicture}
	\begin{pgfonlayer}{nodelayer}
		\node [style=gauge1] (0) at (-1, 0) {};
		\node [style=gauge1] (1) at (0.5, 0) {};
		\node [style=gauge1] (2) at (2, 0) {};
		\node [style=none] (4) at (-1, -0.575) {SU(3)};
		\node [style=none] (5) at (0.5, -0.575) {SU(2)};
		\node [style=none] (6) at (2, -0.575) {SU(1)};
		\node [style=flavour1] (12) at (-2.5, 0) {};
		\node [style=none] (13) at (-2.5, -0.575) {4};
		\node at (0,.5) {};
	\end{pgfonlayer}
	\begin{pgfonlayer}{edgelayer}
		\draw (0) to (1);
		\draw (1) to (2);
		\draw (0) to (12);
	\end{pgfonlayer}
\end{tikzpicture}}

&  \verticalcenter{$
    \dfrac{\left(\begin{array}{c}(1 - t + 13 t^2 + 12 t^3 + 96 t^4 + 172 t^5 + 572 t^6 + 1072 t^7 + 2479 t^8 + 4265 t^9 + 7813 t^{10} + 11874 t^{11}\\ + 18146 t^{12} + 
  24124 t^{13} + 31540 t^{14} + 36640 t^{15} + 41456 t^{16} + 42064 t^{17} +...\text{palindrome}...-t^{33}+ t^{34})\end{array}\right)}
  {(1 - t) (1 - t^2)^5 (1 - t^3)^7 (1 - t^4)^5}$}
&   
\verticalcenter{$\dfrac{(1 - t^2 + t^4 + 4 t^6 + t^8 - t^{10} + t^{12})}
{(1 - t^2) (1 - t^4)^4 (1 - t^6)^2 (1 - t^{12})^{-1}}$}
\\ \hline
    \end{tabular}}
    \caption{Extensions of the $T(SU(4))$ quiver are shown with different combinations of $U/SU$ nodes, along with their Higgs and Coulomb branch unrefined Hilbert series. These correspond to the Coulomb and Higgs branch Hilbert series, respectively, of their mirror quivers shown in Table \ref{tsu4table1}. For brevity, unrefined Hilbert series are shown. Under the appropriate fugacity maps, this correspondence extends to refined Hilbert series.}
    \label{tsu4table}
\end{table}

\end{landscape}
}

\afterpage{
\begin{table}[p]
    \centering
  \scalebox{0.65}{
  \begin{tabular}{|c|c|c|}
    \hline
    \Large Electric Quiver & \Large Higgs branch global symmetry & \Large Coulomb branch global symmetry \\ \hline
\raisebox{-.5\height}{\begin{tikzpicture}
	\begin{pgfonlayer}{nodelayer}
		\node [style=gauge1] (0) at (-1, 0) {};
		\node [style=gauge1] (1) at (0.5, 0) {};
		\node [style=gauge1] (2) at (2, 0) {};
		\node [style=none] (4) at (-1, -0.575) {U(3)};
		\node [style=none] (5) at (0.5, -0.575) {U(2)};
		\node [style=none] (6) at (2, -0.575) {U(1)};
		\node [style=flavour1] (12) at (-2.5, 0) {};
		\node [style=none] (13) at (-2.5, -0.575) {4};
		\node at (0,.5) {};
	\end{pgfonlayer}
	\begin{pgfonlayer}{edgelayer}
		\draw (0) to (1);
		\draw (1) to (2);
		\draw (0) to (12);
	\end{pgfonlayer}
\end{tikzpicture}}
 &   \raisebox{-0.5\height}{\begin{tikzpicture}
    \node at (0,0) {\huge$A_3$};
\end{tikzpicture}}

&    \raisebox{-0.5\height}{\begin{tikzpicture}
    \node at (0,0) {\huge$A_3$};
\end{tikzpicture}}
 \\ \hline
\raisebox{-.5\height}{\begin{tikzpicture}
	\begin{pgfonlayer}{nodelayer}
		\node [style=gauge1] (0) at (-1, 0) {};
		\node [style=gauge1] (1) at (0.5, 0) {};
		\node [style=gauge1] (2) at (2, 0) {};
		\node [style=none] (4) at (-1, -0.575) {SU(3)};
		\node [style=none] (5) at (0.5, -0.575) {U(2)};
		\node [style=none] (6) at (2, -0.575) {U(1)};
		\node [style=flavour1] (12) at (-2.5, 0) {};
		\node [style=none] (13) at (-2.5, -0.575) {4};
		\node at (0,.5) {};
	\end{pgfonlayer}
	\begin{pgfonlayer}{edgelayer}
		\draw (0) to (1);
		\draw (1) to (2);
		\draw (0) to (12);
	\end{pgfonlayer}
\end{tikzpicture}}
&   \multirow{2}{*}{  \rule{0pt}{9ex} \huge$A_3U_1$}
&   \multirow{2}{*}{  \rule{0pt}{9ex}   \huge$A_2$}
     \\ \cline{1-1}
\raisebox{-.5\height}{\begin{tikzpicture}
	\begin{pgfonlayer}{nodelayer}
		\node [style=gauge1] (0) at (-1, 0) {};
		\node [style=gauge1] (1) at (0.5, 0) {};
		\node [style=gauge1] (2) at (2, 0) {};
		\node [style=none] (4) at (-1, -0.575) {U(3)};
		\node [style=none] (5) at (0.5, -0.575) {U(2)};
		\node [style=none] (6) at (2, -0.575) {SU(1)};
		\node [style=flavour1] (12) at (-2.5, 0) {};
		\node [style=none] (13) at (-2.5, -0.575) {4};
		\node at (0,.5) {};
	\end{pgfonlayer}
	\begin{pgfonlayer}{edgelayer}
		\draw (0) to (1);
		\draw (1) to (2);
		\draw (0) to (12);
	\end{pgfonlayer}
\end{tikzpicture}}
 & &
\\ \hline 
\raisebox{-.5\height}{\begin{tikzpicture}
	\begin{pgfonlayer}{nodelayer}
		\node [style=gauge1] (0) at (-1, 0) {};
		\node [style=gauge1] (1) at (0.5, 0) {};
		\node [style=gauge1] (2) at (2, 0) {};
		\node [style=none] (4) at (-1, -0.575) {U(3)};
		\node [style=none] (5) at (0.5, -0.575) {SU(2)};
		\node [style=none] (6) at (2, -0.575) {U(1)};
		\node [style=flavour1] (12) at (-2.5, 0) {};
		\node [style=none] (13) at (-2.5, -0.575) {4};
		\node at (0,.5) {};
	\end{pgfonlayer}
	\begin{pgfonlayer}{edgelayer}
		\draw (0) to (1);
		\draw (1) to (2);
		\draw (0) to (12);
	\end{pgfonlayer}
\end{tikzpicture}}
 &  \raisebox{-0.5\height}{\begin{tikzpicture}
    \node at (0,0) {\huge$A_3U_1$};
\end{tikzpicture}} 
  &   \raisebox{-0.5\height}{\begin{tikzpicture}
    \node at (0,0) {\huge$A_1A_1$};
\end{tikzpicture}}
\\ \hline
\raisebox{-.5\height}{\begin{tikzpicture}
	\begin{pgfonlayer}{nodelayer}
		\node [style=gauge1] (0) at (-1, 0) {};
		\node [style=gauge1] (1) at (0.5, 0) {};
		\node [style=gauge1] (2) at (2, 0) {};
		\node [style=none] (4) at (-1, -0.575) {SU(3)};
		\node [style=none] (5) at (0.5, -0.575) {SU(2)};
		\node [style=none] (6) at (2, -0.575) {U(1)};
		\node [style=flavour1] (12) at (-2.5, 0) {};
		\node [style=none] (13) at (-2.5, -0.575) {4};
		\node at (0,.5) {};
	\end{pgfonlayer}
	\begin{pgfonlayer}{edgelayer}
		\draw (0) to (1);
		\draw (1) to (2);
		\draw (0) to (12);
	\end{pgfonlayer}
\end{tikzpicture}}
 &   \multirow{3}{*}{ \rule{0pt}{13ex} \huge$A_3U_1U_1$}
&    \multirow{3}{*}{ \rule{0pt}{13ex} \huge $A_1$}
 \\ \cline{1-1}
\raisebox{-.5\height}{\begin{tikzpicture}
	\begin{pgfonlayer}{nodelayer}
		\node [style=gauge1] (0) at (-1, 0) {};
		\node [style=gauge1] (1) at (0.5, 0) {};
		\node [style=gauge1] (2) at (2, 0) {};
		\node [style=none] (4) at (-1, -0.575) {SU(3)};
		\node [style=none] (5) at (0.5, -0.575) {U(2)};
		\node [style=none] (6) at (2, -0.575) {SU(1)};
		\node [style=flavour1] (12) at (-2.5, 0) {};
		\node [style=none] (13) at (-2.5, -0.575) {4};
		\node at (0,.5) {};
	\end{pgfonlayer}
	\begin{pgfonlayer}{edgelayer}
		\draw (0) to (1);
		\draw (1) to (2);
		\draw (0) to (12);
	\end{pgfonlayer}
\end{tikzpicture}}
& &
 \\ \cline{1-1}
\raisebox{-.5\height}{\begin{tikzpicture}
	\begin{pgfonlayer}{nodelayer}
		\node [style=gauge1] (0) at (-1, 0) {};
		\node [style=gauge1] (1) at (0.5, 0) {};
		\node [style=gauge1] (2) at (2, 0) {};
		\node [style=none] (4) at (-1, -0.575) {U(3)};
		\node [style=none] (5) at (0.5, -0.575) {SU(2)};
		\node [style=none] (6) at (2, -0.575) {SU(1)};
		\node [style=flavour1] (12) at (-2.5, 0) {};
		\node [style=none] (13) at (-2.5, -0.575) {4};
		\node at (0,.5) {};
	\end{pgfonlayer}
	\begin{pgfonlayer}{edgelayer}
		\draw (0) to (1);
		\draw (1) to (2);
		\draw (0) to (12);
	\end{pgfonlayer}
\end{tikzpicture}}
& &
 \\ \hline
\raisebox{-.5\height}{\begin{tikzpicture}
	\begin{pgfonlayer}{nodelayer}
		\node [style=gauge1] (0) at (-1, 0) {};
		\node [style=gauge1] (1) at (0.5, 0) {};
		\node [style=gauge1] (2) at (2, 0) {};
		\node [style=none] (4) at (-1, -0.575) {SU(3)};
		\node [style=none] (5) at (0.5, -0.575) {SU(2)};
		\node [style=none] (6) at (2, -0.575) {SU(1)};
		\node [style=flavour1] (12) at (-2.5, 0) {};
		\node [style=none] (13) at (-2.5, -0.575) {4};
		\node at (0,.5) {};
	\end{pgfonlayer}
	\begin{pgfonlayer}{edgelayer}
		\draw (0) to (1);
		\draw (1) to (2);
		\draw (0) to (12);
	\end{pgfonlayer}
\end{tikzpicture}}
&    \verticalcenter{\huge$A_3U_1U_1U_1$}
&   
\\ \hline
    \end{tabular}}
    \caption{Extensions of the $T(SU(4))$ quiver are shown with different combinations of $U/SU$ nodes, along with their Higgs and Coulomb branch global symmetry. Notice that the ranks of the global symmetries always add to $6$.}
    \label{tsu4tableGlobal}
\end{table}
}

\afterpage{
\begin{landscape}
\begin{table}[t]
    \centering
    \scalebox{.9}{
    \begin{tabular}{|c|c|c|c|c|c|}
      \hline  \textbf{Partition}  & $[4]$ & $[3,1]$ & $[2,2]$ & $[2,1,1]$ & $[1,1,1,1]$ \\ \hline 
\raisebox{-.5\height}{\begin{tabular}{c}
   \textbf{Hasse}  \\
     \textbf{Diagram}
\end{tabular}}
 & \cellcolor{colors01}\raisebox{-.5\height}{\begin{tikzpicture}[scale=1]
\tikzstyle{hasse} = [circle, fill,inner sep=2pt]
\node[hasse] (0) at (0,0) {};
\node[hasse] (1) at (0,3) {};
\node[hasse] (2) at (0,4) {};
\node[hasse] (3) at (0,5) {};
\node[hasse] (4) at (0,6) {};
\draw (0)--(1)--(2)--(3)--(4);
\node at (-.5,1.5) {$a_3$};
\node at (-.5,3.5) {$a_1$};
\node at (-.5,4.5) {$A_1$};
\node at (-.5,5.5) {$A_3$};
\end{tikzpicture} }
& \cellcolor{colors02}
\raisebox{-.5\height}{\begin{tikzpicture}[scale=1]
\tikzstyle{hasse} = [circle, fill,inner sep=2pt]
\node[hasse] (0) at (0,0) {};
\node[hasse] (1) at (0,3) {};
\node[hasse] (2a) at (-1,4) {};
\node[hasse] (2b) at (1,4) {};
\node[hasse] (3) at (0,6) {};
\node[hasse] (4) at (0,7) {};
\draw (0)--(1)--(2a)--(3)--(4);
\draw (1)--(2b)--(3);
\node at (-.5,1.5) {$a_3$};
\node at (-1,3.3) {$a_1$};
\node at (1,3.3) {$a_1$};
\node at (-1,5) {$a_2$};
\node at (1,5) {$a_2$};
\node at (-.5,6.5) {$A_2$};
\end{tikzpicture} 
}
& \cellcolor{colors03}
\raisebox{-.5\height}{\begin{tikzpicture}[scale=1]
\tikzstyle{hasse} = [circle, fill,inner sep=2pt]
\node[hasse] (0) at (0,0) {};
\node[hasse] (1) at (0,3) {};
\node[hasse] (2) at (0,4) {};
\node[hasse] (3a) at (-1,5) {};
\node[hasse] (3b) at (1,5) {};
\node[hasse] (4a) at (-1,6) {};
\node[hasse] (4b) at (1,6) {};
\node[hasse] (5) at (0,7) {};
\draw (0)--(1)--(2)--(3a)--(4a)--(5);
\draw (2)--(3b)--(4b)--(5);
\draw (3b)--(4a);
\draw (3a)--(4b);
\node at (-.5,1.5) {$a_3$};
\node at (-.5,3.5) {$a_1$};
\node at (-1,4.3) {$a_1$};
\node at (1,4.3) {$a_1$};
\node at (-1.5,5.5) {$a_1$};
\node at (1.5,5.5) {$a_1$};
\node at (-1,6.5) {$a_1$};
\node at (1,6.5) {$a_1$};
\end{tikzpicture} 
      }
       & \cellcolor{colors04} \raisebox{-.5\height}{ \begin{tikzpicture}[scale=1]
\tikzstyle{hasse} = [circle, fill,inner sep=2pt]
\node[hasse] (0) at (0,0) {};
\node[hasse] (1) at (0,3) {};
\node[hasse] (2a) at (-2,4) {};
\node[hasse] (2b) at (0,4) {};
\node[hasse] (2c) at (2,4) {};
\node[hasse] (3) at (0,7) {};
\node[hasse] (4) at (0,8) {};
\draw (0)--(1)--(2a)--(3)--(4);
\draw (1)--(2b)--(3);
\draw (1)--(2c)--(3);
\node at (-.5,1.5) {$a_3$};
\node at (-1.5,3.5) {$a_1$};
\node at (-.3,3.5) {$a_1$};
\node at (1.5,3.5) {$a_1$};
\node at (-1.3,5.5) {$a_3$};
\node at (-.3,5.2) {$a_3$};
\node at (1.3,5.5) {$a_3$};
\node at (-.5,7.5) {$a_1$};
\end{tikzpicture} }
 & \cellcolor{colors05}
   \raisebox{-.5\height}{
   \begin{tikzpicture}[scale=1]
\tikzstyle{hasse} = [circle, fill,inner sep=2pt]
\node[hasse] (0) at (0,0) {};
\node[hasse] (1) at (0,3) {};
\node[hasse] (2a) at (-4,4) {};
\node[hasse] (2b) at (-2,4) {};
\node[hasse] (2c) at (0,4) {};
\node[hasse] (2d) at (2,4) {};
\node[hasse] (2e) at (4,4) {};
\node[hasse] (3) at (0,9) {};
\node at (0,10) {};
\node at (0,-1) {};
\draw (0)--(1)--(2a)--(3);
\draw (1)--(2b)--(3);
\draw (1)--(2c)--(3);
\draw (1)--(2d)--(3);
\draw (1)--(2e)--(3);
\node at (-.5,1.5) {$a_3$};
\node at (-2.8,3.5) {$a_1$};
\node at (-1.4,3.5) {$a_1$};
\node at (-.3,3.5) {$a_1$};
\node at (1.4,3.5) {$a_1$};
\node at (2.8,3.5) {$a_1$};
\node at (-2.5,6.5) {$d_4$};
\node at (-1.3,6.5) {$d_4$};
\node at (-.3,6.5) {$d_4$};
\node at (1.3,6.5) {$d_4$};
\node at (2.5,6.5) {$d_4$};
\end{tikzpicture}}  \\ \hline 
    \end{tabular}}
    \caption{Hasse diagrams of symplectic leaves (black dots) for the Higgs branches of theories shown in Tables \ref{tsu4table1}, \ref{tsu4tableGlobal} and \ref{tsu4table}. The elementary slices between adjacent leaves are labeled $A_n$ for the Klein singularity $\mathbb{C}^2 / \mathbb{Z}_{n+1}$ and $a_n$ (respectively $d_n$) for the closure of the minimal nilpotent orbit of $\mathfrak{sl}(n+1, \mathbb{C})$ (resp. $\mathfrak{so}(2n,\mathbb{C}
    )$). The partition refers to hypers connecting the $\mathrm{U}(3)$ node and the bouquet of $\mathrm{U}(1)$ nodes in the second column of Table \ref{tsu4table1}, as indicated by the coloring of the cells.  }
    \label{tab:tsu4HasseDiagrams}
\end{table}
\end{landscape}
}

\section{Web locking: an algorithm}  
\label{prescription}

In this section we give an algorithm which, for any framed linear quiver $\mathsf{Q}$ with unitary and special unitary nodes, provides a list of quivers $\mathsf{Q}_i$ such that 
\begin{equation}
    \mathcal{H}(\mathsf{Q}) = \bigcup\limits_i \mathcal{C}^{3d} (\mathsf{Q}_i) \, . 
\end{equation}
We state the algorithm in full generality, and illustrate every step of the algorithm with a single example, namely the quiver 
\begin{equation}
\label{quiverExample}
\raisebox{-.5\height}{
    \begin{tikzpicture}
\node (g1) at (10.5,3) [gauge,label=below:{SU(5)}] {};
\node (g2) at (11.5,3) [gauge,label=below:{U(5)}] {};
\node (g3) at (12.5,3) [gauge,label=below:{SU(6)}] {};
\node (f1) at (10.5,4) [flavor,label=above:{2}] {};
\node (f2) at (11.5,4) [flavor,label=above:{3}] {};
\node (f3) at (12.5,4) [flavor,label=above:{4}] {};
\draw (g1)--(g2)--(g3);
\draw (g1)--(f1) (g2)--(f2) (g3)--(f3);
\end{tikzpicture}}
\end{equation}
which illustrates its many intricate features. For more examples, see Section \ref{subsectionTsigmarho}. 

The plan of this section is as follows. In the first subsection, we set up our notation and spell out the combinatorics of locking. Then we define some algorithms about lists and partitions, before turning to the bulk of the algorithm. Then we explain how the algorithm is justified by the brane constructions explored in the rest of the paper. We conclude with a short description of the computer implementation. 

\subsection{Generalities about lockings}

We start from a linear quiver with $n$ gauge nodes, labeled by an index $i \in I = \{ 1 , \dots , n\}$, which are either $\mathrm{SU}(k_i)$ or $\mathrm{U}(k_i)$, and with $k_i\geq1$ and $N_i \geq 0$ flavors. Note that we do \emph{not} make any assumption regarding the balance of the gauge nodes.

We also introduce $I_0 \equiv \{0\} \cup I$. This set labels the $n+1$ NS5 branes in the system (see Figure \ref{fig:BraneWebAlgorithm}). 
We describe how the brane locking technique described in the previous section allows us to compute magnetic quivers for the Higgs variety. In general, the Higgs variety is a union of several hyper-K\"ahler cones, each characterized as the 3d $\mathcal{N}=4$ Coulomb branch of its magnetic quiver. 

A \emph{locking} is a partition\footnote{This is a partition in the set theoretic sense. The ordering of the $L_j$ is irrelevant; for definiteness we order the elements of the partition according to their lowest element.} $\mathbf{L}$ of the set $I_0$ into $c \geq 1$ subsets $\mathbf{L} = (L_j)_{j=1,...,c}$. This means that 
\begin{equation}
    \bigcup\limits_{j=1}^c L_j = I_0 \qquad \textrm{and} \qquad \forall i \neq j \, , \quad L_i \cap L_j = \emptyset \, . 
\end{equation}
Note that a given locking is depicted on the brane webs with a coloring which uses $c$ colors, hence the choice of the letter $c$. 
Let $\mathcal{L}_n$ be the set of all lockings; it depends only on the number of nodes $n$.\footnote{The cardinal of $\mathcal{L}_n$ is known as a Bell number, 
\begin{equation}
\label{bellNumber}
    |\mathcal{L}_n| = B_{n+1} := \frac{1}{e} \sum\limits_{m=1}^{\infty} \frac{m^{n+1}}{m!} \, . 
\end{equation}
Note that this number increases fast with $n$, the first values being (sequence \href{https://oeis.org/A000110}{A000110} in the OEIS)
\begin{equation}
    |\mathcal{L}_1| = 2 \, ,|\mathcal{L}_2| = 5 \, ,|\mathcal{L}_3| = 15 \, ,|\mathcal{L}_4| = 52 \, ,|\mathcal{L}_5| = 203 \, ,|\mathcal{L}_6| = 877  \, . 
\end{equation}} 
The set $\mathcal{L}_n$ can be partially ordered as follows. For two lockings $\mathbf{L} = (L_j)_{j=1,...,c}$ and $\mathbf{L}' = (L'_{j'})_{j'=1,...,c'}$, one says that $\mathbf{L} \leq \mathbf{L}'$ if 
\begin{equation}
    \forall j \in \{ 1,...,c \} , \exists j' \in \{ 1,...,c' \} : L_j \subseteq L'_{j'} \, . 
\end{equation}

\begin{figure}
    \centering
    \begin{tikzpicture}
\node (g1) at (0,7) {$0|1|2|3$};
\node (h1) at (-5,4) {$01|2|3$};
\node (h2) at (-3,4) {$02|1|3$};
\node (h3) at (-1,4) {$03|1|2$};
\node[draw] (h4) at (1,4) {\textcolor{blue}{$0|12|3$}};
\node (h5) at (3,4) {$0|13|2$};
\node (h6) at (5,4) {$0|1|23$};
\node (i1) at (-6,1) {$01|23$};
\node (i2) at (-4,1) {$02|13$};
\node (i3) at (-2,1) {\textcolor{blue}{$03|12$}};
\node (j1) at (0,1) {\textcolor{blue}{$012|3$}};
\node (j2) at (2,1) {$013|2$};
\node (j3) at (4,1) {$023|1$};
\node (j4) at (6,1) {\textcolor{blue}{$0|123$}};
\node (kk) at (0,-2) {\textcolor{blue}{$0123$}};
\draw (g1)--(h1);
\draw (g1)--(h2);
\draw (g1)--(h3);
\draw (g1)--(h4);
\draw (g1)--(h5);
\draw (g1)--(h6);
\draw (h1)--(i1);
\draw (h1)--(j1);
\draw (h1)--(j2);
\draw (h2)--(i2);
\draw (h2)--(j1);
\draw (h2)--(j3);
\draw (h3)--(i3);
\draw (h3)--(j2);
\draw (h3)--(j3);
\draw[thick,blue] (h4)--(i3);
\draw[thick,blue] (h4)--(j1);
\draw[thick,blue] (h4)--(j4);
\draw (h5)--(i2);
\draw (h5)--(j2);
\draw (h5)--(j4);
\draw (h6)--(i1);
\draw (h6)--(j3);
\draw (h6)--(j4);
\draw (kk)--(i1);
\draw (kk)--(i2);
\draw[thick,blue] (kk)--(i3);
\draw[thick,blue] (kk)--(j1);
\draw (kk)--(j2);
\draw (kk)--(j3);
\draw[thick,blue] (kk)--(j4);
\end{tikzpicture}
    \caption{Hasse diagram of lockings $\mathcal{L}_3$. Each locking is denoted using bars $|$. Using the brane web interpretation, this means for instance that $0|13|2$ represents branes 1 and 3 being locked together, while the other two are free to move independently. The locking in the box is $\mathbf{L}(\mathbf{s})$ for example (\ref{quiverExample}). The blue sub-diagram is the diagram of lockings compatible with $\mathbf{L}(\mathbf{s})$ (see definition on page \pageref{compatibility}).   }
    \label{fig:HasseExample}
\end{figure}
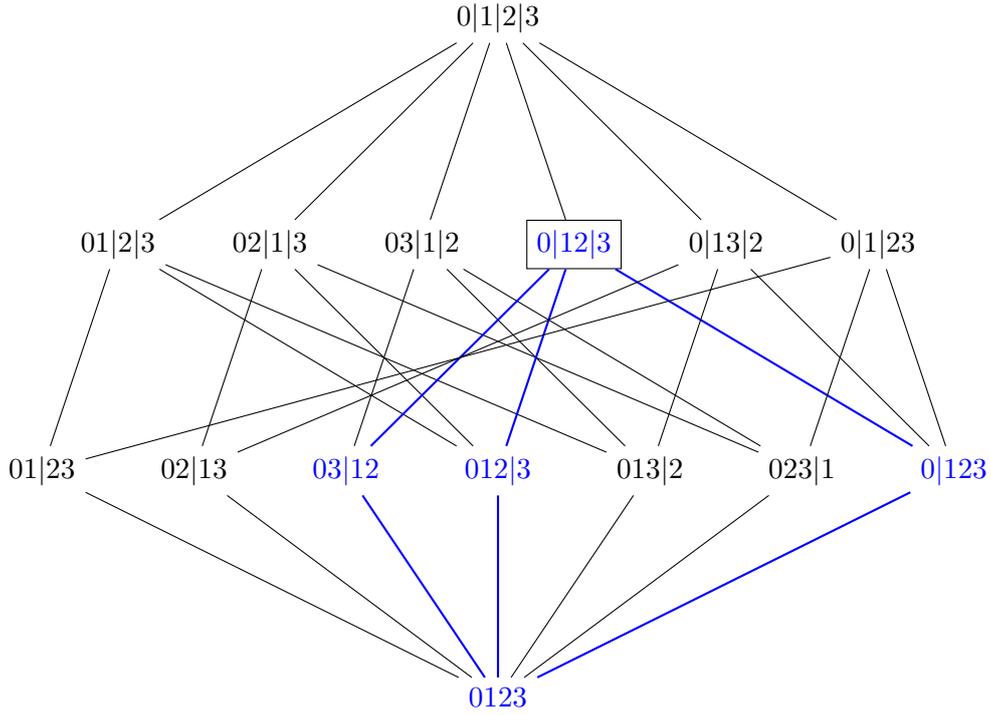

\paragraph{Example}
For our example we have $n=3$, so $I_0 = \{0,1,2,3\}$. There are 15 lockings, which can be arranged in the Hasse diagram shown in Figure \ref{fig:HasseExample}.

\subsection{Partitions and collapse}

A partition $\lambda$ of an integer $\alpha$ is a non-increasing sequence, with length $\ell$, of positive integers $(\lambda_1 , \dots , \lambda_{\ell})$ such that $\lambda_1 + \dots + \lambda_{\ell} = \alpha$. It can be represented as a Young diagram with $\ell$ columns of heights $\lambda_1 , \dots , \lambda_{\ell}$. The lengths of the rows of the diagram give the \emph{transpose} partition, denoted $\lambda^T$. 

Given a list of integers $\mathbf{x} = (x_i)$, we denote by $\Sigma (\mathbf{x})$ the list of partial sums: 
\begin{equation}
    (\Sigma (\mathbf{x}))_i = \sum\limits_{j \geq i} x_j \, . 
\end{equation}
We also call $\mathbf{x}^R$ the list $\mathbf{x}$ in reverse order.

We now describe the \emph{collapse algorithm}. 
This is an algorithm which, given any finite sequence of positive integers, produces a \emph{non-decreasing} sequence of positive integers. Let $\mathbf{x} = (x_1 , \dots , x_{\ell})$ be a sequence of positive integers of length $\ell$. If $x_1 \geq \dots \geq x_{\ell}$ then nothing needs to be done and the sequence defines a partition. Otherwise let $j$ be the first index such that $x_j < x_{j+1}$. One defines a new sequence $(x'_1 , \dots , x'_{\ell})$ where $x'_j = x_j +1$, $x'_{j+1} = x_{j+1} -1$ and $x'_i = x_i$ for $i \neq j,j+1$. One then repeats this operation iteratively on $x'$ until the sequence stabilizes and produces a partition. The resulting partition $\lambda$ is called the \emph{collapse} of the initial sequence $\mathbf{x}$: $\lambda = \mathrm{Collapse}(\mathbf{x})$. 

\paragraph{Example}
Take for instance $\mathbf{x} = (4,7,3,6)$. The collapse algorithm gives in turn 
\begin{equation}
    (4,7,3,6) \rightarrow (5,6,3,6) \rightarrow (6,5,3,6) \rightarrow (6,5,4,5) \rightarrow (6,5,5,4) \, . 
\end{equation}
Therefore $\lambda := (6,5,5,4) = \mathrm{Collapse}((4,7,3,6))$. 
In terms of Young diagrams (where the sequences are encoded in the heights of columns), the collapse corresponds to pushing boxes to the left: 
\begin{equation}
\raisebox{+2\height}{\scalebox{.5}{\begin{ytableau}
    \none  &  &  &  & \\
    \none  &  &  &  & \\
    \none  &  &  &  & \\
    \none  &  &  &  \none  & \\
    \none  &  \none  &  &  \none  & \\
    \none  &  \none  &  &  \none  & \\
    \none  &  \none  &  &  \none  &  \none   \\ 
\end{ytableau}}}  \qquad \longrightarrow  \quad
\raisebox{+2\height}{\scalebox{.5}{\begin{ytableau}
    \none  &  &  &  & \\
    \none  &  &  &  & \\
    \none  &  &  &  & \\
    \none  &  &  &  & \\
    \none  &  &  &  & \none \\
    \none  &  & \none & \none & \none \\
\end{ytableau}}}
\end{equation}
The transpose of $\lambda$ is $\lambda^T = (4,4,4,4,3,1)$. One can evaluate $\Sigma (\lambda) = (20,14,9,4)$ and $\Sigma (\lambda^T) = (20,16,12,8,4,1)$. 

\subsection{Main algorithm}

The inputs of the algorithm are 
\begin{itemize}
    \item The list of non-negative integers $\mathbf{N} = (N_i)_{i=1 , \dots , n}$; we also set $N_0 = N_{n+1}=0$, and $N = \sum N_i$.  
    \item The list of positive integers $\mathbf{k} = (k_i)_{i=1 , \dots , n}$; we also set $k_0 = k_{n+1}=0$. 
    \item The list of symbols $\mathbf{s} = (s_i)_{i=1 , \dots , n}$ with $s_i \in \{ \mathrm{U} , \mathrm{SU}\}$. 
\end{itemize} 
Here $N_i$ denotes the number of flavors on gauge node $i$. A gauge node is $\mathrm{U}(k_i)$ if $s_i = \mathrm{U}$ and $\mathrm{SU}(k_i)$ if $s_i = \mathrm{SU}$. 
To the list of symbols $(s_i)$ one can associate a locking as follows. Let $J \subseteq I$ be the set of indices $i$ such that $s_i = \mathrm{SU}$. We set $c = |J|+1$, and denote $J = \{ i_1 , \dots , i_{c-1}\}$ with $i_1 < \dots < i_{c-1}$. We then define the locking 
\begin{equation}
   \mathbf{L} (s) = (L_1 = \{0 ,\dots , i_1 -1\} , L_2 = \{i_1 , \dots , i_2-1 \} , \dots , L_c = \{i_{c-1} , \dots , n \} ) \, .  
\end{equation}

The algorithm outputs a finite list of quivers, one for each cone of the Higgs variety. The algorithm is divided in two steps: 
\begin{enumerate}
    \item[$\mathbf{\alpha}-$] Determining the list of \emph{locking patterns} corresponding to the cones; 
    \item[$\mathbf{\beta}-$] For each locking pattern, computing the associated magnetic quiver.
\end{enumerate}

We describe the two steps in turn. For clarity it is better to start with Step $\mathbf{\beta}$. In other words, we first explain how, given a locking pattern, we compute a magnetic quiver, and we turn to finding all locking patterns in the next paragraph. 

\subsubsection*{Step $\mathbf{\beta}$}

Let $\mathbf{x}$ be the list defined by\footnote{This is sometimes called the list of linking numbers of the NS5 branes. } 
    \begin{equation}
    \label{definitionListx}
        x_i = k_i - k_{i+1} + \sum\limits_{i' > i} N_{i'}
    \end{equation}
for $i =1 , \dots , n$. 
Let $\mathbf{L} = (L_j)_{j=1,...,c}$ be a locking. For each $L_j$, consider the sublist $\mathbf{x}^j$ of $\mathbf{x}$ defined by $\mathbf{x}^j = (x_i)_{i \in L_j}$. 
    We then define two lists of integers: 
\begin{itemize}
    \item Let 
    \begin{equation}
    \label{definitionRhoTilde}
        \tilde{\rho}^j =  \Sigma \left(  \mathrm{Collapse} (\mathbf{x}^j)^T \right) \, . 
    \end{equation}
    If the length of this list is less than $N$, add $0$s on the right until the length is $N$. Finally, reverse the order of the resulting list. Call $\rho^j$ the resulting list of $N$ integers. 
    \item Define 
    \begin{equation}
        \tilde{\sigma}^j =  \Sigma \left(\mathbf{x}^j \right) \, . 
    \end{equation}
    Similarly, add $0$s on the right until the length is $n+1$, and finally delete the \emph{first} entry. Call $\sigma^j$ the resulting list of $n$ integers. 
\end{itemize}
Define also 
\begin{equation}
\label{definitionRho0}
   \rho^0 := \Sigma (\Sigma (\mathbf{N})^T)^R - \sum\limits_j  \rho^j \, , 
\end{equation}
this is also a list of $N$ integers (the last entry is always 0). 

The elements of $\rho^0$ give the ranks of a linear chain of gauge nodes, that we label by $i = 1 , \dots , N$ (if the rank of the node is 0, the node is dropped). On top of this chain one adds a collection of $c$ $\mathrm{U}(1)$ nodes, that we label by $j = 1 , \dots , c$. One finally needs to describe the connections between those nodes: 
\begin{itemize}
    \item The nodes $i$ and $i+1$ are connected by one link. 
    \item The nodes $j$ and $j'$ are connected by 
    \begin{equation}
    \label{linksJJ}
        \left( \sum\limits_{i=1}^{N-1} \rho^{j}_{i}\rho^{j'}_{i+1} + \rho^{j}_{i+1}\rho^{j'}_{i} -2 \rho^{j}_{i}\rho^{j'}_{i} \right) - \rho^{j}_{N}\rho^{j'}_{N} + \sum\limits_{i \in L_j - \{n\}} \sigma^{j'}_{i+1} + \sum\limits_{i \in L_{j'} - \{n\}} \sigma^{j}_{i+1} 
    \end{equation}
    links. 
    \item The nodes $j$ and $i$ are connected by 
    \begin{equation}
    \label{linksIJ}
        \rho^j_{i+1} + \rho^j_{i-1} - 2 \rho^j_{i}
    \end{equation}
    links. 
\end{itemize}

\paragraph{Example} 
We look at two different lockings. 
\begin{itemize}
    \item Consider the locking $\mathbf{L} = (L_1 = \{ 0,1,2,3 \})$. Then $\mathbf{x} = (4,7,3,6)$ is the one considered in the previous section. Therefore $\tilde{\rho}^1 = (20,16,12,8,4,1)$ and $\rho^1 = (0,0,0,1,4,8,12,16,20)$, and $\sigma^1 = (16,9,6)$. One has $\mathbf{N} = (2,3,4)$, $\Sigma (\mathbf{N}) = (9,7,4)$,  $\Sigma (\mathbf{N})^T = (3,3,3,3,2,2,2,1,1)$, $\Sigma(\Sigma (\mathbf{N})^T)^R = (1,2,4,6,8,11,14,17,20)$. Therefore $\rho^0 = (1,2,4,5,4,3,2,1,0)$. The magnetic quiver then has a chain of eight nodes with the ranks given by $\rho^0$, dropping the last entry which is 0. It also has an additional $\mathrm{U}(1)$ node. Evaluation of (\ref{linksIJ}) for $1 \leq i \leq 8$ gives $(0,0,1,2,1,0,0)$. One deduces the magnetic quiver, which is shown in Table \ref{tab:resultsExample}. 
\item Consider the locking $\mathbf{L} = (L_1 = \{ 0,1,2 \} , L_2 = \{3\})$. One computes similarly $\sigma^1 = (10,3,0)$, $\sigma^2 = (6,6,6)$, $\rho^1 = (0,0,0,1,3,5,8,11,14)$, $\rho^2 = (0, 0, 0, 1, 2, 3, 4, 5, 6)$, and $\rho^0 = (1, 2, 4, 4, 3, 3, 2, 1, 0)$. Again there are eight nodes with ranks given by $\rho^0$ in the chain, and two $\mathrm{U}(1)$ nodes in addition. These two nodes are connected by (\ref{linksJJ}): 
\begin{equation}
    \left( \begin{array}{c}
       2+9+20+40+66 \\ +3+10+24+44+70 \\ -2-12-30-64-110
    \end{array} \right) -84 + (6+6+6)-(0) = 4
\end{equation}
links. Finally (\ref{linksIJ}) for $j=1$ gives $(0,0,1,1,0,1,0,0)$ and for $j=2$ gives $(0,0,1,0,0,0,0,0)$. We then get the quiver which is shown in Table \ref{tab:resultsExample}. 
\end{itemize}

\subsubsection*{Step $\mathbf{\alpha}$}

The algorithm $\beta$ above gives a quiver for each locking $\mathbf{L}$. It remains to determine which lockings contribute cones to the Higgs branch. We identify the lockings  $\mathbf{L} = (L_j)_{j=1,...,c}$ that satisfy three criteria. 
\begin{enumerate}
    \item \textbf{S-rule.} Consider the lists $\tilde{\rho}^j$ defined by (\ref{definitionRhoTilde}) from the list (\ref{definitionListx}) and the locking $\mathbf{L}$. We say that $\mathbf{L}$ satisfies the S-rule if all the $\tilde{\rho}^j$ have length less or equal to $N$, and the list $\rho^0$ defined in (\ref{definitionRho0}) has only non-negative entries. 
    \item \textbf{Compatibility.} \label{compatibility} The locking $\mathbf{L}$ is compatible with the assignment $\mathbf{s}$ if it is lower than $\mathbf{L}(\mathbf{s})$ in the partial order of lockings. See Figure \ref{fig:HasseExample} for an example. 
    \item \textbf{Dominance.} We say that the locking $\mathbf{L}$ dominates the locking $\mathbf{L}' < \mathbf{L}$ if the corresponding vectors $\rho^0$ and $(\rho^0)'$ are such that all the entries of $\rho^0 - (\rho^0)'$ are non-negative. 
\end{enumerate}
The set of lockings which satisfy the S-rule, are compatible with $\mathbf{s}$ and are not dominated by another locking satisfying the same criteria is called \emph{admissible}. 

For each admissible locking, one can then compute the associated magnetic quiver using the algorithm $\beta$. The Higgs branch of the initial quiver is the union of the Coulomb branch of all these quivers.

\begin{table}[]
    \centering
    \begin{tabular}{c|c} \toprule 
    Locking & Quiver \\ \midrule 
        $\mathbf{L}^{\mathrm{I}} =  ( \{0\} ,\{ 1,2\} ,\{ 3 \} )$ & \raisebox{-.5\height}{  \begin{tikzpicture}
\node (g1) at (0,0) [gauge,label=below:{1}] {};
\node (g2) at (1,0) [gauge,label=below:{2}] {};
\node (g3) at (2,0) [gauge,label=below:{3}] {};
\node (g4) at (3,0) [gauge,label=below:{3}] {};
\node (g5) at (4,0) [gauge,label=below:{3}] {};
\node (g6) at (5,0) [gauge,label=below:{3}] {};
\node (g7) at (6,0) [gauge,label=below:{2}] {};
\node (g8) at (7,0) [gauge,label=below:{1}] {};
\node (h1) at (2.5,.8) [gauge,label=below:{1}] {};
\node (h2) at (3.5,.8) [gauge,label=below:{1}] {};
\node (h3) at (3,2) [gauge,label=above:{1}] {};
\draw (g1)--(g2)--(g3)--(g4)--(g5)--(g6)--(g7)--(g8);
\draw (g3)--(h1);
\draw (g5)--(h2);
\draw[very thick] (h2)--(h3);
\draw[very thick] (h1)--(h3);
\draw[transform canvas={yshift=1pt}] (h2)--(h1);
\draw[transform canvas={yshift=-1pt}] (h2)--(h1);
\node at (2.85,1.2) {3};
\node at (3.15,1.2) {3};
\draw (g2)--(h3);
\draw (g6)--(h3);
\end{tikzpicture}} \\
        $\mathbf{L}^{\mathrm{II}}=  ( \{0 ,1,2\} ,\{ 3 \} )$ & \raisebox{-.5\height}{  
            \begin{tikzpicture}
\node (g1) at (0,0) [gauge,label=below:{1}] {};
\node (g2) at (1,0) [gauge,label=below:{2}] {};
\node (g3) at (2,0) [gauge,label=below:{4}] {};
\node (g4) at (3,0) [gauge,label=below:{4}] {};
\node (g5) at (4,0) [gauge,label=below:{3}] {};
\node (g6) at (5,0) [gauge,label=below:{3}] {};
\node (g7) at (6,0) [gauge,label=below:{2}] {};
\node (g8) at (7,0) [gauge,label=below:{1}] {};
\node (h1) at (2,1) [gauge,label=above:{1}] {};
\node (h2) at (3.5,1) [gauge,label=above:{1}] {};
\draw (g1)--(g2)--(g3)--(g4)--(g5)--(g6)--(g7)--(g8);
\draw (g3)--(h1);
\draw (g3)--(h2);
\draw (g4)--(h2);
\draw (g6)--(h2);
\draw[very thick] (h2)--(h1);
\node at (2.75,1.3) {4};
\end{tikzpicture}} \\
        $\mathbf{L}^{\mathrm{III}} =  ( \{0\} ,\{ 1,2, 3 \} )$ & \raisebox{-.5\height}{  \begin{tikzpicture}
\node (g1) at (0,0) [gauge,label=below:{1}] {};
\node (g2) at (1,0) [gauge,label=below:{2}] {};
\node (g3) at (2,0) [gauge,label=below:{3}] {};
\node (g4) at (3,0) [gauge,label=below:{4}] {};
\node (g5) at (4,0) [gauge,label=below:{4}] {};
\node (g6) at (5,0) [gauge,label=below:{3}] {};
\node (g7) at (6,0) [gauge,label=below:{2}] {};
\node (g8) at (7,0) [gauge,label=below:{1}] {};
\node (h1) at (3,1) [gauge,label=left:{1}] {};
\node (h2) at (4.5,1) [gauge,label=right:{1}] {};
\draw (g1)--(g2)--(g3)--(g4)--(g5)--(g6)--(g7)--(g8);
\draw (g4)--(h1);
\draw (g4)--(h2);
\draw (g5)--(h2);
\draw (g7)--(h2);
\draw[very thick] (h2)--(h1);
\node at (3.75,1.3) {4};
\end{tikzpicture}} \\
        $\mathbf{L}^{\mathrm{IV}} =  ( \{0,3\} ,\{ 1,2\} )$ & \raisebox{-.5\height}{ 
            \begin{tikzpicture}
\node (g1) at (0,0) [gauge,label=below:{1}] {};
\node (g2) at (1,0) [gauge,label=below:{2}] {};
\node (g3) at (2,0) [gauge,label=below:{3}] {};
\node (g4) at (3,0) [gauge,label=below:{4}] {};
\node (g5) at (4,0) [gauge,label=below:{3}] {};
\node (g6) at (5,0) [gauge,label=below:{3}] {};
\node (g7) at (6,0) [gauge,label=below:{2}] {};
\node (g8) at (7,0) [gauge,label=below:{1}] {};
\node (h1) at (3,1) [gauge,label=right:{1}] {};
\node (h2) at (3,2) [gauge,label=above:{1}] {};
\draw (g1)--(g2)--(g3)--(g4)--(g5)--(g6)--(g7)--(g8);
\draw[transform canvas={xshift=-1pt}] (g4)--(h1);
\draw[transform canvas={xshift=1pt}] (g4)--(h1);
\draw[very thick] (h2)--(h1);
\node at (2.8,1.5) {6};
\draw (g2)--(h2);
\draw (g6)--(h2);
\end{tikzpicture}} \\
        $\mathbf{L}^{\mathrm{V}} =  ( \{0,1,2, 3 \} )$ & \raisebox{-.5\height}{ 
            \begin{tikzpicture}
\node (g1) at (0,0) [gauge,label=below:{1}] {};
\node (g2) at (1,0) [gauge,label=below:{2}] {};
\node (g3) at (2,0) [gauge,label=below:{4}] {};
\node (g4) at (3,0) [gauge,label=below:{5}] {};
\node (g5) at (4,0) [gauge,label=below:{4}] {};
\node (g6) at (5,0) [gauge,label=below:{3}] {};
\node (g7) at (6,0) [gauge,label=below:{2}] {};
\node (g8) at (7,0) [gauge,label=below:{1}] {};
\node (h1) at (3,1) [gauge,label=above:{1}] {};
\draw (g1)--(g2)--(g3)--(g4)--(g5)--(g6)--(g7)--(g8);
\draw (g3)--(h1);
\draw[transform canvas={xshift=1pt}] (g4)--(h1);
\draw[transform canvas={xshift=-1pt}] (g4)--(h1);
\draw (g5)--(h1);
\end{tikzpicture}} \\ \bottomrule  
    \end{tabular}
    \caption{Quivers for the five cones of the Higgs branch of (\ref{quiverExample}), corresponding to the five admissible lockings, which appear in blue in Figure \ref{fig:HasseExample}. When more than two lines connect two nodes, they are denoted by a thick line, with a number indicating the multiplicity. }
    \label{tab:resultsExample}
\end{table}

\paragraph{Example}
In example (\ref{quiverExample}) we have $\mathbf{s} = (\mathrm{SU} , \mathrm{U} , \mathrm{SU})$ so $J = \{1,3\}$ and
\begin{equation}
    \mathbf{L}(\mathbf{s}) = ( L_1 =  \{0\} , L_2 = \{ 1,2\} , L_3 = \{ 3 \} ) \, . 
\end{equation}
We first compute all the $\tilde{\rho}^j$ that can be involved: 
\begin{itemize}
    \item For $L_1$, $\tilde{\rho} = (4,3,2,1)$. 
    \item For $L_2$, $\tilde{\rho} = (10, 8, 6, 4, 3, 2, 1)$. 
    \item For $L_3$, $\tilde{\rho} = (6, 5, 4, 3, 2, 1)$. 
    \item For $L_1 \cup L_2$, $\tilde{\rho} = (14, 11, 8, 5, 3, 1)$. 
    \item For $L_1 \cup L_3$, $\tilde{\rho} = (10, 8, 6, 4, 2)$. 
    \item For $L_2 \cup L_3$, $\tilde{\rho} = (16, 13, 10, 7, 4, 2, 1)$. 
    \item For $L_1 \cup L_2 \cup L_3$, $\tilde{\rho} = (20, 16, 12, 8, 4, 1)$. 
\end{itemize}
The lengths of all these vectors is $\leq N = 9$, so they satisfy the first part of the S-rule. 
From the Hasse diagram we see that there are five lockings which are compatible: 
\begin{itemize}
    \item $\mathbf{L}^{\mathrm{I}} =  ( \{0\} ,\{ 1,2\} ,\{ 3 \} )$. Here $\rho^0 = (1, 2, 3, 3, 3, 3, 2, 1, 0)$. 
    \item $\mathbf{L}^{\mathrm{II}} =  ( \{0 ,1,2\} ,\{ 3 \} )$. Here $\rho^0 = (1, 2, 4, 4, 3, 3, 2, 1, 0)$. 
    \item $\mathbf{L}^{\mathrm{III}} =  ( \{0\} ,\{ 1,2, 3 \} )$. Here $\rho^0 = (1, 2, 3, 4, 4, 3, 2, 1, 0)$. 
    \item $\mathbf{L}^{\mathrm{IV}} =  ( \{0,3\} ,\{ 1,2\} )$. Here $\rho^0 = (1, 2, 3, 4, 3, 3, 2, 1, 0)$. 
    \item $\mathbf{L}^{\mathrm{V}} =  ( \{0,1,2, 3 \} )$. Here $\rho^0 = (1, 2, 4, 5, 4, 3, 2, 1, 0)$.
\end{itemize}
Therefore all five lockings satisfy the S-rule. We now check that they are all dominant. $\mathbf{L}^{\mathrm{I}}$ does not dominate $\mathbf{L}^{\mathrm{II}}$ as the difference of the $\rho^0$s is $(0,0,-1,-1,0,0,0,0,0)$ which contains negative entries. Similarly  $\mathbf{L}^{\mathrm{I}}$ does not dominate $\mathbf{L}^{\mathrm{III}}$ nor $\mathbf{L}^{\mathrm{IV}}$ nor $\mathbf{L}^{\mathrm{V}}$. One checks in the same way that $\mathbf{L}^{\mathrm{V}}$ is not dominated because of the entry $5$ in its $\rho^0$. We conclude that the Higgs branch is made of five cones, which are the 3d Coulomb branches of the quivers presented in Table \ref{tab:resultsExample}.

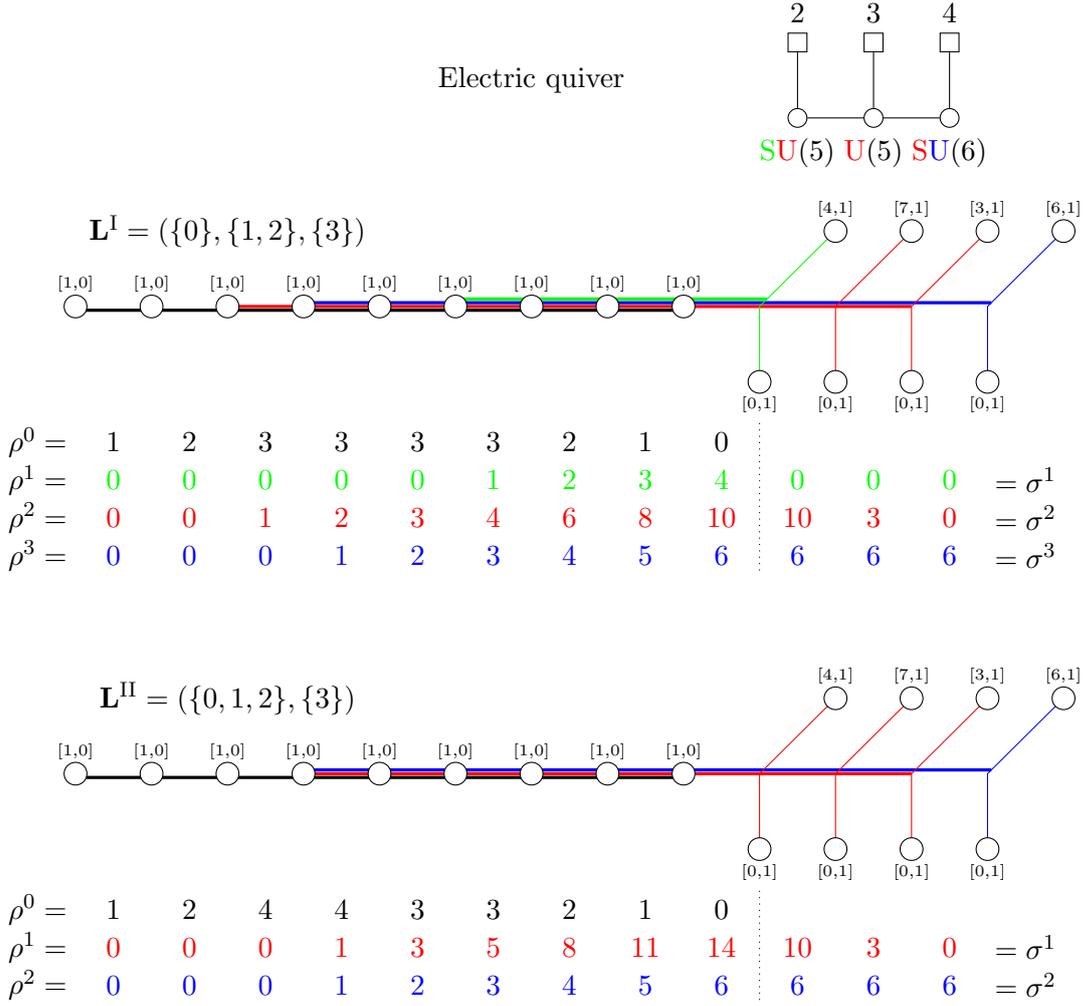
\begin{figure} 
\begin{tikzpicture}
\node (g1) at (10.5,2.5) [gauge,label=below:{\textcolor{green}{S}\textcolor{red}{U}(5)}] {};
\node (g2) at (11.5,2.5) [gauge,label=below:{\textcolor{red}{U}(5)}] {};
\node (g3) at (12.5,2.5) [gauge,label=below:{\textcolor{red}{S}\textcolor{blue}{U}(6)}] {};
\node (f1) at (10.5,3.5) [flavor,label=above:{2}] {};
\node (f2) at (11.5,3.5) [flavor,label=above:{3}] {};
\node (f3) at (12.5,3.5) [flavor,label=above:{4}] {};
\draw (g1)--(g2)--(g3);
\draw (g1)--(f1) (g2)--(f2) (g3)--(f3);
\node at (7,3) {Electric quiver};
\node at (3,1) {$\mathbf{L}^{\mathrm{I}} =  ( \{0\} ,\{ 1,2\} ,\{ 3 \} )$};
\draw[brane,black,very thick] (1,-.05)--(9,-.05); 
\draw[brane,red,very thick] (3,0)--(12,0); 
\draw[brane,blue,very thick] (4,0.05)--(13.05,0.05); 
\draw[brane,green,very thick] (6,0.1)--(10.1,0.1); 
\draw[brane,green] (10,-1)--(10,0)--(11,1); 
\draw[brane,red] (11,-1)--(11,0)--(12,1); 
\draw[brane,red] (12,-1)--(12,0)--(13,1); 
\draw[brane,blue] (13,-1)--(13,0)--(14,1); 
\node[D7] at (1,0) {};
\node[D7] at (2,0) {};
\node[D7] at (3,0) {};
\node[D7] at (4,0) {};
\node[D7] at (5,0) {};
\node[D7] at (6,0) {};
\node[D7] at (7,0) {};
\node[D7] at (8,0) {};
\node[D7] at (9,0) {};
\node[D7] at (10,-1) {};
\node[D7] at (11,-1) {};
\node[D7] at (12,-1) {};
\node[D7] at (13,-1) {};
\node[D7] at (11,1) {};
\node[D7] at (12,1) {};
\node[D7] at (13,1) {};
\node[D7] at (14,1) {};
\node at (11,1.3) {\tiny{[4,1]}};
\node at (12,1.3) {\tiny{[7,1]}};
\node at (13,1.3) {\tiny{[3,1]}};
\node at (14,1.3) {\tiny{[6,1]}};
\foreach \i in {10,...,13}{\node at (\i,-1.3) {\tiny{[0,1]}};}
\foreach \i in {1,...,9}{\node at (\i,.3) {\tiny{[1,0]}};}
\node at (0.5,-3.3) {$\rho^3 = $};
\node at (1.5,-3.3) {\textcolor{blue}{0}};
\node at (2.5,-3.3) {\textcolor{blue}{0}};
\node at (3.5,-3.3) {\textcolor{blue}{0}};
\node at (4.5,-3.3) {\textcolor{blue}{1}};
\node at (5.5,-3.3) {\textcolor{blue}{2}};
\node at (6.5,-3.3) {\textcolor{blue}{3}};
\node at (7.5,-3.3) {\textcolor{blue}{4}};
\node at (8.5,-3.3) {\textcolor{blue}{5}};
\node at (9.5,-3.3) {\textcolor{blue}{6}};
\node at (10.5,-3.3) {\textcolor{blue}{6}};
\node at (11.5,-3.3) {\textcolor{blue}{6}};
\node at (12.5,-3.3) {\textcolor{blue}{6}};
\node at (13.5,-3.3) {$=\sigma^3$};
\node at (0.5,-2.8) {$\rho^2 = $};
\node at (1.5,-2.8) {\textcolor{red}{0}};
\node at (2.5,-2.8) {\textcolor{red}{0}};
\node at (3.5,-2.8) {\textcolor{red}{1}};
\node at (4.5,-2.8) {\textcolor{red}{2}};
\node at (5.5,-2.8) {\textcolor{red}{3}};
\node at (6.5,-2.8) {\textcolor{red}{4}};
\node at (7.5,-2.8) {\textcolor{red}{6}};
\node at (8.5,-2.8) {\textcolor{red}{8}};
\node at (9.5,-2.8) {\textcolor{red}{10}};
\node at (10.5,-2.8) {\textcolor{red}{10}};
\node at (11.5,-2.8) {\textcolor{red}{3}};
\node at (12.5,-2.8) {\textcolor{red}{0}};
\node at (13.5,-2.8) {$=\sigma^2$};
\node at (0.5,-2.3) {$\rho^1 = $};
\node at (1.5,-2.3) {\textcolor{green}{0}};
\node at (2.5,-2.3) {\textcolor{green}{0}};
\node at (3.5,-2.3) {\textcolor{green}{0}};
\node at (4.5,-2.3) {\textcolor{green}{0}};
\node at (5.5,-2.3) {\textcolor{green}{0}};
\node at (6.5,-2.3) {\textcolor{green}{1}};
\node at (7.5,-2.3) {\textcolor{green}{2}};
\node at (8.5,-2.3) {\textcolor{green}{3}};
\node at (9.5,-2.3) {\textcolor{green}{4}};
\node at (10.5,-2.3) {\textcolor{green}{0}};
\node at (11.5,-2.3) {\textcolor{green}{0}};
\node at (12.5,-2.3) {\textcolor{green}{0}};
\node at (13.5,-2.3) {$=\sigma^1$};
\node at (0.5,-1.8) {$\rho^0 = $};
\node at (1.5,-1.8) {\textcolor{black}{1}};
\node at (2.5,-1.8) {\textcolor{black}{2}};
\node at (3.5,-1.8) {\textcolor{black}{3}};
\node at (4.5,-1.8) {\textcolor{black}{3}};
\node at (5.5,-1.8) {\textcolor{black}{3}};
\node at (6.5,-1.8) {\textcolor{black}{3}};
\node at (7.5,-1.8) {\textcolor{black}{2}};
\node at (8.5,-1.8) {\textcolor{black}{1}};
\node at (9.5,-1.8) {\textcolor{black}{0}};
\draw[dotted] (10,-3.5)--(10,-1.5); 
\end{tikzpicture}

\vspace{1cm}

\begin{tikzpicture}
\node at (3,1) {$\mathbf{L}^{\mathrm{II}}=  ( \{0 ,1,2\} ,\{ 3 \} )$};
\draw[brane,black,very thick] (1,-.05)--(9,-.05); 
\draw[brane,red,very thick] (4,0)--(12,0); 
\draw[brane,blue,very thick] (4,0.05)--(13.05,0.05); 
\draw[brane,red] (10,-1)--(10,0)--(11,1); 
\draw[brane,red] (11,-1)--(11,0)--(12,1); 
\draw[brane,red] (12,-1)--(12,0)--(13,1); 
\draw[brane,blue] (13,-1)--(13,0)--(14,1); 
\node[D7] at (1,0) {};
\node[D7] at (2,0) {};
\node[D7] at (3,0) {};
\node[D7] at (4,0) {};
\node[D7] at (5,0) {};
\node[D7] at (6,0) {};
\node[D7] at (7,0) {};
\node[D7] at (8,0) {};
\node[D7] at (9,0) {};
\node[D7] at (10,-1) {};
\node[D7] at (11,-1) {};
\node[D7] at (12,-1) {};
\node[D7] at (13,-1) {};
\node[D7] at (11,1) {};
\node[D7] at (12,1) {};
\node[D7] at (13,1) {};
\node[D7] at (14,1) {};
\node at (11,1.3) {\tiny{[4,1]}};
\node at (12,1.3) {\tiny{[7,1]}};
\node at (13,1.3) {\tiny{[3,1]}};
\node at (14,1.3) {\tiny{[6,1]}};
\foreach \i in {10,...,13}{\node at (\i,-1.3) {\tiny{[0,1]}};}
\foreach \i in {1,...,9}{\node at (\i,.3) {\tiny{[1,0]}};}
\node at (0.5,-2.8) {$\rho^2 = $};
\node at (1.5,-2.8) {\textcolor{blue}{0}};
\node at (2.5,-2.8) {\textcolor{blue}{0}};
\node at (3.5,-2.8) {\textcolor{blue}{0}};
\node at (4.5,-2.8) {\textcolor{blue}{1}};
\node at (5.5,-2.8) {\textcolor{blue}{2}};
\node at (6.5,-2.8) {\textcolor{blue}{3}};
\node at (7.5,-2.8) {\textcolor{blue}{4}};
\node at (8.5,-2.8) {\textcolor{blue}{5}};
\node at (9.5,-2.8) {\textcolor{blue}{6}};
\node at (10.5,-2.8) {\textcolor{blue}{6}};
\node at (11.5,-2.8) {\textcolor{blue}{6}};
\node at (12.5,-2.8) {\textcolor{blue}{6}};
\node at (13.5,-2.8) {$=\sigma^2$};
\node at (0.5,-2.3) {$\rho^1 = $};
\node at (1.5,-2.3) {\textcolor{red}{0}};
\node at (2.5,-2.3) {\textcolor{red}{0}};
\node at (3.5,-2.3) {\textcolor{red}{0}};
\node at (4.5,-2.3) {\textcolor{red}{1}};
\node at (5.5,-2.3) {\textcolor{red}{3}};
\node at (6.5,-2.3) {\textcolor{red}{5}};
\node at (7.5,-2.3) {\textcolor{red}{8}};
\node at (8.5,-2.3) {\textcolor{red}{11}};
\node at (9.5,-2.3) {\textcolor{red}{14}};
\node at (10.5,-2.3) {\textcolor{red}{10}};
\node at (11.5,-2.3) {\textcolor{red}{3}};
\node at (12.5,-2.3) {\textcolor{red}{0}};
\node at (13.5,-2.3) {$=\sigma^1$};
\node at (0.5,-1.8) {$\rho^0 = $};
\node at (1.5,-1.8) {\textcolor{black}{1}};
\node at (2.5,-1.8) {\textcolor{black}{2}};
\node at (3.5,-1.8) {\textcolor{black}{4}};
\node at (4.5,-1.8) {\textcolor{black}{4}};
\node at (5.5,-1.8) {\textcolor{black}{3}};
\node at (6.5,-1.8) {\textcolor{black}{3}};
\node at (7.5,-1.8) {\textcolor{black}{2}};
\node at (8.5,-1.8) {\textcolor{black}{1}};
\node at (9.5,-1.8) {\textcolor{black}{0}};
\draw[dotted] (10,-3)--(10,-1.5); 
\end{tikzpicture}
\caption{Examples of two brane webs corresponding to two different admissible lockings $\mathbf{L}^{\mathrm{I}}$ and $\mathbf{L}^{\mathrm{II}}$. The slopes of the branes of type $[p,1]$ are not respected. Each thick line of a given color corresponds to a stack of D5 branes; the number of D5 branes in each stack is indicated below with a number of the same color. These number correspond to the lists $\rho^j$ and $\sigma^j$ computed in the algorithm (the dotted line shows the limit between $\rho^j$ and $\sigma^j$), for $j = 1 , \dots , c$ running through the colors. The colors other than black denote the locking, while the various black stacks can move independently of each other in the $x^7,x^8,x^9$ direction. The list $\rho^0$ corresponds to the number of D5 branes in those stacks. The electric quiver is depicted on top, with a color code showing the top locking $\mathbf{L}(\mathbf{s}) = \mathbf{L}^{\mathrm{I}}$.  }
\label{fig:BraneWebAlgorithm}
\end{figure}

\paragraph{Number of cones. } The algorithm presented in this section shows that for a linear quiver with $n$ unitary or special unitary gauge nodes, the classical Higgs branch is made up of a finite number of cones. For any given quiver, the number of cones is precisely given by the algorithm, but general statements are still difficult to make. For instance, the number of cones is bounded above by the Bell number (\ref{bellNumber}), but this is a very coarse bound, that can presumably be considerably improved. 

\subsection{Brane Web Interpretation}

The various quantities presented in the algorithm of the previous section have a direct interpretation in terms of brane webs. The idea is to realize the quiver gauge theory on a brane web in the usual way, and to push all the flavor branes to the left using Hanany-Witten (HW) transitions. On our example (\ref{quiverExample}), the starting point is the web
\begin{equation}
    \raisebox{-.5\height}{\begin{tikzpicture}
\draw[brane,black,very thick] (0,0)--(6,0); 
\draw[brane,black] (0,-1)--(0,0)--(-2,.4); 
\draw[brane,black] (2,-1)--(2,1); 
\draw[brane,black] (4,-1)--(4,0)--(3,1); 
\draw[brane,black] (6,-1)--(6,0)--(10,.66); 
\node[D7] at (.8,.3) {};
\node[D7] at (1.2,.3) {};
\node[D7] at (2.4,.3) {};
\node[D7] at (2.8,.3) {};
\node[D7] at (3.2,.3) {};
\node[D7] at (4.4,.3) {};
\node[D7] at (4.8,.3) {};
\node[D7] at (5.2,.3) {};
\node[D7] at (5.6,.3) {};
\node[D7,label=below:{\tiny{[0,1]}}] at (0,-1) {};
\node[D7,label=below:{\tiny{[0,1]}}] at (2,-1) {};
\node[D7,label=below:{\tiny{[0,1]}}] at (4,-1) {};
\node[D7,label=below:{\tiny{[0,1]}}] at (6,-1) {};
\node[D7,label=above:{\tiny{[-5,1]}}] at (-2,.4) {};
\node[D7,label=above:{\tiny{[0,1]}}] at (2,1) {};
\node[D7,label=above:{\tiny{[-1,1]}}] at (3,1) {};
\node[D7,label=above:{\tiny{[6,1]}}] at (10,.66) {};
\node at (1,-.4) {5};
\node at (3,-.4) {5};
\node at (5,-.4) {6};
\end{tikzpicture}}
\end{equation}
where the lines denote fivebranes, and the circles sevenbranes. The charge of the sevenbranes is indicated unless it is $[1,0]$. Thick lines denote stacks of fivebranes, with the number of branes indicated below. Note that this brane web corresponds a priori to the quiver (\ref{quiverExample}) with all gauge nodes made special unitary. The $\mathrm{U}(5)$ gauge node is realized using brane locking. 

After the HW transitions, this creates a stack of $N$ D5 branes which end on D7 branes on the left, and on 5-branes on the right; the way they end can be encoded into partitions, which are the basic objects of the algorithm. More precisely, the $\rho^i$ and $\sigma^i$ sequences of integers correspond to numbers of segments of D5 branes which are bound to move together, as illustrated in Figure \ref{fig:BraneWebAlgorithm}. Crucially, the ability to move is controlled by the locking, and the S-rule. The reader can verify that imposing these conditions is realized by equations (\ref{definitionRhoTilde}) and (\ref{definitionRho0}). Once a partition solution is known, the magnetic quiver can be computed in terms of stable intersections of the tropical curves, supplemented by specific rules for fivebranes ending on sevenbranes. This is encapsulated in formulas (\ref{linksIJ}) and (\ref{linksJJ}).

\begin{figure}[h]
    \centering
    \includegraphics[height=18cm]{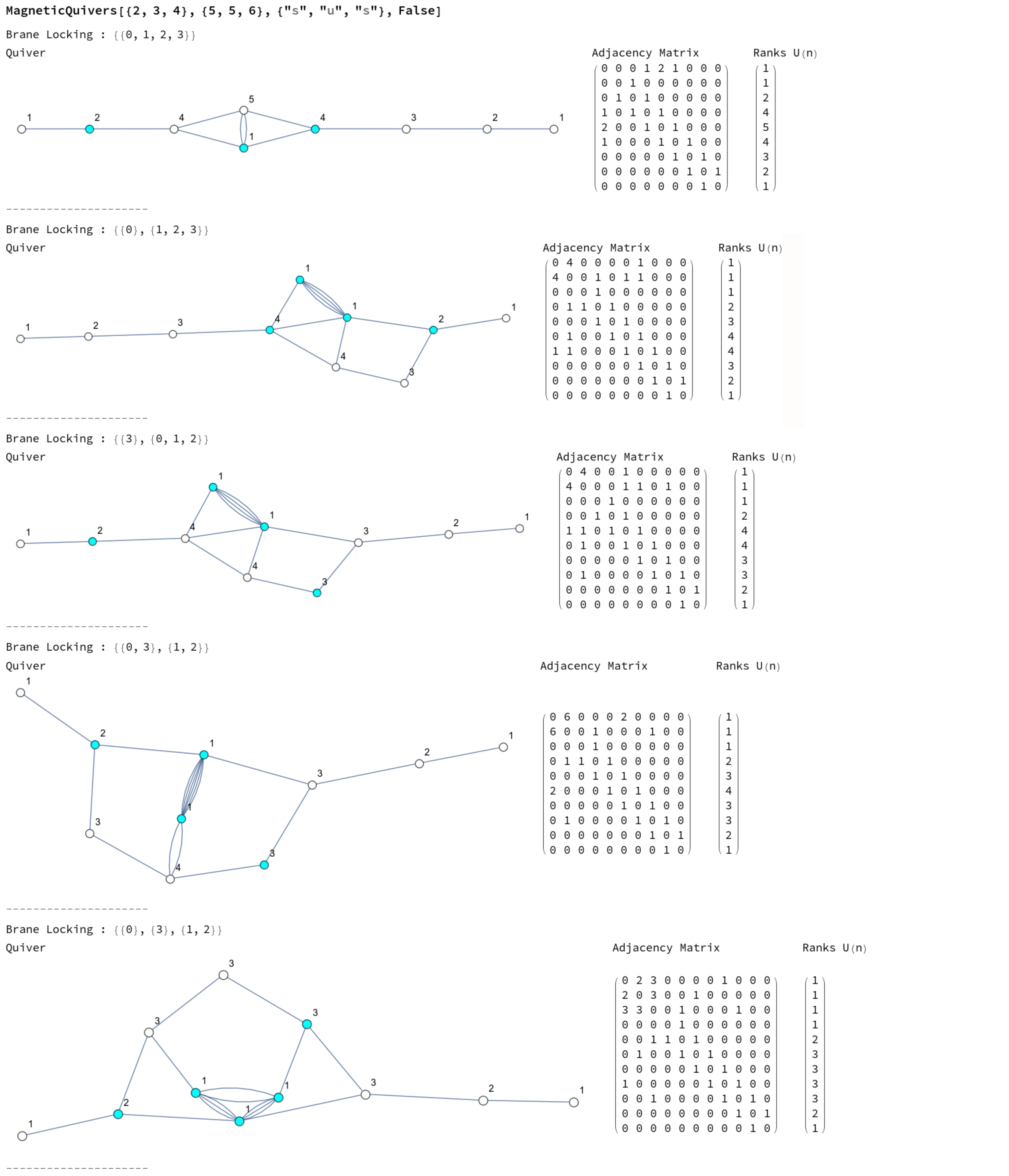}
    \caption{Output of the code for quiver (\ref{quiverExample}), producing the results shown in Table \ref{tab:resultsExample}. Each cone of the Higgs branch, represented by a magnetic quiver, is depicted on the left, and the corresponding adjacency matrix is given on the right, with a vector of ranks for ease of reading. The white nodes are balanced whereas the cyan nodes are overbalanced.}
    \label{fig:mathematica}
\end{figure}

\subsection{Computer Implementation}

A Mathematica code is attached to this paper (see the url in the introduction for the last version), implementing the algorithm presented in the present section. The code is used by calling the function 
\begin{equation}
    \texttt{ MagneticQuivers[ } \textbf{Flavors} \texttt{, } \textbf{Gauges}  \texttt{, } \textbf{U or SU}  \texttt{, } \textbf{Option}  \texttt{ ]} 
\end{equation}
where \textbf{Flavors} is a list of $n$ integers for the flavor nodes (from left to right), \textbf{Gauges} is a list of $n$ integers for the gauge nodes, \textbf{U or SU} is a list of $n$ values among \texttt{"s"} and \texttt{"u"} to denote whether the gauge group is unitary or special unitary, and \textbf{Option} can be \texttt{True} or \texttt{False} depending on if one wants the quivers to be drawn in a linear shape or not (this is a purely aesthetic option, which does not affect the result). For instance, for quiver (\ref{quiverExample}) one inputs 
\begin{equation}
    \texttt{MagneticQuivers[\{2, 3, 4\}, \{5, 5, 6\}, \{"s", "u", "s"\}, False]}
\end{equation}
and the output is shown in Figure \ref{fig:mathematica}. 

In the output is a list of magnetic quivers corresponding exactly to the cones of the Higgs branches. For better readability, the adjacency matrix and the ranks of the gauge nodes are also printed. White nodes are balanced while blue nodes are over-balanced.


\section{Examples and Effects}\label{section4}
In this section, certain families of magnetic quivers are studied in detail. 

\begin{enumerate}
    \item In Section \ref{subsectionTsigmarho} we study electric quivers for which every U$(N_c)$ or SU$(N_c)$ node has at least $2N_c$ hypermultiplets connected to it (i.e.\ the top region of Figure \ref{fig:venn}). These theories are always good. In this case there is only one magnetic quiver which is the 3d mirror of the electric quiver. Furthermore the different magnetic quivers for various choices of U and SU nodes are related to the magnetic quiver for all U nodes. This allows us to make very general statements about the magnetic quivers, based on the $T^\sigma_\rho(SU(n))$ technology of the all U electric quivers.
    \item In Section \ref{sec:negativeBalance} we turn to electric quivers, where some nodes have less hypermultiplets than in the section before. In this case generally, the magnetic quiver is not a 3d mirror, and if there are SU$(N_c)$ nodes with less than $2N_c$ hypers, there are generally multiple magnetic quivers. For magentic quivers that sastisfy a certain set of conditions, we also provide the reversed algorithm which allows us to obtain an electric quiver which is its 3d mirror.
    
    \item In Section \ref{sec:incomplete3d} we explain that for electric theories with incomplete Higgsing the magnetic quiver does not provide a 3d mirror.\footnote{Note that the converse is not true, when there is complete Higgsing there may still be multiple magnetic quivers, as is the case e.g.\ for SU$(3)$ with 4 flavors.} This point is further elaborated in Appendix \ref{bad}.
    \item In Section \ref{sec:AD} we give an example of an Argyres Douglas (AD) theory whose Higgs branch is that of a linear quiver with U and SU nodes. We apply our algorithm to this quiver and produce a magnetic quiver.
    \item Finally in Section \ref{sec:beyondLinear} we leave the scope of this paper, and consider a non-linear electric quiver with SU nodes. Although our methods cannot be applied to produce a magnetic quiver, we can produce one through other means.
\end{enumerate}

\subsection{Linear quivers with nodes of non-negative balance}
\label{subsectionTsigmarho}
From previous sections, we see that replacing an SU node with a U node (or vice versa) in a linear electric quiver can drastically change its magnetic quiver. However, if we focus only on good linear quivers, then many of the features in the magnetic quiver (which is now also a 3d mirror) remain the same.

 Good linear quivers with all nodes U go under the name $T^\sigma_\rho(SU(n))$ theories. Each unitary gauge group is either balanced or overbalanced. 3d mirror symmetry for these theories was studied in \cite{Gaiotto:2008ak} using the classic NS5-D3-D5 HW brane system. If we take the NS5-D3-D5 brane system for any $T^\sigma_\rho(SU(n))$ and we go to the Higgs phase (i.e.\ all D3 branes are suspended between D5 branes, and any D3 branes stuck between a D5 and a NS5 are annihilated by a HW transition) then all the NS5 branes have no D3 branes ending on them.\footnote{Note that we can only reach such a Higgs phase, when there is complete Higgsing. Otherwise there are always some D3 branes suspended between NS5 branes. See Appendix \ref{bad} for more details. }
 T-dualizing this system to a brane web, we obtain the brane system of the electric quiver with all U nodes replaced by SU. The NS5 branes present in this system have no D5 branes ending on them, which means the only 5-branes are NS5 and D5. This greatly simplifies obtaining the magnetic quiver for any choice of locking, i.e.\ any choice of U and SU nodes in the electric quiver. We proceed with some examples.

\subsubsection*{$T(SU(n))$ theories}
The $T(SU(n))$ family has a single SU(n) flavor group and is  3d self mirror. This makes it the simplest example to see how the different arrangements of of U/SU have on the mirror.  The $n=4$ case is already studied in detail in Section \ref{branes}. When all the gauge nodes are unitary, the quiver is balanced and the theory is self-dual:
\begin{equation}
\centering
\scalebox{0.9}{\begin{tikzpicture}
	\begin{pgfonlayer}{nodelayer}
		\node [style=gauge1] (0) at (-3.25, -33) {};
		\node [style=gauge1] (1) at (-2.25, -33) {};
		\node [style=gauge1] (2) at (-1.25, -33) {};
		\node [style=none] (3) at (-0.25, -33) {\dots};
		\node [style=gauge1] (4) at (0.75, -33) {};
		\node [style=flavour1] (5) at (2.25, -33) {};
		\node [style=none] (6) at (-3.25, -33.5) {U(1)};
		\node [style=none] (7) at (-2.25, -33.5) {U(2)};
		\node [style=none] (8) at (-1.25, -33.5) {U(3)};
		\node [style=none] (9) at (0.75, -33.5) {U($n-1$)};
		\node [style=none] (10) at (2.25, -33.5) {$n$};
		\node [style=gauge1] (11) at (12.5, -33) {};
		\node [style=gauge1] (12) at (11.5, -33) {};
		\node [style=gauge1] (13) at (10.5, -33) {};
		\node [style=none] (14) at (9.5, -33) {\dots};
		\node [style=gauge1] (15) at (8.5, -33) {};
		\node [style=none] (17) at (12.5, -33.5) {U(1)};
		\node [style=none] (18) at (11.5, -33.5) {U(2)};
		\node [style=none] (19) at (10.5, -33.5) {U(3)};
		\node [style=none] (20) at (8.5, -33.5) {U($n-1$)};
		\node [style=none] (21) at (7, -33.5) {U(1)};
		\node [style=none] (22) at (3.25, -33) {};
		\node [style=none] (23) at (6, -33) {};
		\node [style=none] (24) at (4.75, -32.5) {3d mirror};
		\node [style=none] (25) at (7.75, -32.75) {$n$};
		\node [style=gauge1] (26) at (7, -33) {};
	\end{pgfonlayer}
	\begin{pgfonlayer}{edgelayer}
		\draw (0) to (2);
		\draw (4) to (5);
		\draw (11) to (13);
		\draw [style=->] (22.center) to (23.center);
		\draw [style=->] (23.center) to (22.center);
		\draw  (26) to (15);
	\end{pgfonlayer}
\end{tikzpicture}}
\label{TSUN}
\end{equation}
As before, we first turn all the gauge nodes from U to SU. In this case, the 3d mirror obtained from the brane web takes the following form: 
\begin{equation}
  \scalebox{0.9}{\begin{tikzpicture}
	\begin{pgfonlayer}{nodelayer}
		\node [style=gauge1] (0) at (0.5, -33) {};
		\node [style=gauge1] (1) at (1.5, -33) {};
		\node [style=gauge1] (2) at (2.5, -33) {};
		\node [style=none] (3) at (3.5, -33) {\dots};
		\node [style=gauge1] (4) at (4.5, -33) {};
		\node [style=flavour1] (5) at (6, -33) {};
		\node [style=none] (6) at (0.5, -33.5) {SU(1)};
		\node [style=none] (7) at (1.5, -33.5) {SU(2)};
		\node [style=none] (8) at (2.5, -33.5) {SU(3)};
		\node [style=none] (9) at (4.5, -33.5) {SU($n-1$)};
		\node [style=none] (10) at (6, -33.5) {$n$};
		\node [style=gauge1] (11) at (17, -33) {};
		\node [style=gauge1] (12) at (16, -33) {};
		\node [style=gauge1] (13) at (15, -33) {};
		\node [style=none] (14) at (14, -33) {\dots};
		\node [style=gauge1] (15) at (13, -33) {};
		\node [style=none] (16) at (17, -33.5) {U(1)};
		\node [style=none] (17) at (16, -33.5) {U(2)};
		\node [style=none] (18) at (15, -33.5) {U(3)};
		\node [style=none] (19) at (13.5, -32.5) {U($n-1$)};
		\node [style=none] (20) at (7, -33) {};
		\node [style=none] (21) at (9.75, -33) {};
		\node [style=none] (22) at (8.5, -32.5) {3d mirror};
		\node [style=gauge1] (23) at (12, -31.5) {};
		\node [style=gauge1] (24) at (12, -32.25) {};
		\node [style=gauge1] (25) at (12, -33.75) {};
		\node [style=gauge1] (26) at (12, -34.5) {};
		\node [style=none] (27) at (12, -33) {\vdots};
		\node [style=none] (28) at (10.75, -31.5) {};
		\node [style=none] (29) at (10.75, -34.5) {};
		\node [style=none] (30) at (10.25, -33) {$n$};
		\node [style=none] (31) at (11.25, -31.5) {U(1)};
		\node [style=none] (32) at (11.25, -32.25) {U(1)};
		\node [style=none] (33) at (11.25, -33.75) {U(1)};
		\node [style=none] (34) at (11.25, -34.5) {U(1)};
	\end{pgfonlayer}
	\begin{pgfonlayer}{edgelayer}
		\draw (0) to (2);
		\draw (4) to (5);
		\draw (11) to (13);
		\draw [style=->] (20.center) to (21.center);
		\draw [style=->] (21.center) to (20.center);
		\draw (23) to (15);
		\draw (24) to (15);
		\draw (25) to (15);
		\draw (26) to (15);
		\draw [style=brace1] (28.center) to (29.center);
	\end{pgfonlayer}
\end{tikzpicture}}
\label{TSUNSU}
\end{equation}
which appeared in \cite{Dancer:2020wll}. Comparing \eqref{TSUN} and \eqref{TSUNSU} we see that the only difference in the mirror quivers is the $\mathrm{U}(n-1)$ connected to a $\mathrm{U}(1)$ with $n$ links exploded into a bouquet of $n$ $\mathrm{U}(1)$s.  As pointed out in previous sections, regardless of the choice of U/SU groups, the $\mathrm{U}(N_i)$ gauge nodes with $1 \leq N_i \leq n-1$, in the mirror theory remain the same because they correspond to D5 branes in the brane web and not NS5 branes, hence are not affected by locking. When all gauge nodes are SU, the brane web has $n$ independent (unlocked) NS5 branes, each corresponding to a $\mathrm{U}(1)$ in the bouquet. 

Starting from \eqref{TSUNSU}, we then turn some of the SU into U. The only change to the 3d mirror is in the $\mathrm{U}(1)$ bouquet which is connected to the $\mathrm{U}(n-1)$ node. The number of $\mathrm{U}(1)$ nodes in the new bouquet and the multiplicity of the edges connected to the $\mathrm{U}(n-1)$ can be determined solely from the different ways the NS5 branes are locked. To illustrate this, it is sufficient to draw an \textit{incomplete} brane diagram with only NS5 branes. When all the gauge groups are SU, the NS5 branes are unlocked and is denoted by different colors:
\begin{equation}
\raisebox{-.5\height}{\begin{tikzpicture}
	\begin{pgfonlayer}{nodelayer}
		\node [style=none] (0) at (0.5, 0) {};
		\node [style=none] (1) at (0.5, -3) {};
		\node [style=none] (2) at (2.5, 0) {};
		\node [style=none] (3) at (2.5, -3) {};
		\node [style=none] (4) at (1.5, 0) {};
		\node [style=none] (5) at (1.5, -3) {};
		\node [style=none] (6) at (5, 0) {};
		\node [style=none] (7) at (5, -3) {};
		\node [style=none] (8) at (7, 0) {};
		\node [style=none] (9) at (7, -3) {};
		\node [style=none] (10) at (6, 0) {};
		\node [style=none] (11) at (6, -3) {};
		\node [style=none] (12) at (4, 0) {};
		\node [style=none] (13) at (4, -3) {};
		\node [style=none] (14) at (7, -4) {};
		\node [style=none] (15) at (0.25, -4) {};
		\node [style=none] (16) at (3.55, -4.75) {$n$};
		\node [style=none] (17) at (3.25, -1.25) {$\dots$};
		\node [style=none] (18) at (1, -3.5) {$S$};
		\node [style=none] (19) at (6.5, -3.5) {$S$};
		\node [style=none] (20) at (5.5, -3.5) {$S$};
		\node [style=none] (21) at (4.5, -3.5) {$S$};
		\node [style=none] (22) at (2, -3.5) {$S$};
	\end{pgfonlayer}
	\begin{pgfonlayer}{edgelayer}
		\draw [style=rede] (0.center) to (1.center);
		\draw [style=magicmintline] (4.center) to (5.center);
		\draw [style=brownline] (2.center) to (3.center);
		\draw [style={line width=0.5mm,yellow}] (12.center) to (13.center);
		\draw [style=bluee] (6.center) to (7.center);
		\draw [style={line width=0.5mm,green}] (10.center) to (11.center);
		\draw [style=pinkline] (8.center) to (9.center);
		\draw [style=brace2] (14.center) to (15.center);
	\end{pgfonlayer}
\end{tikzpicture}}
\label{simplebrane}
\end{equation}
where $S$ stand for special unitary group in the electric quiver. For $n-1$ balanced nodes in the electric quiver, there are $n$ NS5 branes in the brane web.
The dictionary between the figure and the electric quiver is as follows.
For adjacent NS5s with different/same color, a D brane stretched between them has a special unitary/unitary gauge group, respectively.
Changing the gauge groups in the original quiver from special unitary to unitary is equivalent to setting the adjacent branes to the same color.
The adjacent branes are locked, hence corresponding to a single $\mathrm{U}(1)$ node in the bouquet of the mirror quiver. The multiplicity of the edge is then the number of NS5 branes that move together. This is because in the full brane web, this number is the intersection number between the locked NS5s and the D5 branes. 

The changes in the mirror quiver are dictated soley by the arrangement  of U/SU nodes in the electric quiver. The ranks of the gauge and flavor groups are irrelevant here. For example, if the electric quiver has a $USUSUU$ structure (where $U/S$ stand for unitary/special unitary group, respectively), then the NS5 branes takes the form:
\begin{equation}
\raisebox{-.5\height}{\begin{tikzpicture}
	\begin{pgfonlayer}{nodelayer}
		\node [style=none] (0) at (1, 0) {};
		\node [style=none] (1) at (1, -3) {};
		\node [style=none] (2) at (3, 0) {};
		\node [style=none] (3) at (3, -3) {};
		\node [style=none] (4) at (2, 0) {};
		\node [style=none] (5) at (2, -3) {};
		\node [style=none] (6) at (5, 0) {};
		\node [style=none] (7) at (5, -3) {};
		\node [style=none] (8) at (7, 0) {};
		\node [style=none] (9) at (7, -3) {};
		\node [style=none] (10) at (6, 0) {};
		\node [style=none] (11) at (6, -3) {};
		\node [style=none] (14) at (4, 0) {};
		\node [style=none] (15) at (4, -3) {};
		\node [style=none] (19) at (1, 0) {};
		\node [style=none] (20) at (1, -3) {};
		\node [style=none] (21) at (2, 0) {};
		\node [style=none] (22) at (2, -3) {};
		\node [style=none] (23) at (5.5, -3.25) {$U$};
		\node [style=none] (24) at (3.5, -3.25) {$U$};
		\node [style=none] (25) at (6.5, -3.25) {$U$};
		\node [style=none] (26) at (4.5, -3.25) {$S$};
		\node [style=none] (27) at (2.5, -3.25) {$S$};
		\node [style=none] (28) at (1.5, -3.25) {$U$};
	\end{pgfonlayer}
	\begin{pgfonlayer}{edgelayer}
		\draw [style=rede] (19.center) to (20.center);
		\draw [style=rede] (21.center) to (22.center);
		\draw [style=bluee] (2.center) to (3.center);
		\draw [style=bluee] (14.center) to (15.center);
		\draw [style=darkgreenline] (8.center) to (9.center);
		\draw [style=darkgreenline] (11.center) to (10.center);
		\draw [style=darkgreenline] (6.center) to (7.center);
	\end{pgfonlayer}
\end{tikzpicture}}
\label{balancedguyyo}
\end{equation}
   The mirror quiver has the following bouquet:
\begin{equation}
\raisebox{-.5\height}{\begin{tikzpicture}
	\begin{pgfonlayer}{nodelayer}
		\node [style=gauge1] (0) at (4.075, -3) {};
		\node [style=none] (1) at (3.075, -3) {};
		\node [style=gauge1] (3) at (5.325, -1.5) {};
		\node [style=gauge1] (4) at (4.075, -1.5) {};
		\node [style=gauge1] (5) at (2.8, -1.5) {};
		\node [style=none] (6) at (4.2, -1.5) {};
		\node [style=none] (7) at (3.95, -1.5) {};
		\node [style=none] (8) at (4.2, -2.925) {};
		\node [style=none] (9) at (3.95, -2.925) {};
		\node [style=none] (10) at (5.375, -1.675) {};
		\node [style=none] (11) at (5.125, -1.525) {};
		\node [style=none] (12) at (2.975, -1.5) {};
		\node [style=none] (13) at (2.7, -1.7) {};
		\node [style=none] (14) at (4.3, -3.1) {};
		\node [style=none] (15) at (4.1, -2.875) {};
		\node [style=none] (16) at (4.05, -2.875) {};
		\node [style=none] (17) at (3.85, -3.125) {};
		\node [style=none] (18) at (5.325, -1) {1};
		\node [style=none] (19) at (4.075, -1) {1};
		\node [style=none] (20) at (2.825, -1) {1};
		\node [style=none] (21) at (3.975, -3) {};
		\node [style=none] (22) at (2.825, -1.6) {};
	\end{pgfonlayer}
	\begin{pgfonlayer}{edgelayer}
		\draw [style=new edge style 1] (1.center) to (0);
		\draw (6.center) to (8.center);
		\draw (9.center) to (7.center);
		\draw (10.center) to (14.center);
		\draw (11.center) to (15.center);
		\draw (12.center) to (16.center);
		\draw (17.center) to (13.center);
		\draw (22.center) to (21.center);
	\end{pgfonlayer}
\end{tikzpicture}}
\label{firstreducedexample}
\end{equation}
where the rest of the quiver remains the same. Notice that the order of the links with multiplicities (triple, double, double) in \eqref{firstreducedexample} is in reverse to the multiplicities read from  \eqref{balancedguyyo}. This order reversal is just to be consistent with the way the brane webs are drawn throughout this paper and in the Mathematica code. The reverse order does not make a difference here since all gauge nodes are balanced. However, this becomes important below when there are overbalanced nodes. We can illustrate this with $T(SU(7))$ with a particular choice of U/SU:
\begin{equation}
\raisebox{-.5\height}{\begin{tikzpicture}
	\begin{pgfonlayer}{nodelayer}
		\node [style=gauge1] (0) at (-3, 0) {};
		\node [style=gauge1] (1) at (-1.5, 0) {};
		\node [style=gauge1] (2) at (0, 0) {};
		\node [style=gauge1] (3) at (1.5, 0) {};
		\node [style=none] (12) at (-2, -2.5) {};
		\node [style=none] (13) at (0, -2.5) {};
		\node [style=none] (25) at (-3, -0.5) {U(1)};
		\node [style=none] (26) at (-1.5, -0.5) {SU(2)};
		\node [style=none] (27) at (0, -0.5) {U(3)};
		\node [style=none] (28) at (1.5, -0.5) {SU(4)};
		\node [style=gauge1] (33) at (3, 0) {};
		\node [style=gauge1] (34) at (4.5, 0) {};
		\node [style=flavour1] (35) at (6, 0) {};
		\node [style=none] (36) at (3, -0.5) {U(5)};
		\node [style=none] (37) at (4.5, -0.5) {U(6)};
		\node [style=none] (38) at (6, -0.5) {7};
		\node [style=none] (39) at (-1, -3.5) {3d mirror};
		\node [style=gauge1] (40) at (2.575, -3.5) {};
		\node [style=gauge1] (41) at (4.25, -3.5) {};
		\node [style=gauge1] (42) at (5.75, -3.5) {};
		\node [style=gauge1] (43) at (7.25, -3.5) {};
		\node [style=none] (44) at (2.575, -4) {6};
		\node [style=none] (45) at (4.25, -4) {5};
		\node [style=none] (46) at (5.75, -4) {4};
		\node [style=none] (47) at (7.25, -4) {3};
		\node [style=gauge1] (48) at (8.75, -3.5) {};
		\node [style=gauge1] (49) at (10.25, -3.5) {};
		\node [style=none] (51) at (8.75, -4) {2};
		\node [style=none] (52) at (10.25, -4) {1};
		\node [style=gauge1] (64) at (2.575, -3.5) {};
		\node [style=gauge1] (67) at (3.825, -2) {};
		\node [style=gauge1] (68) at (2.575, -2) {};
		\node [style=gauge1] (69) at (1.3, -2) {};
		\node [style=none] (70) at (2.7, -2) {};
		\node [style=none] (71) at (2.45, -2) {};
		\node [style=none] (72) at (2.7, -3.425) {};
		\node [style=none] (73) at (2.45, -3.425) {};
		\node [style=none] (74) at (3.875, -2.175) {};
		\node [style=none] (75) at (3.625, -2.025) {};
		\node [style=none] (76) at (1.475, -2) {};
		\node [style=none] (77) at (1.2, -2.2) {};
		\node [style=none] (78) at (2.8, -3.6) {};
		\node [style=none] (79) at (2.6, -3.375) {};
		\node [style=none] (80) at (2.55, -3.375) {};
		\node [style=none] (81) at (2.35, -3.625) {};
		\node [style=none] (82) at (3.825, -1.5) {1};
		\node [style=none] (83) at (2.575, -1.5) {1};
		\node [style=none] (84) at (1.325, -1.5) {1};
		\node [style=none] (85) at (2.475, -3.5) {};
		\node [style=none] (86) at (1.325, -2.1) {};
	\end{pgfonlayer}
	\begin{pgfonlayer}{edgelayer}
		\draw (0) to (1);
		\draw (2) to (3);
		\draw (1) to (2);
		\draw (3) to (33);
		\draw (33) to (34);
		\draw (34) to (35);
		\draw [style=->] (13.center) to (12.center);
		\draw [style=->] (12.center) to (13.center);
		\draw (40) to (41);
		\draw (42) to (43);
		\draw (41) to (42);
		\draw (43) to (48);
		\draw (48) to (49);
		\draw (70.center) to (72.center);
		\draw (73.center) to (71.center);
		\draw (74.center) to (78.center);
		\draw (75.center) to (79.center);
		\draw (76.center) to (80.center);
		\draw (81.center) to (77.center);
		\draw (86.center) to (85.center);
	\end{pgfonlayer}
\end{tikzpicture}}
\end{equation}
For $T(SU(n))$ theories, we can easily write down the prescription for any general U/SU combination:
\begin{equation}
\scalebox{0.75}[1]{
    \begin{tikzpicture}
	\begin{pgfonlayer}{nodelayer}
		\node [style=mini] (0) at (0.5, 0) {};
		\node [style=mini] (1) at (1.5, 0) {};
		\node [style=mini] (2) at (3.5, 0) {};
		\node [style=miniG] (3) at (4.5, 0) {};
		\node [style=mini] (4) at (5.5, 0) {};
		\node [style=mini] (5) at (7.5, 0) {};
		\node [style=miniG] (6) at (8.5, 0) {};
		\node [style=mini] (7) at (9.5, 0) {};
		\node [style=none] (8) at (14.5, 0) {};
		\node [style=mini] (9) at (15.5, 0) {};
		\node [style=miniG] (10) at (16.5, 0) {};
		\node [style=mini] (11) at (17.5, 0) {};
		\node [style=mini] (12) at (19.5, 0) {};
		\node [style=mini] (13) at (11.5, 0) {};
		\node [style=miniG] (14) at (12.25, 0) {};
		\node [style=none] (15) at (13.25, 0) {};
		\node [style=none] (16) at (0.5, 0.45) {};
		\node [style=none] (17) at (3.5, 0.45) {};
		\node [style=none] (18) at (5.5, 0.3) {};
		\node [style=none] (19) at (7.5, 0.3) {};
		\node [style=none] (22) at (14, 0.3) {};
		\node [style=none] (23) at (15.5, 0.3) {};
		\node [style=none] (26) at (2, 0.75) {$a_1-1$};
		\node [style=none] (27) at (6.5, 0.75) {$a_2-1$};
		\node [style=none] (29) at (14.75, 0.75) {$a_{k-1}-1$};
		\node [style=flavour1] (31) at (21, 0) {};
		\node [style=none] (32) at (21, -0.75) {$n$};
		\node [style=mini] (33) at (4.5, 0) {};
		\node [style=mini] (34) at (8.5, 0) {};
		\node [style=mini] (35) at (12.25, 0) {};
		\node [style=mini] (36) at (16.5, 0) {};
		\node [style=none] (37) at (0.5, -0.5) {U};
		\node [style=none] (38) at (1.5, -0.5) {U};
		\node [style=none] (39) at (3.5, -0.5) {U};
		\node [style=none] (40) at (4.5, -0.5) {S};
		\node [style=none] (41) at (5.5, -0.5) {U};
		\node [style=none] (42) at (7.5, -0.5) {U};
		\node [style=none] (43) at (8.5, -0.5) {S};
		\node [style=none] (44) at (9.5, -0.5) {U};
		\node [style=none] (45) at (11.5, -0.5) {U};
		\node [style=none] (46) at (12.25, -0.5) {S};
		\node [style=none] (47) at (15.5, -0.5) {U};
		\node [style=none] (48) at (16.5, -0.5) {S};
		\node [style=none] (49) at (9.5, 0.3) {};
		\node [style=none] (50) at (11.5, 0.3) {};
		\node [style=none] (51) at (10.5, 0.75) {$a_3-1$};
		\node [style=none] (52) at (17.5, 0.3) {};
		\node [style=none] (53) at (19.5, 0.3) {};
		\node [style=none] (54) at (18.5, 0.75) {$a_{k}-1$};
		\node [style=none] (55) at (17.5, -0.5) {U};
		\node [style=none] (56) at (19.5, -0.5) {U};
	\end{pgfonlayer}
	\begin{pgfonlayer}{edgelayer}
		\draw (0) to (1);
		\draw [style=new edge style 1] (1) to (2);
		\draw (3) to (4);
		\draw (3) to (2);
		\draw [style=new edge style 1] (4) to (5);
		\draw [style=new edge style 1] (8.center) to (9);
		\draw [style=new edge style 1] (11) to (12);
		\draw (11) to (10);
		\draw (9) to (10);
		\draw (6) to (7);
		\draw (5) to (6);
		\draw [style=new edge style 1] (7) to (13);
		\draw [style=new edge style 1] (7) to (13);
		\draw [style=new edge style 1] (14) to (15.center);
		\draw (13) to (14);
		\draw [style=brace2] (18.center) to (19.center);
		\draw [style=brace2] (22.center) to (23.center);
		\draw (12) to (31);
		\draw [style=brace2] (16.center) to (17.center);
		\draw [style=brace2] (49.center) to (50.center);
		\draw [style=brace2] (52.center) to (53.center);
	\end{pgfonlayer}
\end{tikzpicture}}
\end{equation}
where $a_i-1$ counts the number of unitary nodes in between neighboring special unitary nodes such that
\begin{equation}
    \sum^k_{i=1}a_i = n
\end{equation}
and the gauge node starts with $\mathrm{(U \; or\; SU)}(1),\mathrm{(U \; or\; SU)}(2),\dots, \mathrm{(U \; or\; SU)}(n-1)$. The mirror quiver takes the following form:
\begin{equation}
\raisebox{-.5\height}{
\begin{tikzpicture}
	\begin{pgfonlayer}{nodelayer}
		\node [style=gauge1] (0) at (-4.5, 0) {};
		\node [style=gauge1] (1) at (-3.25, 0) {};
		\node [style=gauge1] (2) at (-2, 0) {};
		\node [style=gauge1] (3) at (0, 0) {};
		\node [style=none] (5) at (-4.5, -0.5) {1};
		\node [style=none] (6) at (-3.25, -0.5) {2};
		\node [style=none] (7) at (-2, -0.5) {3};
		\node [style=none] (8) at (-0.45, -0.5) {$n{-}1$};
		\node [style=gauge1] (9) at (-0.75, 1.5) {};
		\node [style=gauge1] (10) at (0.75, 1.5) {};
		\node [style=gauge1] (11) at (1.5, -1) {};
		\node [style=gauge1] (12) at (0.25, -1.75) {};
		\node [style=gauge1] (13) at (1.75, 1) {};
		\node [style=none] (14) at (-0.675, 0.725) {$a_1$};
		\node [style=none] (15) at (0.2, 1.05) {$a_2$};
		\node [style=none] (16) at (0.95, 0.825) {$a_3$};
		\node [style=none] (17) at (1.25, -0.5) {$a_{k-1}$};
		\node [style=none] (18) at (0.5, -1.25) {$a_k$};
		\node [style=none] (19) at (-1, 2) {1};
		\node [style=none] (20) at (1, 2) {1};
		\node [style=none] (21) at (2.25, 1.25) {1};
		\node [style=none] (22) at (1.85, -1.4) {1};
		\node [style=none] (23) at (0.35, -2.25) {1};
		\node [style=dotsize] (24) at (2.1, 0.5) {};
		\node [style=dotsize] (25) at (2.225, -0.125) {};
		\node [style=dotsize] (26) at (1.975, -0.725) {};
	\end{pgfonlayer}
	\begin{pgfonlayer}{edgelayer}
		\draw (0) to (1);
		\draw (1) to (2);
		\draw [style=new edge style 1] (2) to (3);
		\draw (9) to (3);
		\draw (3) to (10);
		\draw (3) to (11);
		\draw (3) to (12);
		\draw (13) to (3);
	\end{pgfonlayer}
\end{tikzpicture}}
\end{equation}

\subsubsection*{Linear quiver where all gauge nodes are balanced}
A linear quiver where \textit{all} $k$ gauge nodes are unitary and balanced has a 3d mirror with only one $\mathrm{SU}(k+1)$ flavor node. Written as an unframed quiver, this means there is a $\mathrm{U}(1)$ gauge group connected to \emph{one} of the other gauge groups with $k+1$ links. This is because the Coulomb branch global symmetry of an electric theory with $k$ consecutive balanced nodes is $\mathrm{SU}(k+1)$. The Higgs branch global symmetry of the mirror theory is $\mathrm{SU}(k+1)$ as well, which translates to a single flavor node\footnote{Recall, the global symmetry of the Coulomb branch can be read off from the number of balanced nodes in the quiver. The global symmetry of the Higgs branch is the same as the flavor symmetry.}. If $k$ unitary gauge nodes are replaced with special unitary nodes, then the single flavor node in the mirror will become a bouquet of $k$ $\mathrm{U}(1)$s with multiplicities $a_i$ such that $\sum_i^ka_i=n$. The procedure described above in this subsection can then be straightforwardly applied to any linear quiver where all gauge nodes are balanced. 
For example, consider the following balanced quiver and its mirror:
\begin{equation}
  \scalebox{0.9}{ \begin{tikzpicture}
	\begin{pgfonlayer}{nodelayer}
		\node [style=gauge1] (0) at (-9.5, -3.75) {};
		\node [style=gauge1] (1) at (-8, -3.75) {};
		\node [style=gauge1] (2) at (-6.5, -3.75) {};
		\node [style=gauge1] (3) at (-5, -3.75) {};
		\node [style=none] (12) at (-2.25, -3.75) {};
		\node [style=none] (13) at (-0.25, -3.75) {};
		\node [style=none] (25) at (-9.5, -4.25) {U(3)};
		\node [style=none] (27) at (-6.5, -4.25) {U(3)};
		\node [style=gauge1] (33) at (-3.5, -3.75) {};
		\node [style=none] (39) at (-1.25, -4.5) {3d mirror};
		\node [style=gauge1] (41) at (1, -3.75) {};
		\node [style=gauge1] (42) at (2.5, -3.75) {};
		\node [style=gauge1] (43) at (4, -3.75) {};
		\node [style=none] (45) at (1, -4.25) {1};
		\node [style=none] (46) at (2.5, -4.25) {2};
		\node [style=none] (47) at (4, -4.25) {3};
		\node [style=gauge1] (48) at (5.5, -3.75) {};
		\node [style=gauge1] (49) at (7, -3.75) {};
		\node [style=none] (51) at (5.5, -4.25) {2};
		\node [style=none] (52) at (7, -4.25) {1};
		\node [style=none] (87) at (-8, -4.25) {U(3)};
		\node [style=none] (88) at (-5, -4.25) {U(3)};
		\node [style=none] (89) at (-3.5, -4.25) {U(3)};
		\node [style=flavour1] (90) at (-9.5, -2.25) {};
		\node [style=flavour1] (91) at (-3.5, -2.25) {};
		\node [style=none] (92) at (-3.5, -1.75) {3};
		\node [style=none] (93) at (-9.5, -1.75) {3};
		\node [style=gauge1] (94) at (4, -2.25) {};
		\node [style=none] (95) at (4.25, -3) {6};
		\node [style=none] (96) at (4, -1.75) {1};
	\end{pgfonlayer}
	\begin{pgfonlayer}{edgelayer}
		\draw (0) to (1);
		\draw (2) to (3);
		\draw (1) to (2);
		\draw (3) to (33);
		\draw [style=->] (13.center) to (12.center);
		\draw [style=->] (12.center) to (13.center);
		\draw (42) to (43);
		\draw (41) to (42);
		\draw (43) to (48);
		\draw (48) to (49);
		\draw (90) to (0);
		\draw (91) to (33);
		\draw (94) to (43);
	\end{pgfonlayer}
\end{tikzpicture}
}
\end{equation}
and when all gauge nodes are SU\footnote{The quiver on the right is also the 3d mirror of the $A_2$ class $\mathcal{S}$ theory with 2 maximal and 6 minimal punctures}:
\begin{equation}
   \scalebox{0.9}{\begin{tikzpicture}
	\begin{pgfonlayer}{nodelayer}
		\node [style=gauge1] (0) at (-9.5, -3.75) {};
		\node [style=gauge1] (1) at (-8, -3.75) {};
		\node [style=gauge1] (2) at (-6.5, -3.75) {};
		\node [style=gauge1] (3) at (-5, -3.75) {};
		\node [style=none] (12) at (-2.25, -3.75) {};
		\node [style=none] (13) at (-0.25, -3.75) {};
		\node [style=none] (25) at (-9.5, -4.25) {SU(3)};
		\node [style=none] (27) at (-6.5, -4.25) {SU(3)};
		\node [style=gauge1] (33) at (-3.5, -3.75) {};
		\node [style=none] (39) at (-1.25, -4.5) {3d mirror};
		\node [style=gauge1] (41) at (1, -3.75) {};
		\node [style=gauge1] (42) at (2.5, -3.75) {};
		\node [style=gauge1] (43) at (4, -3.75) {};
		\node [style=none] (45) at (1, -4.25) {1};
		\node [style=none] (46) at (2.5, -4.25) {2};
		\node [style=none] (47) at (4, -4.25) {3};
		\node [style=gauge1] (48) at (5.5, -3.75) {};
		\node [style=gauge1] (49) at (7, -3.75) {};
		\node [style=none] (51) at (5.5, -4.25) {2};
		\node [style=none] (52) at (7, -4.25) {1};
		\node [style=none] (87) at (-8, -4.25) {SU(3)};
		\node [style=none] (88) at (-5, -4.25) {SU(3)};
		\node [style=none] (89) at (-3.5, -4.25) {SU(3)};
		\node [style=flavour1] (90) at (-9.5, -2.25) {};
		\node [style=flavour1] (91) at (-3.5, -2.25) {};
		\node [style=none] (92) at (-3.5, -1.75) {3};
		\node [style=none] (93) at (-9.5, -1.75) {3};
		\node [style=gauge1] (94) at (2.5, -2.5) {};
		\node [style=gauge1] (95) at (3.125, -2.5) {};
		\node [style=gauge1] (96) at (3.675, -2.5) {};
		\node [style=gauge1] (97) at (4.325, -2.5) {};
		\node [style=gauge1] (98) at (4.875, -2.5) {};
		\node [style=gauge1] (99) at (5.5, -2.5) {};
		\node [style=none] (100) at (2.5, -2) {1};
		\node [style=none] (101) at (3.125, -2) {1};
		\node [style=none] (102) at (3.675, -2) {1};
		\node [style=none] (103) at (4.325, -2) {1};
		\node [style=none] (104) at (4.875, -2) {1};
		\node [style=none] (105) at (5.5, -2) {1};
	\end{pgfonlayer}
	\begin{pgfonlayer}{edgelayer}
		\draw (0) to (1);
		\draw (2) to (3);
		\draw (1) to (2);
		\draw (3) to (33);
		\draw [style=->] (13.center) to (12.center);
		\draw [style=->] (12.center) to (13.center);
		\draw (42) to (43);
		\draw (41) to (42);
		\draw (43) to (48);
		\draw (48) to (49);
		\draw (90) to (0);
		\draw (91) to (33);
		\draw (94) to (43);
		\draw (43) to (95);
		\draw (96) to (43);
		\draw (43) to (97);
		\draw (98) to (43);
		\draw (99) to (43);
	\end{pgfonlayer}
\end{tikzpicture}}
\end{equation}
An arbitrary selection of U/SU gives a bouquet of $\mathrm{U}(1)$s connected to the $\mathrm{U}(3)$ with links of different multiplicities. For instance, 
\begin{equation}
\scalebox{0.9}{\begin{tikzpicture}
	\begin{pgfonlayer}{nodelayer}
		\node [style=gauge1] (0) at (-9.5, -3.75) {};
		\node [style=gauge1] (1) at (-8, -3.75) {};
		\node [style=gauge1] (2) at (-6.5, -3.75) {};
		\node [style=gauge1] (3) at (-5, -3.75) {};
		\node [style=none] (12) at (-2.25, -3.75) {};
		\node [style=none] (13) at (-0.25, -3.75) {};
		\node [style=none] (25) at (-9.5, -4.25) {U(3)};
		\node [style=none] (27) at (-6.5, -4.25) {SU(3)};
		\node [style=gauge1] (33) at (-3.5, -3.75) {};
		\node [style=none] (39) at (-1.25, -4.5) {3d mirror};
		\node [style=gauge1] (41) at (7, -3.75) {};
		\node [style=gauge1] (42) at (5.5, -3.75) {};
		\node [style=gauge1] (43) at (4, -3.75) {};
		\node [style=none] (45) at (7, -4.25) {1};
		\node [style=none] (46) at (5.5, -4.25) {2};
		\node [style=none] (47) at (4, -4.25) {3};
		\node [style=gauge1] (48) at (2.5, -3.75) {};
		\node [style=gauge1] (49) at (1, -3.75) {};
		\node [style=none] (51) at (2.5, -4.25) {2};
		\node [style=none] (52) at (1, -4.25) {1};
		\node [style=none] (87) at (-8, -4.25) {SU(3)};
		\node [style=none] (88) at (-5, -4.25) {U(3)};
		\node [style=none] (89) at (-3.5, -4.25) {SU(3)};
		\node [style=flavour1] (90) at (-9.5, -2.25) {};
		\node [style=flavour1] (91) at (-3.5, -2.25) {};
		\node [style=none] (92) at (-3.5, -1.75) {3};
		\node [style=none] (93) at (-9.5, -1.75) {3};
		\node [style=gauge1] (97) at (4, -3.75) {};
		\node [style=gauge1] (98) at (4, -3.75) {};
		\node [style=gauge1] (99) at (5.5, -2.25) {};
		\node [style=gauge1] (100) at (3.475, -2.025) {};
		\node [style=none] (101) at (3.6, -2.025) {};
		\node [style=none] (102) at (3.35, -2.025) {};
		\node [style=none] (103) at (4.125, -3.675) {};
		\node [style=none] (104) at (4, -3.75) {};
		\node [style=none] (105) at (5.55, -2.425) {};
		\node [style=none] (106) at (5.3, -2.275) {};
		\node [style=none] (107) at (4.125, -3.875) {};
		\node [style=none] (108) at (4.025, -3.625) {};
		\node [style=none] (109) at (5.5, -1.75) {1};
		\node [style=none] (110) at (3.5, -1.5) {1};
		\node [style=gauge1] (111) at (4.5, -2) {};
		\node [style=gauge1] (112) at (2.6, -2.325) {};
		\node [style=none] (113) at (4.5, -1.5) {1};
		\node [style=none] (114) at (2.6, -1.825) {1};
		\node [style=none] (115) at (3.925, -3.825) {};
	\end{pgfonlayer}
	\begin{pgfonlayer}{edgelayer}
		\draw (0) to (1);
		\draw (2) to (3);
		\draw (1) to (2);
		\draw (3) to (33);
		\draw [style=->] (13.center) to (12.center);
		\draw [style=->] (12.center) to (13.center);
		\draw (42) to (43);
		\draw (41) to (42);
		\draw (43) to (48);
		\draw (48) to (49);
		\draw (90) to (0);
		\draw (91) to (33);
		\draw (101.center) to (103.center);
		\draw (105.center) to (107.center);
		\draw (106.center) to (108.center);
		\draw (111) to (104.center);
		\draw (112) to (104.center);
		\draw (102.center) to (115.center);
	\end{pgfonlayer}
\end{tikzpicture}}
\end{equation}

\subsubsection*{One or more overbalanced nodes}
As mentioned above, for good linear quivers the different combinations of U/SU only affect the way $\mathrm{U}(1)$ bouquets behave
in the mirror.  When all gauge nodes are balanced, there is only a single bouquet. When there are one or more overbalanced gauge nodes then there are more bouquets in the mirror. For the remainder of this subsection, it is sufficient just to focus on the different kinds of bouquets that can arise under different combinations of U/SU and balanced/overbalanced nodes.

A good linear quiver with only unitary gauge nodes with one or more being overbalanced, has a mirror quiver with more than one flavor node. Written as an unframed quiver, this means the $\mathrm{U}(1)$ node connects to several other gauge nodes. Let us start with a linear theory with gauge nodes $US\textcolor{cyan}{U}SUU$ where the cyan node is overbalanced. Once again, the rank of the gauge nodes and the flavor nodes do not affect the results. The configuration of NS5 branes takes the form:
\begin{equation}
 \raisebox{-.5\height}{\begin{tikzpicture}
	\begin{pgfonlayer}{nodelayer}
		\node [style=none] (0) at (1, 0) {};
		\node [style=none] (1) at (1, -3) {};
		\node [style=none] (2) at (3, 0) {};
		\node [style=none] (3) at (3, -3) {};
		\node [style=none] (4) at (2, 0) {};
		\node [style=none] (5) at (2, -3) {};
		\node [style=none] (6) at (5, 0) {};
		\node [style=none] (7) at (5, -3) {};
		\node [style=none] (8) at (7, 0) {};
		\node [style=none] (9) at (7, -3) {};
		\node [style=none] (10) at (6, 0) {};
		\node [style=none] (11) at (6, -3) {};
		\node [style=none] (14) at (4, 0) {};
		\node [style=none] (15) at (4, -3) {};
		\node [style=none] (19) at (1, 0) {};
		\node [style=none] (20) at (1, -3) {};
		\node [style=none] (21) at (2, 0) {};
		\node [style=none] (22) at (2, -3) {};
		\node [style=none] (23) at (2.5, -3.25) {$S$};
		\node [style=none] (24) at (4.5, -3.25) {$S$};
		\node [style=none] (25) at (1.5, -3.25) {$U$};
		\node [style=none] (26) at (3.5, -3.25) {$\textcolor{cyan}{U}$};
		\node [style=none] (27) at (5.5, -3.25) {$U$};
		\node [style=none] (28) at (6.5, -3.25) {$U$};
	\end{pgfonlayer}
	\begin{pgfonlayer}{edgelayer}
		\draw [style=rede] (19.center) to (20.center);
		\draw [style=rede] (21.center) to (22.center);
		\draw [style=bluee] (2.center) to (3.center);
		\draw [style=bluee] (14.center) to (15.center);
		\draw [style=darkgreenline] (8.center) to (9.center);
		\draw [style=darkgreenline] (11.center) to (10.center);
		\draw [style=darkgreenline] (6.center) to (7.center);
	\end{pgfonlayer}
\end{tikzpicture}}
\label{unbalancedex}
\end{equation}
It is now important to pay attention to the interval between the two blue NS branes.
This is now an additional information beyond the color coding of the NS5s that needs to be taken into consideration. 
The number of gauge nodes in the mirror theory that have $\mathrm{U}(1)$ bouquets  is $u+1$ where $u$ is the number of overbalanced nodes in the electric quiver. Here, we have two bouquets in the mirror quiver. The novelty compared to the all balanced case is that the $\mathrm{U}(1)$ nodes in the bouquet may have edges connected to more than one gauge node. For (\ref{unbalancedex}) the mirror quiver always has the following bouquets:
\begin{equation}
\raisebox{-.5\height}{\begin{tikzpicture}
	\begin{pgfonlayer}{nodelayer}
		\node [style=gauge1] (0) at (5, -3) {};
		\node [style=none] (1) at (4, -3) {};
		\node [style=none] (2) at (6, -3) {};
		\node [style=gauge1] (3) at (6.25, -1.5) {};
		\node [style=gauge1] (4) at (2.8, -1.5) {};
		\node [style=none] (5) at (5.125, -2.925) {};
		\node [style=none] (6) at (6.3, -1.675) {};
		\node [style=none] (7) at (6.05, -1.525) {};
		\node [style=none] (8) at (2.975, -1.5) {};
		\node [style=none] (9) at (2.7, -1.7) {};
		\node [style=none] (10) at (5.2, -3.1) {};
		\node [style=none] (11) at (5.025, -2.875) {};
		\node [style=none] (12) at (4.05, -2.825) {};
		\node [style=none] (13) at (3.85, -3.075) {};
		\node [style=none] (14) at (6.25, -1) {1};
		\node [style=none] (15) at (4.5, -1) {1};
		\node [style=none] (16) at (3.975, -2.95) {};
		\node [style=none] (17) at (2.825, -1.6) {};
		\node [style=gauge1] (18) at (4, -3) {};
		\node [style=none] (19) at (3, -3) {};
		\node [style=gauge1] (20) at (4.5, -1.5) {};
		\node [style=none] (21) at (2.7, -1) {1};
	\end{pgfonlayer}
	\begin{pgfonlayer}{edgelayer}
		\draw [style=new edge style 1] (0) to (2.center);
		\draw [style=new edge style 1] (1.center) to (0);
		\draw (6.center) to (10.center);
		\draw (7.center) to (11.center);
		\draw (8.center) to (12.center);
		\draw (13.center) to (9.center);
		\draw (17.center) to (16.center);
		\draw [style=new edge style 1] (18) to (19.center);
		\draw (20) to (0);
		\draw (20) to (18);
	\end{pgfonlayer}
\end{tikzpicture}}
\label{generalmirrorstructure}
\end{equation}
Depending on the ranks of the gauge groups and flavor groups in the electric theory, there can be many gauge nodes between the two unmarked nodes in (\ref{generalmirrorstructure}) but they will not have any links to the three $\mathrm{U}(1)$s. In other words, they won't have any $\mathrm{U}(1)$ bouquets irrespective of the U/SU combination in the electric quiver.

Next up, we place two unbalanced unitary nodes next to each other. For a quiver with  $US\textcolor{cyan}{U}\textcolor{cyan}{U}UU$, the NS5 configuration is:
\begin{equation}
\raisebox{-.5\height}{\begin{tikzpicture}
	\begin{pgfonlayer}{nodelayer}
		\node [style=none] (0) at (1, 0) {};
		\node [style=none] (1) at (1, -3) {};
		\node [style=none] (2) at (3, 0) {};
		\node [style=none] (3) at (3, -3) {};
		\node [style=none] (4) at (2, 0) {};
		\node [style=none] (5) at (2, -3) {};
		\node [style=none] (6) at (5, 0) {};
		\node [style=none] (7) at (5, -3) {};
		\node [style=none] (8) at (7, 0) {};
		\node [style=none] (9) at (7, -3) {};
		\node [style=none] (10) at (6, 0) {};
		\node [style=none] (11) at (6, -3) {};
		\node [style=none] (14) at (4, 0) {};
		\node [style=none] (15) at (4, -3) {};
		\node [style=none] (19) at (1, 0) {};
		\node [style=none] (20) at (1, -3) {};
		\node [style=none] (21) at (2, 0) {};
		\node [style=none] (22) at (2, -3) {};
		\node [style=none] (23) at (2.5, -3.25) {$S$};
		\node [style=none] (25) at (1.5, -3.25) {$U$};
		\node [style=none] (26) at (3.5, -3.25) {$\textcolor{cyan}{U}$};
		\node [style=none] (27) at (5.5, -3.25) {$U$};
		\node [style=none] (28) at (6.5, -3.25) {$U$};
		\node [style=none] (29) at (4.5, -3.25) {$\textcolor{cyan}{U}$};
	\end{pgfonlayer}
	\begin{pgfonlayer}{edgelayer}
		\draw [style=rede] (19.center) to (20.center);
		\draw [style=rede] (21.center) to (22.center);
		\draw [style=bluee] (2.center) to (3.center);
		\draw [style=bluee] (14.center) to (15.center);
		\draw [style=bluee] (8.center) to (9.center);
		\draw [style=bluee] (11.center) to (10.center);
		\draw [style=bluee] (6.center) to (7.center);
	\end{pgfonlayer}
\end{tikzpicture}}
\end{equation}
Following the same set of rules, there are three gauge nodes in the mirror with bouquets. The mirror quiver has the structure:
\begin{equation}
\raisebox{-.5\height}{\begin{tikzpicture}
	\begin{pgfonlayer}{nodelayer}
		\node [style=gauge1] (0) at (5.7, -3) {};
		\node [style=none] (1) at (3.7, -3) {};
		\node [style=none] (2) at (6.7, -3) {};
		\node [style=gauge1] (3) at (6.95, -1.5) {};
		\node [style=none] (8) at (5.825, -2.925) {};
		\node [style=none] (10) at (7, -1.675) {};
		\node [style=none] (11) at (6.75, -1.525) {};
		\node [style=none] (14) at (5.9, -3.1) {};
		\node [style=none] (15) at (5.725, -2.875) {};
		\node [style=none] (18) at (6.95, -1) {1};
		\node [style=none] (20) at (4.7, -1) {1};
		\node [style=gauge1] (23) at (3.7, -3) {};
		\node [style=none] (24) at (2.7, -3) {};
		\node [style=gauge1] (25) at (4.7, -1.5) {};
		\node [style=gauge1] (27) at (4.7, -3) {};
		\node [style=gauge1] (28) at (3.7, -3) {};
		\node [style=none] (29) at (3.75, -3.15) {};
		\node [style=none] (30) at (3.475, -2.95) {};
		\node [style=none] (31) at (4.825, -1.575) {};
		\node [style=none] (32) at (4.625, -1.325) {};
		\node [style=none] (33) at (4.75, -1.45) {};
		\node [style=none] (34) at (3.6, -3.05) {};
	\end{pgfonlayer}
	\begin{pgfonlayer}{edgelayer}
		\draw [style=new edge style 1] (0) to (2.center);
		\draw [style=new edge style 1] (1.center) to (0);
		\draw (10.center) to (14.center);
		\draw (11.center) to (15.center);
		\draw [style=new edge style 1] (23) to (24.center);
		\draw (25) to (0);
		\draw (25) to (27);
		\draw (29.center) to (31.center);
		\draw (32.center) to (30.center);
		\draw (34.center) to (33.center);
	\end{pgfonlayer}
\end{tikzpicture}}
\end{equation}
\subsubsection*{Unbalanced special unitary gauge group}
If we look at $US\textcolor{cyan}{S}SUU$ where the overbalanced node is now special unitary, the NS5s read:
\begin{equation}
   \raisebox{-.5\height}{\begin{tikzpicture}
	\begin{pgfonlayer}{nodelayer}
		\node [style=none] (0) at (1, 0) {};
		\node [style=none] (1) at (1, -3) {};
		\node [style=none] (2) at (3, 0) {};
		\node [style=none] (3) at (3, -3) {};
		\node [style=none] (4) at (2, 0) {};
		\node [style=none] (5) at (2, -3) {};
		\node [style=none] (6) at (5, 0) {};
		\node [style=none] (7) at (5, -3) {};
		\node [style=none] (8) at (7, 0) {};
		\node [style=none] (9) at (7, -3) {};
		\node [style=none] (10) at (6, 0) {};
		\node [style=none] (11) at (6, -3) {};
		\node [style=none] (14) at (4, 0) {};
		\node [style=none] (15) at (4, -3) {};
		\node [style=none] (19) at (1, 0) {};
		\node [style=none] (20) at (1, -3) {};
		\node [style=none] (21) at (2, 0) {};
		\node [style=none] (22) at (2, -3) {};
		\node [style=none] (23) at (2.5, -3.25) {$S$};
		\node [style=none] (25) at (1.5, -3.25) {$U$};
		\node [style=none] (26) at (3.5, -3.25) {$\textcolor{cyan}{S}$};
		\node [style=none] (27) at (5.5, -3.25) {$U$};
		\node [style=none] (28) at (6.5, -3.25) {$U$};
		\node [style=none] (29) at (4.5, -3.25) {$S$};
	\end{pgfonlayer}
	\begin{pgfonlayer}{edgelayer}
		\draw [style=rede] (19.center) to (20.center);
		\draw [style=rede] (21.center) to (22.center);
		\draw [style=bluee] (2.center) to (3.center);
		\draw [style=darkgreenline] (8.center) to (9.center);
		\draw [style=darkgreenline] (11.center) to (10.center);
		\draw [style=darkgreenline] (6.center) to (7.center);
		\draw [style=oliveline] (14.center) to (15.center);
	\end{pgfonlayer}
\end{tikzpicture}}
\end{equation}
Like before, the special unitary node means the nearby NS5 branes will move independently from each other, yielding a mirror quiver with the following bouquets:
\begin{equation}
\raisebox{-.5\height}{\begin{tikzpicture}
	\begin{pgfonlayer}{nodelayer}
		\node [style=gauge1] (0) at (5, -3) {};
		\node [style=none] (1) at (4, -3) {};
		\node [style=none] (2) at (6, -3) {};
		\node [style=gauge1] (3) at (6.25, -1.5) {};
		\node [style=gauge1] (5) at (2.8, -1.5) {};
		\node [style=none] (8) at (5.125, -2.925) {};
		\node [style=none] (10) at (6.3, -1.675) {};
		\node [style=none] (11) at (6.05, -1.525) {};
		\node [style=none] (12) at (2.975, -1.5) {};
		\node [style=none] (13) at (2.7, -1.7) {};
		\node [style=none] (14) at (5.2, -3.1) {};
		\node [style=none] (15) at (5.025, -2.875) {};
		\node [style=none] (16) at (4.05, -2.825) {};
		\node [style=none] (17) at (3.85, -3.075) {};
		\node [style=none] (18) at (6.25, -1) {1};
		\node [style=none] (20) at (5, -1) {1};
		\node [style=none] (21) at (3.975, -2.95) {};
		\node [style=none] (22) at (2.825, -1.6) {};
		\node [style=gauge1] (23) at (4, -3) {};
		\node [style=none] (24) at (3, -3) {};
		\node [style=gauge1] (25) at (5, -1.5) {};
		\node [style=none] (26) at (2.7, -1) {1};
		\node [style=none] (27) at (4, -1) {1};
		\node [style=gauge1] (28) at (4, -1.5) {};
	\end{pgfonlayer}
	\begin{pgfonlayer}{edgelayer}
		\draw [style=new edge style 1] (0) to (2.center);
		\draw [style=new edge style 1] (1.center) to (0);
		\draw (10.center) to (14.center);
		\draw (11.center) to (15.center);
		\draw (12.center) to (16.center);
		\draw (17.center) to (13.center);
		\draw (22.center) to (21.center);
		\draw [style=new edge style 1] (23) to (24.center);
		\draw (25) to (0);
		\draw (28) to (23);
	\end{pgfonlayer}
\end{tikzpicture}}
\end{equation}
Therefore, an unbalanced special unitary node results in the two $\mathrm{U}(1)$ nodes being connected to two separate gauge nodes. 

\subsubsection*{Generalization}
In general, the procedure of finding the mirror pair of a good linear quiver is the following: 
\begin{enumerate}
    \item Set all the gauge nodes in the electric quiver to be unitary. This is now a $T^\sigma_\rho(SU(n))$ quiver whose mirror quiver is $T^\rho_\sigma(SU(n))$ and can be easily obtained following \cite{Gaiotto:2008ak,Cremonesi:2014uva}. 
    \item Switch all the U nodes in the electric quiver to SU which translates to exploding all the $U(N_i)$ flavor nodes in the mirror into bouquets of $N_i$ $\mathrm{U}(1)$s. 
    \item Identify which gauge groups in the electric quiver are overbalanced. 
    \item Draw the \emph{incomplete} brane configurations introduced above with only NS5 branes and alter the bouquets depending on the U/SU and balanced/overbalanced conditions outlined above.  This reproduces the mirror quiver of the mixed U/SU electric theory. 
\end{enumerate} 

\subsubsection*{Reverse algorithm}
To reiterate, the procedure above is completely included in the algorithm in Section \ref{prescription}. Nevertheless, working with only linear electric quivers that are good allows us to simplify the algorithm immensely using incomplete brane configurations with only NS5s. Another advantage is that the algorithm for a good linear quiver can be reversed: given a quiver, one can decide whether it is the mirror of a good linear quiver, and if so we can find it.  
\begin{enumerate}
    \item The reverse algorithm only works if all the gauge groups are either balanced or overbalanced (we are now talking about the mirror quiver but this still needs to be true).
    Look for a set of U(1) gauge nodes, each not connected to any other in that set, such that ungauging all of them (i.e turning $\mathrm{U}(1)$ gauge groups into $\mathrm{U}(1)$ flavors) produces a framed linear quiver with only multiplicity 1 links. 
    If this is not possible, i.e. (a) the remaining gauge groups do not form a linear quiver, or (b) there are multiple links between the remaining gauge groups, then there is no mirror that is a  good linear quiver and the algorithm stops here.
    \footnote{ If there are links between two $\mathrm{U}(1)$s with multiplicity $k>0$,  then ungauging both of them gives rise to \scalebox{0.7}
    {\begin{tikzpicture}
	\begin{pgfonlayer}{nodelayer}
		\node [style=flavour1] (0) at (3, 0) {};
		\node [style=flavour1] (1) at (4, 0) {};
		\node [style=none] (2) at (2.5, 0) {1};
		\node [style=none] (3) at (3.5, 0.5) {k};
		\node [style=none] (4) at (4.5, 0) {1};
		\node [style=none] (5) at (3.025, 0.125) {};
		\node [style=none] (6) at (3.025, -0.125) {};
		\node [style=none] (7) at (4.025, 0.125) {};
		\node [style=none] (8) at (4.025, -0.125) {};
	\end{pgfonlayer}
	\begin{pgfonlayer}{edgelayer}
		\draw (0) to (1);
		\draw (5.center) to (7.center);
		\draw (6.center) to (8.center);
	\end{pgfonlayer}
\end{tikzpicture}}.
If such a feature arises when creating a linear quiver, the electric quiver will be an ugly/bad quiver and the reverse algorithm will not work. Another way to think about this is that the information contained in the links between the $\mathrm{U}(1)$s is lost after the ungauging.}
    
\item Whenever multiple $U(1)$ flavor nodes are attached to a single gauge node they should be aggregated into a single $U(k)$ flavor, taking account of linking multiplicities.

\item The resulting quiver will be a linear chain of unitary gauge nodes with flavors. If it is a good linear quiver, then it is a $T^\rho_\sigma(SU(n))$ theory. The mirror theory $T^\sigma_\rho(SU(n))$ is straightforward to obtain once the quiver is expressed using partitions $(n,\rho,\sigma)$. $T^\sigma_\rho(SU(n))$ is identical to the desired electric quiver under the reverse algorithm, but with all its gauge groups set to unitary. Therefore, the next step is to figure out which arrangement of U/SU nodes in $T^\sigma_\rho(SU(n))$ will reproduce our mirror quiver under the forward algorithm.

\item Return to the mirror quiver in the beginning but highlight the gauge nodes that form the linear chain in step 1. The $\mathrm{U}(1)$s attached to this chain will be the bouquets. This brings the quiver into a familiar form which we see throughout the paper. By studying how the bouquets connect to the rest of the quiver, we can reconstruct the incomplete brane diagram where the number of NS5s is equal to the total number of links to the $\mathrm{U}(1)$s in the  bouquets. With the incomplete brane diagram, we can now identify which of the gauge groups in the electric quiver are unitary or special unitary. Replacing the electric quiver in step 3 with the correct U/SU arrangement recovers the electric quiver. 

\end{enumerate}

We demonstrate this with the following example: 
\begin{equation}
\scalebox{0.9}{\begin{tikzpicture}
	\begin{pgfonlayer}{nodelayer}
		\node [style=gauge1] (35) at (2.975, -4.7) {};
		\node [style=none] (36) at (2.925, -4.55) {};
		\node [style=none] (37) at (3.1, -4.75) {};
		\node [style=none] (38) at (1.85, -6.15) {};
		\node [style=none] (39) at (2.05, -6.4) {};
		\node [style=none] (40) at (1.95, -6.25) {};
		\node [style=none] (41) at (3.025, -4.625) {};
		\node [style=gauge1] (42) at (2, -6.2) {};
		\node [style=gauge1] (43) at (4, -6.2) {};
		\node [style=gauge1] (44) at (2.975, -4.7) {};
		\node [style=none] (45) at (3.875, -6.125) {};
		\node [style=none] (46) at (2.925, -4.875) {};
		\node [style=none] (47) at (3.075, -4.725) {};
		\node [style=none] (48) at (3.8, -6.3) {};
		\node [style=none] (49) at (3.975, -6.075) {};
		\node [style=gauge1] (50) at (6.025, -4.675) {};
		\node [style=gauge1] (51) at (4, -6.2) {};
		\node [style=none] (52) at (3.95, -6.025) {};
		\node [style=none] (53) at (4.2, -6.175) {};
		\node [style=none] (54) at (5.925, -4.575) {};
		\node [style=none] (55) at (6.1, -4.8) {};
		\node [style=none] (56) at (5.825, -4.6) {};
		\node [style=none] (57) at (5.825, -6.35) {};
		\node [style=none] (58) at (5.95, -4.6) {};
		\node [style=none] (59) at (5.95, -6.35) {};
		\node [style=none] (60) at (6.075, -4.6) {};
		\node [style=none] (61) at (6.075, -6.35) {};
		\node [style=none] (62) at (6.2, -4.6) {};
		\node [style=none] (63) at (6.2, -6.35) {};
		\node [style=gauge1] (65) at (6, -6.2) {};
		\node [style=gauge1] (66) at (8, -6.2) {};
		\node [style=gauge1] (67) at (10, -6.2) {};
		\node [style=gauge1] (68) at (12, -6.2) {};
		\node [style=gauge1] (69) at (9.975, -4.625) {};
		\node [style=gauge1] (74) at (9.975, -4.625) {};
		\node [style=none] (75) at (9.85, -4.6) {};
		\node [style=none] (76) at (9.85, -6.35) {};
		\node [style=none] (77) at (10.1, -4.6) {};
		\node [style=none] (78) at (10.1, -6.35) {};
		\node [style=none] (79) at (9.925, -4.675) {};
		\node [style=none] (80) at (10.175, -4.525) {};
		\node [style=none] (81) at (11.875, -6.275) {};
		\node [style=none] (82) at (12.05, -6.05) {};
		\node [style=gauge1] (83) at (12, -4.7) {};
		\node [style=gauge1] (85) at (12, -6.2) {};
		\node [style=none] (86) at (2.025, -6.75) {3};
		\node [style=none] (87) at (4, -6.725) {6};
		\node [style=none] (88) at (6, -6.7) {5};
		\node [style=none] (89) at (8, -6.7) {2};
		\node [style=none] (90) at (10, -6.7) {4};
		\node [style=none] (91) at (12, -6.7) {4};
		\node [style=none] (92) at (3, -3.975) {1};
		\node [style=none] (93) at (6, -4.2) {1};
		\node [style=none] (94) at (10, -4.2) {1};
		\node [style=none] (95) at (12, -4.2) {1};
		\node [style=none] (96) at (11.875, -4.7) {};
		\node [style=none] (97) at (12.125, -4.7) {};
		\node [style=none] (98) at (11.875, -6.2) {};
		\node [style=none] (99) at (12.125, -6.2) {};
	\end{pgfonlayer}
	\begin{pgfonlayer}{edgelayer}
		\draw (36.center) to (38.center);
		\draw (39.center) to (37.center);
		\draw (41.center) to (40.center);
		\draw (46.center) to (48.center);
		\draw (47.center) to (49.center);
		\draw (52.center) to (54.center);
		\draw (53.center) to (55.center);
		\draw (56.center) to (57.center);
		\draw (58.center) to (59.center);
		\draw (60.center) to (61.center);
		\draw (62.center) to (63.center);
		\draw (42) to (51);
		\draw (51) to (65);
		\draw (65) to (66);
		\draw (66) to (67);
		\draw (67) to (68);
		\draw (74) to (66);
		\draw (75.center) to (76.center);
		\draw (77.center) to (78.center);
		\draw (79.center) to (81.center);
		\draw (80.center) to (82.center);
		\draw (96.center) to (98.center);
		\draw (97.center) to (99.center);
	\end{pgfonlayer}
\end{tikzpicture}}
\label{backwards}
\end{equation}
Step 1-2: Ungauge the $\mathrm{U}(1)$s until we have a linear quiver without links with multiplicity:
\begin{equation}\scalebox{0.9}{ \begin{tikzpicture}
	\begin{pgfonlayer}{nodelayer}
		\node [style=gauge1] (3) at (2, -6.2) {};
		\node [style=gauge1] (4) at (4, -6.2) {};
		\node [style=gauge1] (8) at (4, -6.2) {};
		\node [style=gauge1] (15) at (6, -6.2) {};
		\node [style=gauge1] (16) at (8, -6.2) {};
		\node [style=gauge1] (17) at (10, -6.2) {};
		\node [style=gauge1] (18) at (12, -6.2) {};
		\node [style=gauge1] (23) at (12, -6.2) {};
		\node [style=none] (24) at (2.025, -6.75) {3};
		\node [style=none] (25) at (4, -6.725) {6};
		\node [style=none] (26) at (6, -6.7) {5};
		\node [style=none] (27) at (8, -6.7) {2};
		\node [style=none] (28) at (10, -6.7) {4};
		\node [style=none] (29) at (12, -6.7) {4};
		\node [style=flavour1] (30) at (2, -5) {};
		\node [style=flavour1] (31) at (4, -5) {};
		\node [style=flavour1] (32) at (6, -5) {};
		\node [style=flavour1] (33) at (8, -5) {};
		\node [style=flavour1] (34) at (10, -5) {};
		\node [style=flavour1] (35) at (12, -5) {};
		\node [style=none] (36) at (2, -4.5) {3};
		\node [style=none] (37) at (4, -4.5) {4};
		\node [style=none] (38) at (6, -4.5) {4};
		\node [style=none] (39) at (8, -4.5) {1};
		\node [style=none] (40) at (10, -4.5) {2};
		\node [style=none] (41) at (12, -4.5) {4};
	\end{pgfonlayer}
	\begin{pgfonlayer}{edgelayer}
		\draw (3) to (8);
		\draw (8) to (15);
		\draw (15) to (16);
		\draw (16) to (17);
		\draw (17) to (18);
		\draw (30) to (3);
		\draw (8) to (31);
		\draw (32) to (15);
		\draw (16) to (33);
		\draw (34) to (17);
		\draw (23) to (35);
	\end{pgfonlayer}
\end{tikzpicture}}
\label{ungaugedtrial}
\end{equation}
Step 3. \eqref{ungaugedtrial} is a linear quiver where all gauge nodes are unitary and are either balanced or overbalanced and thus a $T^\rho_\sigma(SU(N))$ theory. The 3d mirror (which can be obtained either by following \cite{Gaiotto:2008ak,Cremonesi:2014uva} or using the Mathematica code) is:
\begin{equation}
  \scalebox{0.8}{\begin{tikzpicture}
	\begin{pgfonlayer}{nodelayer}
		\node [style=gauge1] (0) at (2, -4) {};
		\node [style=gauge1] (1) at (3, -4) {};
		\node [style=gauge1] (2) at (4, -4) {};
		\node [style=gauge1] (3) at (5, -4) {};
		\node [style=gauge1] (4) at (6, -4) {};
		\node [style=gauge1] (5) at (7, -4) {};
		\node [style=gauge1] (6) at (8, -4) {};
		\node [style=gauge1] (7) at (9, -4) {};
		\node [style=gauge1] (8) at (10, -4) {};
		\node [style=gauge1] (9) at (11, -4) {};
		\node [style=gauge1] (10) at (12, -4) {};
		\node [style=gauge1] (11) at (13, -4) {};
		\node [style=gauge1] (12) at (14, -4) {};
		\node [style=gauge1] (13) at (15, -4) {};
		\node [style=gauge1] (14) at (16, -4) {};
		\node [style=gauge1] (15) at (17, -4) {};
		\node [style=gauge1] (16) at (18, -4) {};
		\node [style=none] (17) at (2, -4.5) {U(1)};
		\node [style=none] (18) at (3, -4.5) {U(2)};
		\node [style=none] (19) at (4, -4.5) {U(3)};
		\node [style=none] (20) at (5, -4.5) {U(4)};
		\node [style=none] (21) at (6, -4.5) {U(3)};
		\node [style=none] (22) at (7, -4.5) {U(2)};
		\node [style=none] (23) at (8, -4.5) {U(2)};
		\node [style=none] (24) at (9, -4.5) {U(3)};
		\node [style=none] (25) at (10, -4.5) {U(4)};
		\node [style=none] (26) at (11, -4.5) {U(5)};
		\node [style=none] (27) at (12, -4.5) {U(5)};
		\node [style=none] (28) at (13, -4.5) {U(6)};
		\node [style=none] (29) at (14, -4.5) {U(5)};
		\node [style=none] (30) at (15, -4.5) {U(4)};
		\node [style=none] (31) at (16, -4.5) {U(3)};
		\node [style=none] (32) at (17, -4.5) {U(2)};
		\node [style=none] (33) at (18, -4.5) {U(1)};
		\node [style=flavour1] (35) at (13, -3) {};
		\node [style=flavour1] (36) at (11, -3) {};
		\node [style=flavour1] (37) at (5, -3) {};
		\node [style=none] (38) at (5, -2.5) {3};
		\node [style=none] (39) at (11, -2.5) {1};
		\node [style=none] (40) at (13, -2.5) {2};
		\node [style=flavour1] (41) at (16, -3) {};
		\node [style=none] (42) at (16, -2.5) {1};
	\end{pgfonlayer}
	\begin{pgfonlayer}{edgelayer}
		\draw (37) to (3);
		\draw (36) to (9);
		\draw (35) to (11);
		\draw (16) to (0);
		\draw (41) to (14);
	\end{pgfonlayer}
\end{tikzpicture}}  
\label{firstelectric}
\end{equation}
Step 4: From the structure of the four bouquets in \eqref{backwards}, we read off the following incomplete brane diagram with only NS5s:
\begin{equation}
\scalebox{0.9}{\begin{tikzpicture}
	\begin{pgfonlayer}{nodelayer}
		\node [style=none] (1) at (10.25, -1.5) {};
		\node [style=none] (2) at (10.25, -3.75) {};
		\node [style=none] (3) at (9.25, -1.5) {};
		\node [style=none] (4) at (9.25, -3.75) {};
		\node [style=none] (5) at (8.25, -1.5) {};
		\node [style=none] (6) at (8.25, -3.75) {};
		\node [style=none] (7) at (7.25, -1.5) {};
		\node [style=none] (8) at (7.25, -3.75) {};
		\node [style=none] (9) at (6.25, -1.5) {};
		\node [style=none] (10) at (6.25, -3.75) {};
		\node [style=none] (11) at (5.25, -1.5) {};
		\node [style=none] (12) at (5.25, -3.75) {};
		\node [style=none] (13) at (4.25, -1.5) {};
		\node [style=none] (14) at (4.25, -3.75) {};
		\node [style=none] (15) at (3.25, -1.5) {};
		\node [style=none] (16) at (3.25, -3.75) {};
		\node [style=none] (17) at (2.25, -1.5) {};
		\node [style=none] (18) at (2.25, -3.75) {};
		\node [style=none] (19) at (1.25, -1.5) {};
		\node [style=none] (20) at (1.25, -3.75) {};
		\node [style=none] (21) at (0.25, -1.5) {};
		\node [style=none] (22) at (0.25, -3.75) {};
		\node [style=none] (23) at (-0.75, -1.5) {};
		\node [style=none] (24) at (-0.75, -3.75) {};
		\node [style=none] (25) at (-1.75, -1.5) {};
		\node [style=none] (26) at (-1.75, -3.75) {};
		\node [style=none] (27) at (-2.75, -1.5) {};
		\node [style=none] (28) at (-2.75, -3.75) {};
		\node [style=none] (29) at (-3.75, -1.5) {};
		\node [style=none] (30) at (-3.75, -3.75) {};
		\node [style=none] (31) at (-4.75, -1.5) {};
		\node [style=none] (32) at (-4.75, -3.75) {};
		\node [style=none] (33) at (-5.75, -1.5) {};
		\node [style=none] (34) at (-5.75, -3.75) {};
		\node [style=none] (35) at (-6.75, -1.5) {};
		\node [style=none] (36) at (-6.75, -3.75) {};
		\node [style=none] (37) at (9.75, -4.25) {U};
		\node [style=none] (38) at (8.75, -4.25) {U};
		\node [style=none] (39) at (6.75, -4.25) {U};
		\node [style=none] (40) at (5.75, -4.25) {S};
		\node [style=none] (41) at (2.75, -4.25) {U};
		\node [style=none] (42) at (1.75, -4.25) {U};
		\node [style=none] (43) at (0.75, -4.25) {U};
		\node [style=none] (44) at (-2.25, -4.25) {U};
		\node [style=none] (45) at (-4.25, -4.25) {U};
		\node [style=none] (46) at (-5.25, -4.25) {S};
		\node [style=none] (47) at (-6.25, -4.25) {U};
		\node [style=none] (48) at (7.75, -4.25) {\textcolor{cyan}{U}};
		\node [style=none] (49) at (-0.25, -4.25) {\textcolor{cyan}{S}};
		\node [style=none] (50) at (3.75, -4.25) {\textcolor{cyan}{U}};
		\node [style=none] (51) at (-3.25, -4.25) {\textcolor{cyan}{U}};
		\node [style=none] (52) at (-1.25, -4.25) {\textcolor{cyan}{U}};
		\node [style=none] (53) at (4.75, -4.25) {U};
	\end{pgfonlayer}
	\begin{pgfonlayer}{edgelayer}
		\draw [style=rede] (1.center) to (2.center);
		\draw [style=rede] (4.center) to (3.center);
		\draw [style=rede] (5.center) to (6.center);
		\draw [style=rede] (8.center) to (7.center);
		\draw [style=rede] (9.center) to (10.center);
		\draw [style=bluee] (11.center) to (12.center);
		\draw [style=bluee] (14.center) to (13.center);
		\draw [style=bluee] (15.center) to (16.center);
		\draw [style=bluee] (18.center) to (17.center);
		\draw [style=bluee] (19.center) to (20.center);
		\draw [style=bluee] (21.center) to (22.center);
		\draw [style=pinkline] (23.center) to (24.center);
		\draw [style=pinkline] (25.center) to (26.center);
		\draw [style=pinkline] (27.center) to (28.center);
		\draw [style=pinkline] (29.center) to (30.center);
		\draw [style=pinkline] (31.center) to (32.center);
		\draw [style=brownline] (33.center) to (34.center);
		\draw [style=brownline] (36.center) to (35.center);
	\end{pgfonlayer}
\end{tikzpicture}}
\label{electricquiverarrangement}
\end{equation}
This information specifies the U/SU nodes in \eqref{firstelectric} that makes it the electric quiver of \eqref{backwards}:
\begin{equation}
\scalebox{0.8}{
\begin{tikzpicture}
	\begin{pgfonlayer}{nodelayer}
		\node [style=gauge1] (0) at (2, -4) {};
		\node [style=gauge1] (1) at (3, -4) {};
		\node [style=gauge1] (2) at (4, -4) {};
		\node [style=gauge1] (3) at (5, -4) {};
		\node [style=gauge1] (4) at (6, -4) {};
		\node [style=gauge1] (5) at (7, -4) {};
		\node [style=gauge1] (6) at (8, -4) {};
		\node [style=gauge1] (7) at (9, -4) {};
		\node [style=gauge1] (8) at (10, -4) {};
		\node [style=gauge1] (9) at (11, -4) {};
		\node [style=gauge1] (10) at (12, -4) {};
		\node [style=gauge1] (11) at (13, -4) {};
		\node [style=gauge1] (12) at (14, -4) {};
		\node [style=gauge1] (13) at (15, -4) {};
		\node [style=gauge1] (14) at (16, -4) {};
		\node [style=gauge1] (15) at (17, -4) {};
		\node [style=gauge1] (16) at (18, -4) {};
		\node [style=none] (17) at (2, -4.5) {U(1)};
		\node [style=none] (18) at (3, -4.5) {SU(2)};
		\node [style=none] (19) at (4, -4.5) {U(3)};
		\node [style=none] (20) at (5, -4.5) {U(4)};
		\node [style=none] (21) at (6, -4.5) {U(3)};
		\node [style=none] (22) at (7, -4.5) {U(2)};
		\node [style=none] (23) at (8, -4.5) {SU(2)};
		\node [style=none] (24) at (9, -4.5) {U(3)};
		\node [style=none] (25) at (10, -4.5) {U(4)};
		\node [style=none] (26) at (11, -4.5) {U(5)};
		\node [style=none] (27) at (12, -4.5) {U(5)};
		\node [style=none] (28) at (13, -4.5) {U(6)};
		\node [style=none] (29) at (14, -4.5) {SU(5)};
		\node [style=none] (30) at (15, -4.5) {U(4)};
		\node [style=none] (31) at (16, -4.5) {U(3)};
		\node [style=none] (32) at (17, -4.5) {U(2)};
		\node [style=none] (33) at (18, -4.5) {U(1)};
		\node [style=flavour1] (34) at (13, -3) {};
		\node [style=flavour1] (35) at (11, -3) {};
		\node [style=flavour1] (36) at (5, -3) {};
		\node [style=none] (37) at (5, -2.5) {3};
		\node [style=none] (38) at (11, -2.5) {1};
		\node [style=none] (39) at (13, -2.5) {2};
		\node [style=flavour1] (40) at (16, -3) {};
		\node [style=none] (41) at (16, -2.5) {1};
	\end{pgfonlayer}
	\begin{pgfonlayer}{edgelayer}
		\draw (36) to (3);
		\draw (35) to (9);
		\draw (34) to (11);
		\draw (16) to (0);
		\draw (40) to (14);
	\end{pgfonlayer}
\end{tikzpicture}}
\label{finalelectricquiver}
\end{equation}
As a consistency check, the position of the overbalanced nodes (cyan) predicted in \eqref{electricquiverarrangement} matches with those in \eqref{finalelectricquiver}. \eqref{finalelectricquiver} is indeed the 3d mirror of \eqref{backwards} which can now be checked by putting it as an input in the Mathematica code for the forward algorithm.

\subsection{Some underbalanced nodes} \label{sec:negativeBalance}
The Higgs branch of a 3d $\mathcal{N}=4$ quiver where all gauge nodes are either balanced or overbalanced is a single cone. If one or more gauge groups are underbalanced, the Higgs branch could be the union of several cones as first observed for SQCD in \cite{Argyres:1996eh}. As a result, we have one magnetic quiver for each of the cones. The concept of 3d mirror pairs is ill defined in this case and we will therefore only speak of magnetic quivers. 

The multitude of hyper-K\"ahler cones and hence magnetic quivers come from inequivalent choices of maximal decomposition of our brane web into subwebs \cite{Cabrera:2018jxt}. Consider the following quiver:
\begin{equation}
\raisebox{-.5\height}{\begin{tikzpicture}
	\begin{pgfonlayer}{nodelayer}
		\node [style=gauge1] (38) at (0.5, -0.25) {};
		\node [style=none] (40) at (0.5, -0.75) {SU(5)};
		\node [style=none] (55) at (-1, -0.75) {SU(5)};
		\node [style=gauge1] (58) at (-1, -0.25) {};
		\node [style=flavour1] (59) at (-1, 1.25) {};
		\node [style=flavour1] (60) at (0.5, 1.25) {};
		\node [style=none] (61) at (-1, 1.75) {2};
		\node [style=none] (62) at (0.5, 1.75) {3};
	\end{pgfonlayer}
	\begin{pgfonlayer}{edgelayer}
		\draw (38) to (58);
		\draw (60) to (38);
		\draw (59) to (58);
	\end{pgfonlayer}
\end{tikzpicture}}
\end{equation}
We see that both gauge nodes are underbalanced. When drawing the brane web, we follow the convention of the algorithm in Section \ref{prescription} and  move all the D7 branes to the left:
\begin{equation}
    \raisebox{-.5\height}{\begin{tikzpicture}
	\begin{pgfonlayer}{nodelayer}
		\node [style=gauge1] (0) at (-3, 0) {};
		\node [style=gauge1] (1) at (-2, 0) {};
		\node [style=gauge1] (2) at (-1, 0) {};
		\node [style=gauge1] (3) at (0, 0) {};
		\node [style=gauge1] (4) at (1, 0) {};
		\node [style=none] (5) at (-2.5, 0.25) {1};
		\node [style=none] (6) at (-1.5, 0.25) {2};
		\node [style=none] (7) at (-0.5, 0.25) {4};
		\node [style=none] (8) at (0.5, 0.25) {6};
		\node [style=none] (10) at (1.5, 0.25) {8};
		\node [style=none] (11) at (3, 0) {};
		\node [style=none] (12) at (2, 0) {};
		\node [style=none] (13) at (2.5, 0.25) {8};
		\node [style=none] (14) at (2, 1) {};
		\node [style=none] (15) at (2, -1) {};
		\node [style=gauge1] (16) at (2, 1) {};
		\node [style=gauge1] (17) at (2, -1) {};
		\node [style=none] (18) at (4, 1) {};
		\node [style=none] (19) at (3, -1) {};
		\node [style=gauge1] (20) at (4, 1) {};
		\node [style=gauge1] (21) at (3, -1) {};
		\node [style=none] (22) at (5, 0) {};
		\node [style=none] (23) at (6.75, 0.75) {};
		\node [style=none] (24) at (5, -1) {};
		\node [style=gauge1] (25) at (5, -1) {};
		\node [style=gauge1] (26) at (6.75, 0.75) {};
		\node [style=none] (27) at (6.75, 1.25) {[5,1]};
		\node [style=none] (28) at (5, -1.5) {[0,1]};
		\node [style=none] (29) at (3, -1.5) {[0,1]};
		\node [style=none] (30) at (2, -1.5) {[0,1]};
		\node [style=none] (31) at (2, 1.5) {[0,1]};
		\node [style=none] (32) at (4, 1.5) {[3,1]};
		\node [style=none] (33) at (4.25, 0.25) {5};
	\end{pgfonlayer}
	\begin{pgfonlayer}{edgelayer}
		\draw (0) to (4);
		\draw (4) to (12.center);
		\draw (12.center) to (11.center);
		\draw (14.center) to (15.center);
		\draw (18.center) to (11.center);
		\draw (11.center) to (19.center);
		\draw (11.center) to (22.center);
		\draw (22.center) to (24.center);
		\draw (22.center) to (23.center);
	\end{pgfonlayer}
\end{tikzpicture}}
\end{equation}
The brane web has four inequivalent maximal decompositions, each giving rise to a magnetic quiver, listed in the first row of 
Table \ref{badquiver}. (This statement, and subsequent statements in this section, can be obtained immediately using the attached code). An immediate observation is that the subwebs associated with NS5s now have non-trivial intersection number between them. In other words, the bouquets of $\mathrm{U}(1)$ nodes in the magnetic quiver can now have edges between them, which is something we do not observe in good linear quivers where all nodes are balanced or overbalanced.

The next step is to turn  the $\mathrm{SU}(5)$s into $\mathrm{U}(5)$s by locking the branes. In contrast to  good quivers where locking makes minor changes to the magnetic quivers, for bad quivers it can also leave the magnetic quiver unchanged or it can change the structure drastically. 

As already stated in this section, for a good linear quiver there is always a set of HW transitions such that there is an unbound state of D5 and NS5 branes. As a corollary, the NS5 branes  move independently from each other. If the electric quiver contains  underbalanced nodes, this no longer holds and there may be bound states one cannot get rid of. This, in return, can allow for more than one maximal decomposition with some of the 5-branes forced to move together. Note, we are not doing any locking here, but this is a natural feature of a maximal decomposition of a brane web, even when all the gauge nodes in the electric quiver are SU. In particular, for an SQCD electric quiver, the hyper-K\"ahler cone where all the NS5 move independently from each other is called the baryonic branch. If some of the NS5 move together, it is the mesonic branch. 
In Table \ref{badquiver}, the baryonic cone is given by Magnetic Quiver 4 in the first row. The remaining magnetic quivers all have some or all of the NS5 branes moving together. As a result, if the NS5 branes that are already moving together are locked when going from SU to U, the resulting magnetic quiver will stay the same. For example, Magnetic Quiver 1 remains the same when the second SU(5) in the electric quiver is turned to U(5).

\afterpage{
\begin{landscape}
\begin{table}[]
    \centering
    \begin{tabular}{|c|c|c|c|c|} \hline 
        Electric Quiver & Magnetic Quiver 1  & Magnetic Quiver 2 & Magnetic Quiver 3 & Magnetic Quiver 4  \\ \hline
       \begin{tikzpicture}
	\begin{pgfonlayer}{nodelayer}
		\node [style=gauge1] (38) at (0.5, -0.25) {};
		\node [style=none] (40) at (0.5, -0.75) {SU(5)};
		\node [style=none] (55) at (-1, -0.75) {SU(5)};
		\node [style=gauge1] (58) at (-1, -0.25) {};
		\node [style=flavour1] (59) at (-1, 1.25) {};
		\node [style=flavour1] (60) at (0.5, 1.25) {};
		\node [style=none] (61) at (-1, 1.75) {2};
		\node [style=none] (62) at (0.5, 1.75) {3};
	\end{pgfonlayer}
	\begin{pgfonlayer}{edgelayer}
		\draw (38) to (58);
		\draw (60) to (38);
		\draw (59) to (58);
	\end{pgfonlayer}
\end{tikzpicture}
  & \begin{tikzpicture}
	\begin{pgfonlayer}{nodelayer}
		\node [style=gauge1] (0) at (0, 0) {};
		\node [style=gauge1] (1) at (1, 0) {};
		\node [style=gauge1] (2) at (2, 0) {};
		\node [style=none] (3) at (0, -0.5) {1};
		\node [style=none] (4) at (1, -0.5) {1};
		\node [style=none] (5) at (2, -0.5) {1};
		\node [style=none] (6) at (0, 0.125) {};
		\node [style=none] (7) at (0, -0.125) {};
		\node [style=none] (8) at (1, 0.125) {};
		\node [style=none] (9) at (1, -0.125) {};
		\node [style=none] (10) at (1, 0.125) {};
		\node [style=none] (11) at (1, -0.125) {};
		\node [style=none] (12) at (2, 0.125) {};
		\node [style=none] (13) at (2, -0.125) {};
		\node [style=none] (14) at (0.5, 0.5) {8};
		\node [style=none] (15) at (1.5, 0.5) {2};
	\end{pgfonlayer}
	\begin{pgfonlayer}{edgelayer}
		\draw (6.center) to (8.center);
		\draw (7.center) to (9.center);
		\draw (10.center) to (12.center);
		\draw (11.center) to (13.center);
	\end{pgfonlayer}
\end{tikzpicture}
 & \begin{tikzpicture}
	\begin{pgfonlayer}{nodelayer}
		\node [style=gauge1] (0) at (0.5, 0) {};
		\node [style=gauge1] (1) at (1.5, 0) {};
		\node [style=gauge1] (2) at (2.5, 0) {};
		\node [style=none] (3) at (0.5, -0.5) {1};
		\node [style=none] (4) at (1.5, -0.5) {2};
		\node [style=none] (5) at (2.5, -0.5) {2};
		\node [style=gauge1] (6) at (3.5, 0) {};
		\node [style=none] (7) at (3.5, -0.5) {1};
		\node [style=gauge1] (8) at (1.5, 1) {};
		\node [style=gauge1] (9) at (2.5, 1) {};
		\node [style=none] (10) at (2, 1.5) {3};
		\node [style=none] (11) at (1.5, 1.125) {};
		\node [style=none] (12) at (1.5, 0.875) {};
		\node [style=none] (13) at (2.5, 1.125) {};
		\node [style=none] (14) at (2.5, 0.875) {};
		\node [style=none] (15) at (1, 1) {1};
		\node [style=none] (16) at (3, 1) {1};
	\end{pgfonlayer}
	\begin{pgfonlayer}{edgelayer}
		\draw (0) to (1);
		\draw (1) to (2);
		\draw (2) to (6);
		\draw (8) to (1);
		\draw (8) to (2);
		\draw (9) to (2);
		\draw (11.center) to (13.center);
		\draw (12.center) to (14.center);
	\end{pgfonlayer}
\end{tikzpicture}
 & \begin{tikzpicture}
	\begin{pgfonlayer}{nodelayer}
		\node [style=gauge1] (0) at (0, 0) {};
		\node [style=gauge1] (1) at (1, 0) {};
		\node [style=none] (2) at (0, 0.125) {};
		\node [style=none] (3) at (0, -0.125) {};
		\node [style=none] (4) at (1, 0.125) {};
		\node [style=none] (5) at (1, -0.125) {};
		\node [style=none] (6) at (1, 0.125) {};
		\node [style=none] (7) at (1.025, 0) {};
		\node [style=gauge1] (8) at (2, 0.75) {};
		\node [style=gauge1] (9) at (2, -0.75) {};
		\node [style=none] (10) at (2.5, 0.75) {1};
		\node [style=none] (11) at (2.5, -0.75) {1};
		\node [style=none] (12) at (1, -0.5) {1};
		\node [style=none] (13) at (0, -0.5) {1};
		\node [style=none] (14) at (0.5, 0.5) {7};
	\end{pgfonlayer}
	\begin{pgfonlayer}{edgelayer}
		\draw (2.center) to (4.center);
		\draw (3.center) to (5.center);
		\draw (8) to (7.center);
		\draw (9) to (7.center);
		\draw (8) to (9);
	\end{pgfonlayer}
\end{tikzpicture}
 & \begin{tikzpicture}
	\begin{pgfonlayer}{nodelayer}
		\node [style=gauge1] (8) at (-0.8, -0.35) {};
		\node [style=gauge1] (9) at (-0.3, 0.4) {};
		\node [style=none] (10) at (-0.675, -0.35) {};
		\node [style=none] (11) at (-0.925, -0.35) {};
		\node [style=none] (12) at (-0.175, 0.4) {};
		\node [style=none] (13) at (-0.425, 0.4) {};
		\node [style=none] (14) at (-0.175, 0.4) {};
		\node [style=none] (15) at (-0.3, 0.425) {};
		\node [style=gauge1] (16) at (-0.3, 0.4) {};
		\node [style=gauge1] (17) at (0.2, -0.35) {};
		\node [style=none] (18) at (-0.175, 0.425) {};
		\node [style=none] (19) at (-0.425, 0.425) {};
		\node [style=none] (20) at (0.325, -0.35) {};
		\node [style=none] (21) at (0.075, -0.35) {};
		\node [style=none] (22) at (0.325, -0.35) {};
		\node [style=none] (23) at (0.2, -0.375) {};
		\node [style=gauge1] (24) at (0.2, -0.35) {};
		\node [style=gauge1] (25) at (-0.8, -0.35) {};
		\node [style=none] (26) at (0.2, -0.475) {};
		\node [style=none] (27) at (0.2, -0.225) {};
		\node [style=none] (28) at (-0.8, -0.475) {};
		\node [style=none] (29) at (-0.8, -0.225) {};
		\node [style=none] (30) at (-0.8, -0.475) {};
		\node [style=none] (31) at (-0.825, -0.35) {};
		\node [style=none] (32) at (0.25, 0.25) {2};
		\node [style=none] (33) at (-1, 0.25) {3};
		\node [style=none] (34) at (-0.25, -0.75) {5};
		\node [style=none] (35) at (-0.3, 0.925) {1};
		\node [style=none] (36) at (0.75, -0.375) {1};
		\node [style=none] (37) at (-1.3, -0.375) {1};
	\end{pgfonlayer}
	\begin{pgfonlayer}{edgelayer}
		\draw (10.center) to (12.center);
		\draw (11.center) to (13.center);
		\draw (18.center) to (20.center);
		\draw (19.center) to (21.center);
		\draw (26.center) to (28.center);
		\draw (27.center) to (29.center);
	\end{pgfonlayer}
\end{tikzpicture}
 \\ \hline
 \begin{tikzpicture}
	\begin{pgfonlayer}{nodelayer}
		\node [style=gauge1] (38) at (0.5, -0.25) {};
		\node [style=none] (40) at (0.5, -0.75) {U(5)};
		\node [style=none] (55) at (-1, -0.75) {SU(5)};
		\node [style=gauge1] (58) at (-1, -0.25) {};
		\node [style=flavour1] (59) at (-1, 1.25) {};
		\node [style=flavour1] (60) at (0.5, 1.25) {};
		\node [style=none] (61) at (-1, 1.75) {2};
		\node [style=none] (62) at (0.5, 1.75) {3};
	\end{pgfonlayer}
	\begin{pgfonlayer}{edgelayer}
		\draw (38) to (58);
		\draw (60) to (38);
		\draw (59) to (58);
	\end{pgfonlayer}
\end{tikzpicture}
& \cellcolor{blue!05} \begin{tikzpicture}
	\begin{pgfonlayer}{nodelayer}
		\node [style=gauge1] (0) at (0, 0) {};
		\node [style=gauge1] (1) at (1, 0) {};
		\node [style=gauge1] (2) at (2, 0) {};
		\node [style=none] (3) at (0, -0.5) {1};
		\node [style=none] (4) at (1, -0.5) {1};
		\node [style=none] (5) at (2, -0.5) {1};
		\node [style=none] (6) at (0, 0.125) {};
		\node [style=none] (7) at (0, -0.125) {};
		\node [style=none] (8) at (1, 0.125) {};
		\node [style=none] (9) at (1, -0.125) {};
		\node [style=none] (10) at (1, 0.125) {};
		\node [style=none] (11) at (1, -0.125) {};
		\node [style=none] (12) at (2, 0.125) {};
		\node [style=none] (13) at (2, -0.125) {};
		\node [style=none] (14) at (0.5, 0.5) {8};
		\node [style=none] (15) at (1.5, 0.5) {2};
	\end{pgfonlayer}
	\begin{pgfonlayer}{edgelayer}
		\draw (6.center) to (8.center);
		\draw (7.center) to (9.center);
		\draw (10.center) to (12.center);
		\draw (11.center) to (13.center);
	\end{pgfonlayer}
\end{tikzpicture}
 & \multicolumn{2}{c|}{ \begin{tikzpicture}
	\begin{pgfonlayer}{nodelayer}
		\node [style=gauge1] (0) at (0, 0) {};
		\node [style=gauge1] (1) at (1, 0) {};
		\node [style=gauge1] (2) at (2, 0) {};
		\node [style=none] (3) at (0, -0.5) {1};
		\node [style=none] (4) at (1, -0.5) {2};
		\node [style=none] (5) at (2, -0.5) {2};
		\node [style=gauge1] (6) at (3, 0) {};
		\node [style=none] (7) at (3, -0.5) {1};
		\node [style=gauge1] (8) at (1.5, 1) {};
		\node [style=none] (9) at (0.75, 0.75) {2};
		\node [style=none] (10) at (1.35, 0.9) {};
		\node [style=none] (11) at (0.85, 0.15) {};
		\node [style=none] (12) at (1.6, 0.9) {};
		\node [style=none] (13) at (1.1, 0.15) {};
		\node [style=none] (14) at (1.5, 1.5) {1};
	\end{pgfonlayer}
	\begin{pgfonlayer}{edgelayer}
		\draw (0) to (1);
		\draw (1) to (2);
		\draw (2) to (6);
		\draw (8) to (2);
		\draw (10.center) to (11.center);
		\draw (12.center) to (13.center);
	\end{pgfonlayer}
\end{tikzpicture}}
& \cellcolor{blue!05} 
 \\ \hline 
 \begin{tikzpicture}
	\begin{pgfonlayer}{nodelayer}
		\node [style=gauge1] (38) at (0.5, -0.25) {};
		\node [style=none] (40) at (0.5, -0.75) {SU(5)};
		\node [style=none] (55) at (-1, -0.75) {U(5)};
		\node [style=gauge1] (58) at (-1, -0.25) {};
		\node [style=flavour1] (59) at (-1, 1.25) {};
		\node [style=flavour1] (60) at (0.5, 1.25) {};
		\node [style=none] (61) at (-1, 1.75) {2};
		\node [style=none] (62) at (0.5, 1.75) {3};
	\end{pgfonlayer}
	\begin{pgfonlayer}{edgelayer}
		\draw (38) to (58);
		\draw (60) to (38);
		\draw (59) to (58);
	\end{pgfonlayer}
\end{tikzpicture}
 &  \multicolumn{2}{c|}{\begin{tikzpicture}
	\begin{pgfonlayer}{nodelayer}
		\node [style=gauge1] (0) at (0.5, 0) {};
		\node [style=gauge1] (1) at (1.5, 0) {};
		\node [style=gauge1] (2) at (2.5, 0) {};
		\node [style=none] (3) at (0.5, -0.5) {1};
		\node [style=none] (4) at (1.5, -0.5) {2};
		\node [style=none] (5) at (2.5, -0.5) {2};
		\node [style=gauge1] (6) at (3.5, 0) {};
		\node [style=none] (7) at (3.5, -0.5) {1};
		\node [style=gauge1] (8) at (2, 1) {};
		\node [style=none] (9) at (1.25, 0.75) {2};
		\node [style=none] (10) at (1.85, 0.9) {};
		\node [style=none] (11) at (1.35, 0.15) {};
		\node [style=none] (12) at (2.1, 0.9) {};
		\node [style=none] (13) at (1.6, 0.15) {};
		\node [style=none] (14) at (2, 1.5) {1};
	\end{pgfonlayer}
	\begin{pgfonlayer}{edgelayer}
		\draw (0) to (1);
		\draw (1) to (2);
		\draw (2) to (6);
		\draw (8) to (2);
		\draw (10.center) to (11.center);
		\draw (12.center) to (13.center);
	\end{pgfonlayer}
\end{tikzpicture} }
 &  \multicolumn{2}{c|}{ \begin{tikzpicture}
	\begin{pgfonlayer}{nodelayer}
		\node [style=gauge1] (0) at (0, 0) {};
		\node [style=gauge1] (1) at (1, 0) {};
		\node [style=none] (2) at (0, 0.125) {};
		\node [style=none] (3) at (0, -0.125) {};
		\node [style=none] (4) at (1, 0.125) {};
		\node [style=none] (5) at (1, -0.125) {};
		\node [style=none] (6) at (1, 0.125) {};
		\node [style=none] (7) at (1.025, 0) {};
		\node [style=gauge1] (8) at (2, 0.75) {};
		\node [style=gauge1] (9) at (2, -0.75) {};
		\node [style=none] (10) at (2.5, 0.75) {1};
		\node [style=none] (11) at (2.5, -0.75) {1};
		\node [style=none] (12) at (1, -0.5) {1};
		\node [style=none] (13) at (0, -0.5) {1};
		\node [style=none] (14) at (0.5, 0.5) {7};
	\end{pgfonlayer}
	\begin{pgfonlayer}{edgelayer}
		\draw (2.center) to (4.center);
		\draw (3.center) to (5.center);
		\draw (8) to (7.center);
		\draw (9) to (7.center);
		\draw (8) to (9);
	\end{pgfonlayer}
\end{tikzpicture}}
 \\ \hline 
 \begin{tikzpicture}
	\begin{pgfonlayer}{nodelayer}
		\node [style=gauge1] (38) at (0.5, -0.25) {};
		\node [style=none] (40) at (0.5, -0.75) {U(5)};
		\node [style=none] (55) at (-1, -0.75) {U(5)};
		\node [style=gauge1] (58) at (-1, -0.25) {};
		\node [style=flavour1] (59) at (-1, 1.25) {};
		\node [style=flavour1] (60) at (0.5, 1.25) {};
		\node [style=none] (61) at (-1, 1.75) {2};
		\node [style=none] (62) at (0.5, 1.75) {3};
	\end{pgfonlayer}
	\begin{pgfonlayer}{edgelayer}
		\draw (38) to (58);
		\draw (60) to (38);
		\draw (59) to (58);
	\end{pgfonlayer}
\end{tikzpicture}
 & \multicolumn{4}{c|}{\begin{tikzpicture}
	\begin{pgfonlayer}{nodelayer}
		\node [style=gauge1] (0) at (0, 0) {};
		\node [style=gauge1] (1) at (1, 0) {};
		\node [style=gauge1] (2) at (2, 0) {};
		\node [style=none] (3) at (0, -0.5) {1};
		\node [style=none] (4) at (1, -0.5) {2};
		\node [style=none] (5) at (2, -0.5) {2};
		\node [style=gauge1] (6) at (3, 0) {};
		\node [style=none] (7) at (3, -0.5) {1};
		\node [style=gauge1] (8) at (1.5, 1) {};
		\node [style=none] (9) at (0.75, 0.75) {2};
		\node [style=none] (10) at (1.35, 0.9) {};
		\node [style=none] (11) at (0.85, 0.15) {};
		\node [style=none] (12) at (1.6, 0.9) {};
		\node [style=none] (13) at (1.1, 0.15) {};
		\node [style=none] (14) at (1.5, 1.5) {1};
	\end{pgfonlayer}
	\begin{pgfonlayer}{edgelayer}
		\draw (0) to (1);
		\draw (1) to (2);
		\draw (2) to (6);
		\draw (8) to (2);
		\draw (10.center) to (11.center);
		\draw (12.center) to (13.center);
	\end{pgfonlayer}
\end{tikzpicture}} \\ \hline
    \end{tabular}
    \caption{The first row displays the electric theory and the four corresponding magnetic quivers. The next few rows show how the magnetic quivers change as the SU nodes in the electric theory are turned into U nodes. We observe how distinct subdivisions of the brane web (and hence their magnetic quivers) become identical when some of the SU nodes are turned to U nodes. The light blue colored box indicates the same magnetic quiver.  }
    \label{badquiver}
\end{table}
\end{landscape}
}

In the case of good linear quivers, we have seen that changing nodes in the electric quiver from SU to U simply translates to merging some of the $\mathrm{U}(1)$s in the bouquet. For a quiver with bad nodes, the $s$-rule plays a crucial role which can result in a complete change to  the  structure of the magnetic quiver. For example, Magnetic Quiver 3 changes drastically when the second $\mathrm{SU}(5)$ in the electric quiver is changed to $\mathrm{U}(5)$. 

When all the SU nodes are changed to U in the electric quiver, all four magnetic quivers become identical. In other words, the four hyper-K\"ahler cones coalesce into a single cone. This is expected as the Higgs branch of a bad quiver with only unitary gauge nodes should be a single hyper-K\"ahler cone as observed in \cite{Bourget:2019rtl}. This is shown in the fourth line of Table \ref{badquiver}. As a consistency check, the Higgs branch of any linear bad quiver with only unitary gauge nodes is equivalent to the Higgs branch of a good quiver. The good quiver can be obtained through a set of operations outlined in \cite{Assel:2017jgo} and reviewed in Appendix \ref{bad}. Basically, assuming vanishing FI parameters, a bad node of $\mathrm{U}(k)$ with $N_f$ flavor is replaced with $\mathrm{U}(\lfloor N_f/2 \rfloor)$. This process is repeated until all  gauge nodes are good. For our electric quiver, the following equivalence in Higgs branches hold:
\begin{equation}
 \raisebox{-.5\height}{\scalebox{.8}{\begin{tikzpicture}
	\begin{pgfonlayer}{nodelayer}
		\node [style=gauge1] (38) at (0.5, -0.25) {};
		\node [style=none] (40) at (0.5, -0.75) {U(5)};
		\node [style=none] (55) at (-1, -0.75) {U(5)};
		\node [style=gauge1] (58) at (-1, -0.25) {};
		\node [style=flavour1] (59) at (-1, 1.25) {};
		\node [style=flavour1] (60) at (0.5, 1.25) {};
		\node [style=none] (61) at (-1, 1.75) {2};
		\node [style=none] (62) at (0.5, 1.75) {3};
		\node [style=none] (63) at (-2.5, 0.5) {$\mathcal{H}$};
		\node [style=none] (64) at (-1.5, 2) {};
		\node [style=none] (65) at (-1.5, -0.75) {};
		\node [style=none] (66) at (1, 2) {};
		\node [style=none] (67) at (1, -0.75) {};
		\node [style=gauge1] (68) at (6, -0.25) {};
		\node [style=none] (69) at (6, -0.75) {U(2)};
		\node [style=none] (70) at (4.5, -0.75) {U(2)};
		\node [style=gauge1] (71) at (4.5, -0.25) {};
		\node [style=flavour1] (72) at (4.5, 1.25) {};
		\node [style=flavour1] (73) at (6, 1.25) {};
		\node [style=none] (74) at (4.5, 1.75) {2};
		\node [style=none] (75) at (6, 1.75) {3};
		\node [style=none] (76) at (3, 0.5) {$\mathcal{H}$};
		\node [style=none] (77) at (4, 2) {};
		\node [style=none] (78) at (4, -0.75) {};
		\node [style=none] (79) at (6.5, 2) {};
		\node [style=none] (80) at (6.5, -0.75) {};
		\node [style=none] (81) at (2, 0.5) {=};
		\node [style=none] (82) at (7.25, 0.5) {=};
		\node [style=none] (83) at (8.25, 0.5) {$\mathcal{C}$};
		\node [style=none] (84) at (9.25, 2) {};
		\node [style=none] (85) at (9.25, -0.75) {};
		\node [style=gauge1] (86) at (9.75, 0.5) {};
		\node [style=gauge1] (87) at (10.75, 0.5) {};
		\node [style=gauge1] (88) at (11.75, 0.5) {};
		\node [style=none] (89) at (9.75, 0) {1};
		\node [style=none] (90) at (10.75, 0) {2};
		\node [style=none] (91) at (11.75, 0) {2};
		\node [style=gauge1] (92) at (12.75, 0.5) {};
		\node [style=none] (93) at (12.75, 0) {1};
		\node [style=gauge1] (94) at (11.25, 1.5) {};
		\node [style=none] (95) at (10.5, 1.25) {2};
		\node [style=none] (96) at (11.1, 1.4) {};
		\node [style=none] (97) at (10.6, 0.65) {};
		\node [style=none] (98) at (11.35, 1.4) {};
		\node [style=none] (99) at (10.85, 0.65) {};
		\node [style=none] (100) at (13, 2) {};
		\node [style=none] (101) at (13, -0.75) {};
	\end{pgfonlayer}
	\begin{pgfonlayer}{edgelayer}
		\draw (38) to (58);
		\draw (60) to (38);
		\draw (59) to (58);
		\draw [bend right=15] (64.center) to (65.center);
		\draw [bend left=15] (66.center) to (67.center);
		\draw (68) to (71);
		\draw (73) to (68);
		\draw (72) to (71);
		\draw [bend right=15] (77.center) to (78.center);
		\draw [bend left=15] (79.center) to (80.center);
		\draw [bend right=15] (84.center) to (85.center);
		\draw (86) to (87);
		\draw (87) to (88);
		\draw (88) to (92);
		\draw (94) to (88);
		\draw (96.center) to (97.center);
		\draw (98.center) to (99.center);
		\draw [bend left=15] (100.center) to (101.center);
	\end{pgfonlayer}
\end{tikzpicture}}}
\end{equation}
where the right side is the known 3d mirror of the good theory in the middle.

For good linear quivers with different combinations of U/SU nodes, we can check the conjectured mirror pairs through explicit Hilbert series computations. For electric quivers with bad nodes, however, computational difficulties prevent us from doing the same explicit checks. Nevertheless, we find consistency when comparing to the magnetic quivers found using other methods. 

\subsection{3d Mirror Symmetry vs incomplete Higgsing}
\label{sec:incomplete3d}

In some cases, the algorithm produces a 3d mirror pair. A necessary (but not sufficient) condition for this is that all the nodes satisfy the 3d mirror condition in Table \ref{tab:effects}. Another necessary condition for this to happen is that a single magnetic quiver is produced, but this is again not a sufficient condition. 
In many cases, the gauge groups in the linear U/SU quivers can not be completely Higgsed using the available matter content. In this case, it is not possible to have a mirror pair, even if only one magnetic quiver describes the Higgs branch, as the Higgs branch is entirely contained in a larger mixed branch. 
We illustrate these phenomena with several examples. 

\paragraph{Example}
Consider the pair of quivers  
\begin{equation}
\raisebox{-.5\height}{
    \begin{tikzpicture}
\node (g1) at (10.5,3) [gauge,label=below:{SU(2)}] {};
\node (g2) at (11.5,3) [gauge,label=below:{U(2)}] {};
\node (g3) at (12.5,3) [gauge,label=below:{SU(2)}] {};
\node (f1) at (10.5,4) [flavor,label=above:{1}] {};
\node (f2) at (11.5,4) [flavor,label=above:{2}] {};
\node (f3) at (12.5,4) [flavor,label=above:{2}] {};
\draw (g1)--(g2)--(g3);
\draw (g1)--(f1) (g2)--(f2) (g3)--(f3);
\end{tikzpicture}} \qquad \longleftrightarrow  \qquad 
\raisebox{-.5\height}{
    \begin{tikzpicture}
\node (g1) at (-2,1) [gauge,label=above:{1}] {};
\node (g2) at (-1,.5) [gauge,label=above:{2}] {};
\node (g3) at (0,0) [gauge,label=below:{2}] {};
\node (g4) at (-2,-1) [gauge,label=below:{1}] {};
\node (g5) at (-1,-.5) [gauge,label=below:{1}] {};
\node (g6) at (1,1) [gauge,label=right:{1}] {};
\node (g7) at (1,-1) [gauge,label=right:{1}] {};
\draw (g1)--(g2)--(g3)--(g6);
\draw (g4)--(g5)--(g3)--(g7);
\draw (g1)--(g5) (g2)--(g4);
\end{tikzpicture}} 
\label{rudolphexample}
\end{equation}
where the right hand side is the magnetic quiver for the left hand side produced by the algorithm in Section \ref{prescription}. The necessary conditions of Table \ref{tab:effects} are met for this to be a 3d mirror pair, and indeed this is confirmed with a Hilbert series computation. The Coulomb branch unrefined Hilbert series of the magnetic quiver on the right hand side of \eqref{rudolphexample} is:
\begin{equation}
    \mathcal{C}\eqref{rudolphexample}= \frac{ {\small \left( \begin{array}{c} 1+3 t+14 t^2+51 t^3+170 t^4+511 t^5+1424 t^6+3621 t^7+8555 t^8+18760 t^9\\+38410 t^{10}+73586
   t^{11}+132502 t^{12}+224680 t^{13}+359934 t^{14}+545730 t^{15}\\+784778 t^{16}+1071828
   t^{17}+1392469 t^{18}+1722351 t^{19}+2030391 t^{20}+2282454 t^{21}\\ +2448132 t^{22}+2505814
   t^{23}+\dots\mathrm{palindromic}\dots+t^{46}\end{array} \right)}}{(1-t)^3(1-t^2)^4(1-t^3)^7(1-t^4)^4(1-t^5)^4}
\end{equation}
The Higgs branch Hilbert series of the magnetic quiver is:
\begin{equation}
   \mathcal{H}\eqref{rudolphexample}= \frac{ \left(\begin{array}{c}1+t^2+7 t^4+15 t^6+37 t^8+59 t^{10}+97 t^{12}+117 t^{14}+134 t^{16}+117 t^{18}\\+97 t^{20}+59 t^{22}+37 t^{24}+15
   t^{26}+7 t^{28}+t^{30}+t^{32}\end{array} \right)}{(1-t^2)(1-t^4)^3(1-t^6)^3(1-t^8)}
\end{equation}
These are equal to the Higgs / Coulomb branches Hilbert series of the left hand side of (\ref{rudolphexample}). Therefore one can claim that (\ref{rudolphexample}) is a 3d mirror pair.

\begin{table}[]
    \centering
    \begin{tabular}{cccccc}
    \toprule
        SQCD & Condition & 3d MS & \# of Cones & $\mathcal{HB}=\mathcal{HV}$ & Complete Higgsing \\ \midrule
        \multirow{3}{*}{\raisebox{-.5\height}{
    \begin{tikzpicture}
\node (g1) at (0,0) [gauge,label=left:{U($k$)}] {};
\node (f1) at (0,1) [flavor,label=left:{$N$}] {};
\draw (g1)--(f1) ;
\end{tikzpicture}}} & $N \geq 2 k$ &  \color{goodgreen}{$\cmark$}  & 1 &  \color{goodgreen}{$\cmark$}  &  \color{goodgreen}{$\cmark$}  \\
        & $N = 2 k-1$ &  ({\color{goodgreen}{$\cmark$}})  & 1 &  \color{goodgreen}{$\cmark$}  &  \color{red}{\xmark}  \\
        & $N \leq 2 k-2$ &  \color{red}{\xmark}  & 1 &  \color{goodgreen}{$\cmark$}  &  \color{red}{\xmark}  \\ \hline
        \multirow{4}{*}{\raisebox{-.5\height}{
    \begin{tikzpicture}
\node (g1) at (0,0) [gauge,label=left:{SU($k$)}] {};
\node (f1) at (0,1) [flavor,label=left:{$N$}] {};
\draw (g1)--(f1) ;
\end{tikzpicture}}} & $N \geq 2 k - 1$ &  \color{goodgreen}{$\cmark$}  & 1 &  \color{goodgreen}{$\cmark$}  &  \color{goodgreen}{$\cmark$}  \\
        & $N = 2 k-2$ &  \color{red}{\xmark}  & 2 &  \color{goodgreen}{$\cmark$}  &  \color{goodgreen}{$\cmark$}  \\
        & $k \leq N \leq 2 k-3$ &  \color{red}{\xmark}  & 2 &  \color{red}{\xmark}  &  \color{red}{\xmark}  \\
        & $N < k$ &  \color{red}{\xmark}  & 1 &  \color{red}{\xmark}  &  \color{red}{\xmark}  \\
        \bottomrule
    \end{tabular}
    \caption{Summary of some properties of SQCD theories with unitary or special unitary gauge group. The unitary case is discussed e.g. in \cite{Assel:2017jgo}, and Appendix \ref{bad}. The special unitary case is taken from \cite{Bourget:2019rtl}. The 3d mirror of U$(k)$ with $2k-1$ fundamental hypers consists of its magnetic quiver plus a free hyper, hence the ({\color{goodgreen}{$\cmark$}}) in the table. $\mathcal{HB}$ denotes the (possibly non-reduced) Higgs branch, while $\mathcal{HV}$ denotes the Higgs variety.}
    \label{tab:effects}
\end{table}

\paragraph{Example}
Consider now the very simple example 
\begin{equation}
\label{quiverExamplebad}
\raisebox{-.5\height}{
    \begin{tikzpicture}
\node (g1) at (10.5,3) [gauge,label=below:{U(3)}] {};
\node (f1) at (10.5,4) [flavor,label=above:{4}] {};
\draw (g1)--(f1);
\end{tikzpicture}} \qquad \longrightarrow  \qquad 
\raisebox{-.5\height}{
    \begin{tikzpicture}
\node (g1) at (-1,0) [gauge,label=below:{1}] {};
\node (g2) at (0,0) [gauge,label=below:{2}] {};
\node (g3) at (1,0) [gauge,label=below:{1}] {};
\node (g4) at (0,1) [gauge,label=right:{1}] {};
\draw (g1)--(g2)--(g3);
\draw[transform canvas={xshift=-1pt}] (g2)--(g4);
\draw[transform canvas={xshift=1pt}] (g2)--(g4);
\end{tikzpicture}} 
\end{equation}
The right-hand side quiver is a magnetic quiver for the Higgs branch of the left-hand side quiver. However the quivers do not form a mirror pair. Indeed, as explained in detail in Appendix \ref{bad}, the $\mathrm{U}(3)$ gauge group can not be fully Higgsed. On a generic point of the Higgs branch there is a residual U$(1)$ gauge symmetry. Therefore the Higgs branch is a subvariety of a (quaternionic five dimensional) mixed branch,\footnote{Whenever there is incomplete Higgsing, the Higgs branch is included in such a mixed branch. The Coulomb part of this mixed branch is the Coulomb branch of the remaining gauge theory.} sometimes called \emph{Kibble branch} \cite{Hanany:2010qu}, or \emph{enhanced Higgs branch} \cite{Beem:2021jnm} as a dual notion to the enhanced Coulomb branch \cite{Argyres:2016xmc}. The Higgs part of this branch has quaternionic dimension $3 \times 4 - (3^3 - 1^2)= 4$ (see (\ref{HasseDiagramU(3)})). In summary, 
\begin{equation}
    \mathrm{dim}_{\mathbb{H}} \, \mathcal{H} \left(  \raisebox{-.5\height}{
    \begin{tikzpicture}
\node (g1) at (10.5,3) [gauge,label=below:{U(3)}] {};
\node (f1) at (10.5,4) [flavor,label=above:{4}] {};
\draw (g1)--(f1);
\end{tikzpicture}} \right) = 4 =  \mathrm{dim}_{\mathbb{H}} \, \mathcal{C} \left( \raisebox{-.5\height}{
    \begin{tikzpicture}
\node (g1) at (-1,0) [gauge,label=below:{1}] {};
\node (g2) at (0,0) [gauge,label=below:{2}] {};
\node (g3) at (1,0) [gauge,label=below:{1}] {};
\node (g4) at (0,1) [flavourJ,label=right:{2}] {};
\draw (g1)--(g2)--(g3) (g2)--(g4);
\end{tikzpicture}}   \right) 
\label{equality1}
\end{equation}
but 
\begin{equation}
    \mathrm{dim}_{\mathbb{H}} \, \mathcal{C} \left(  \raisebox{-.5\height}{
    \begin{tikzpicture}
\node (g1) at (10.5,3) [gauge,label=below:{U(3)}] {};
\node (f1) at (10.5,4) [flavor,label=above:{4}] {};
\draw (g1)--(f1);
\end{tikzpicture}} \right) = 3 \neq 2 =  \mathrm{dim}_{\mathbb{H}} \, \mathcal{H} \left( \raisebox{-.5\height}{
    \begin{tikzpicture}
\node (g1) at (-1,0) [gauge,label=below:{1}] {};
\node (g2) at (0,0) [gauge,label=below:{2}] {};
\node (g3) at (1,0) [gauge,label=below:{1}] {};
\node (g4) at (0,1) [flavourJ,label=right:{2}] {};
\draw (g1)--(g2)--(g3) (g2)--(g4);
\end{tikzpicture}}   \right)  \, . 
\end{equation}

Running the algorithm iteratively we find
\begin{equation}
\raisebox{-.5\height}{
    \begin{tikzpicture}
\node (g1) at (10.5,3) [gauge,label=below:{U(3)}] {};
\node (f1) at (10.5,4) [flavor,label=above:{4}] {};
\draw (g1)--(f1);
\end{tikzpicture}}  \longrightarrow   
\raisebox{-.5\height}{
    \begin{tikzpicture}
\node (g1) at (-1,0) [gauge,label=below:{1}] {};
\node (g2) at (0,0) [gauge,label=below:{2}] {};
\node (g3) at (1,0) [gauge,label=below:{1}] {};
\node (g4) at (0,1) [flavourJ,label=right:{2}] {};
\draw (g1)--(g2)--(g3) (g2)--(g4);
\end{tikzpicture}}  \longrightarrow   
\raisebox{-.5\height}{
    \begin{tikzpicture}
\node (g1) at (10.5,3) [gauge,label=below:{U(2)}] {};
\node (f1) at (10.5,4) [flavor,label=above:{4}] {};
\draw (g1)--(f1);
\end{tikzpicture}}   \longrightarrow   
\raisebox{-.5\height}{
    \begin{tikzpicture}
\node (g1) at (-1,0) [gauge,label=below:{1}] {};
\node (g2) at (0,0) [gauge,label=below:{2}] {};
\node (g3) at (1,0) [gauge,label=below:{1}] {};
\node (g4) at (0,1) [flavourJ,label=right:{2}] {};
\draw (g1)--(g2)--(g3) (g2)--(g4);
\end{tikzpicture}}  \longrightarrow   \dots 
\end{equation}
The cycle on which the algorithm stabilizes is a well-known mirror pair. 

Note that adding free hypers on the right-hand side in (\ref{quiverExamplebad}) does not restore mirror symmetry, as this does not survive turning on Fayet-Iliopoulos (FI) parameters -- see Appendix \ref{bad} for details on this point. 

\vspace{.5cm}

The effects discussed in Table \ref{tab:effects} concern SQCD theories, i.e. theories with a single gauge group (unitary or special unitary). When considering a quiver with several gauge nodes, all the effects shown in the table can occur, but it is much more difficult to diagnose exactly when they occur: in particular, it is \emph{not} enough to check the bounds of Table \ref{tab:effects} at each gauge node to guarantee that there is a single cone, or that there is complete Higgsing, etc. 

\paragraph{Example}
To illustrate this point, consider the quiver 
\begin{equation}
\raisebox{-.5\height}{
    \begin{tikzpicture}
\node (g1) at (10,3) [gauge,label=below:{SU(2)}] {};
\node (g2) at (11.5,3) [gauge,label=below:{SU(3)}] {};
\node (g3) at (13,3) [gauge,label=below:{SU(2)}] {};
\node (f2) at (11.5,4) [flavor,label=above:{1}] {};
\draw (g1)--(g2)--(g3);
\draw (g2)--(f2);
\end{tikzpicture}}
\end{equation}
All the gauge nodes satisfy the equality $N_f = 2N_c - 1$, i.e.\ the balanced condition for SU nodes, but the algorithm of Section \ref{prescription} shows that the Higgs branch is made up of two cones, furthermore the theory is bad, as can be seen by the divergence of the monopole formula.

\subsection{Argyres-Douglas theories}
\label{sec:AD}

The Higgs branch of certain Argyres-Douglas (AD) theories has been argued to coincide with the Higgs branch of U/SU linear quivers where one or more nodes have negative balance \cite{Closset:2020afy,Giacomelli:2020ryy,Xie:2021ewm,Dey:2021rxw}. 
These linear quivers can be used in the algorithm presented in Section \ref{prescription} in order to produce
magnetic quivers for the Higgs branches of the AD theories. The results are consistent with the computations given in \cite{Closset:2020afy,Giacomelli:2020ryy,Xie:2021ewm,Dey:2021rxw}, which use different techniques.

As an illustration, let us consider the $D_{14}(\mathrm{SU}(8))$ theory, studied in \cite[Example 4]{Closset:2020afy}. We neglect free (twisted) hypermultiplets in this discussion. There, the Higgs branch of $D_{14}(\mathrm{SU}(8))$ is proposed to be the Higgs branch of the quiver
\begin{equation}
\raisebox{-.5\height}{
    \begin{tikzpicture}
\node (g1) at (1,3) [gauge,label=below:{U(7)}] {};
\node (g2) at (2,3) [gauge,label=below:{U(6)}] {};
\node (g3) at (3,3) [gauge,label=below:{U(6)}] {};
\node (g4) at (4,3) [gauge,label=below:{U(5)}] {};
\node (g5) at (5,3) [gauge,label=below:{U(5)}] {};
\node (g6) at (6,3) [gauge,label=below:{U(4)}] {};
\node (g7) at (7,3) [gauge,label=below:{SU(4)}] {};
\node (g8) at (8,3) [gauge,label=below:{U(3)}] {};
\node (g9) at (9,3) [gauge,label=below:{U(2)}] {};
\node (g10) at (10,3) [gauge,label=below:{U(2)}] {};
\node (g11) at (11,3) [gauge,label=below:{U(1)}] {};
\node (g12) at (12,3) [gauge,label=below:{U(1)}] {};
\node (f1) at (1,4) [flavor,label=above:{8}] {};
\draw (f1)--(g1)--(g2)--(g3)--(g4)--(g5)--(g6)--(g7)--(g8)--(g9)--(g10)--(g11)--(g12);
\end{tikzpicture}}\,.
\end{equation}
Inputting this quiver in our algorithm, yields the magnetic quiver
\begin{equation}
\label{ADquiver1}
\raisebox{-.5\height}{
    \begin{tikzpicture}
\node (g1) at (1,3) [gauge,label=below:{1}] {};
\node (g2) at (2,3) [gauge,label=below:{2}] {};
\node (g3) at (3,3) [gauge,label=below:{3}] {};
\node (g4) at (4,3) [gauge,label=below:{4}] {};
\node (g5) at (5,3) [gauge,label=below:{5}] {};
\node (g6) at (6,3) [gauge,label=below:{6}] {};
\node (g7) at (7,3) [gauge,label=below:{7}] {};
\node (g8) at (8.5,4) [gauge,label=right:{1}] {};
\node (g9) at (8.5,2) [gauge,label=right:{1}] {};
\draw (g1)--(g2)--(g3)--(g4)--(g5)--(g6)--(g7);
\draw[very thick] (g7)--(g8)--(g9)--(g7);
\node at (8.8,3) {12};
\node at (7.65,3.75) {4};
\node at (7.65,2.25) {4};
\end{tikzpicture}}\,,
\end{equation}
which agrees with \cite[(3.63)]{Giacomelli:2020ryy}.

The theory $D_{14}(\mathrm{SU}(8))$ can be decomposed into a ``generalized" quiver, or ``quiverine" \cite{Closset:2020scj,Closset:2020afy}, which contains other Argyres-Douglas theories \cite{Closset:2020afy}:
\begin{equation}
\label{electricAD}
    D_{14}(\mathrm{SU}(8))\cong\mathcal{D}_7 (8,4) - \mathrm{SU}(4) - D_7 (\mathrm{SU}(4)) \, ,
\end{equation}
where the right hand side denotes the gauging of a diagonal SU$(4)$ flavour symmetry of the $\mathcal{D}_7(8,4)$ and $D_7(\mathrm{SU}(4))$ theories. Magnetic quivers for the $\mathcal{D}_7 (8,4)$ theory and the $D_7 (\mathrm{SU}(4) )$ theory can be computed similarly to the case before, yielding the quiver on top of Figure \ref{fig:ADquiver2} and the quiver for the nilpotent cone of $SL(4)$, respectively. In Figure \ref{fig:ADquiver2} we comment on the magnetic quiver side of the SU$(4)$ gauging process.

\begin{figure}
    \centering
\makebox[\textwidth][c]{\begin{tikzpicture}
\node (g1) at (1,3) [gauge,label=below:{1}] {};
\node (g2) at (2,3) [gauge,label=below:{2}] {};
\node (g3) at (3,3) [gauge,label=below:{3}] {};
\node (g4) at (4,3) [gauge,label=below:{4}] {};
\node (g5) at (5,3) [gauge,label=below:{5}] {};
\node (g6) at (6,3) [gauge,label=below:{6}] {};
\node (g7) at (7,3) [gauge,label=below:{7}] {};
\node (g8) at (8.5,4) [gauge,label=right:{1}] {};
\node (g9) at (8.5,2) [gauge,fill=red,label=below:{4}] {};
\node (g10) at (9.5,2) [gauge,label=below:{3}] {};
\node (g11) at (10.5,2) [gauge,label=below:{2}] {};
\node (g12) at (11.5,2) [gauge,label=below:{1}] {};
\draw (g1)--(g2)--(g3)--(g4)--(g5)--(g6)--(g7)--(g9)--(g10)--(g11)--(g12);
\draw[very thick] (g7)--(g8)--(g9);
\node at (8.8,3) {3};
\node at (7.65,3.75) {4};
\node (h1) at (9.5,1) [gauge,label=below:{1}] {};
\node (h2) at (10.5,1) [gauge,label=below:{2}] {};
\node (h3) at (11.5,1) [gauge,label=below:{3}] {};
\node (h4) at (12.5,1) [gauge,fill=blue,label=below:{1}] {};
\draw (h1)--(h2)--(h3);
\draw[very thick] (h3)--(h4);
\node at (12,.5) {4};
\node at (10.5,2.5) {\textcolor{goodgreen}{SU$(4)$}};
        \draw[goodgreen] \convexpath{h1,h3}{6pt};
        \draw[goodgreen] \convexpath{g10,g12}{6pt};
\node (gg1) at (1,-2) [gauge,label=below:{1}] {};
\node (gg2) at (2,-2) [gauge,label=below:{2}] {};
\node (gg3) at (3,-2) [gauge,label=below:{3}] {};
\node (gg4) at (4,-2) [gauge,label=below:{4}] {};
\node (gg5) at (5,-2) [gauge,label=below:{5}] {};
\node (gg6) at (6,-2) [gauge,label=below:{6}] {};
\node (gg7) at (7,-2) [gauge,label=below:{7}] {};
\node (gg8) at (8.5,-1) [gauge,label=right:{1}] {};
\node (gg9) at (8.5,-3) [gauge,fill=purple,label=right:{1}] {};
\draw (gg1)--(gg2)--(gg3)--(gg4)--(gg5)--(gg6)--(gg7);
\draw[very thick] (gg7)--(gg8)--(gg9)--(gg7);
\node at (9,-2) {$4 \times 3$};
\node at (7.65,-1.25) {4};
\node at (7.35,-2.75) {$4 \times 1$};
\draw [thick,goodgreen,->] (10.5,0) -- (10.5,-3) -- (9.5,-3);
\node at (4,4) {$\mathsf{MQ}\left(\mathcal{D}_7(8,4)\right)$};
\node at (14.5,1) {$\mathsf{MQ}\left(D_7(\mathrm{SU}(4))\right)$};
\node at (5,-3.8) {$\mathsf{MQ}\left(D_{14}(\mathrm{SU}(8))\right)=\mathsf{MQ}\left(\mathcal{D}_7 (8,4)- SU(4) -D_7 (\mathrm{SU}(4) )\right)$};
\node at (12,-1.5) {gauge \color{goodgreen}SU$(4)$};
\end{tikzpicture}}
    \caption{Magnetic quivers ($\mathsf{MQ}$) involved in gauging $\mathcal{D}_7 (8,4)$ and $D_7 (\mathrm{SU}(4) )$ to $\mathcal{D}_7 (8,4)- SU(4) -D_7 (\mathrm{SU}(4) )\cong D_{14}(\mathrm{SU}(8))$. The SU$(4)$ gauging seems to be realized in the following way as an action on the magnetic quivers: The {\color{goodgreen}SU$(4)$} which is gauged is the diagonal subgroup of the SU$(4)$ global symmetries stemming from the balanced nodes highlighted in green in both original magnetic quivers. These two `tails' are destroyed by gauging the SU$(4)$. Furthermore the red {\color{red}U$(4)$} node, connected to the tail in the first quiver, and the blue {\color{blue}U$(1)$} node, connected to the tail in the second quiver, are identified. Since $\mathrm{gcd}(4,1)=1$ the resulting purple node should be a {\color{purple}U$(1)$}. However the nodes connected to the U$(4)$ in the top quiver should keep their balance, hence some edge numbers are multiplied by $4$, resulting in the quiver at the bottom of the figure, which matches \eqref{ADquiver1}. Note that free (twisted) hypermultiplets are neglected in those computations. See \cite[Section 3.5]{Giacomelli:2020ryy} for more information.}
    \label{fig:ADquiver2}
\end{figure}

\subsection{Beyond linear quivers}
\label{sec:beyondLinear}

We end this section with an example outside the scope of this paper, namely a non-linear quiver (i.e. the gauge nodes do not form an $A$-type Dynkin diagram).

We consider the following electric quiver, where the gauge nodes form a $D_4$ Dynkin diagram: 
\begin{equation}
  \raisebox{-.5\height}{ \begin{tikzpicture}
        \node[flavourJ,label=below:{1}] (0) at (0,0) {};
        \node[gaugeJ,label=below:{U$(2)$}] (1) at (1,0) {};
        \node[gaugeJ,label=below:{U$(3)$}] (2) at (2,0) {};
        \node[gaugeJ,label=below:{U$(2)$}] (3) at (3,0) {};
        \node[flavourJ,label=below:{1}] (4) at (4,0) {};
        \node[gaugeJ,label=left:{U$(2)$}] (2u) at (2,1) {};
        \node[flavourJ,label=left:{$1$}] (2uu) at (2,2) {};
        \draw (0)--(1)--(2)--(3)--(4) (2)--(2u)--(2uu);
    \end{tikzpicture}}
\label{D4_1}
\end{equation}
It is known that this quiver has a 3d mirror quiver
\begin{equation}
    \begin{tikzpicture}
        \node[gaugeJ,label=below:{U$(1)$}] (0) at (0,0) {};
        \node[gaugeJ,label=below:{USp$(4)$}] (1) at (1.5,0) {};
        \node[flavourJ,label=below:{$D_4$}] (2) at (3,0) {};
        \draw (0)--(1)--(2);
    \end{tikzpicture}\,,
\end{equation}
where the U$(1)$ can be viewed as an SO$(2)$. Both of these quivers can be obtained from a NS5-D3-D5 system in the presence of an ON${}^-$ plane, see e.g.\ \cite{Bourget:2021siw}.

Let us now turn all U nodes in \eqref{D4_1} into SU nodes, yielding
\begin{equation}
  \raisebox{-.5\height}{ \begin{tikzpicture}
        \node[flavourJ,label=below:{1}] (0) at (0,0) {};
        \node[gaugeJ,label=below:{SU$(2)$}] (1) at (1,0) {};
        \node[gaugeJ,label=below:{SU$(3)$}] (2) at (2,0) {};
        \node[gaugeJ,label=below:{SU$(2)$}] (3) at (3,0) {};
        \node[flavourJ,label=below:{1}] (4) at (4,0) {};
        \node[gaugeJ,label=left:{SU$(2)$}] (2u) at (2,1) {};
        \node[flavourJ,label=left:{$1$}] (2uu) at (2,2) {};
        \draw (0)--(1)--(2)--(3)--(4) (2)--(2u)--(2uu);
    \end{tikzpicture}}
\label{D4_2}
\end{equation}
By `exploding'\footnote{Explosion denotes the inverse process of hyper-K\"ahler implosion.} the flavour node \cite{Dancer:2020wll,Bourget:2021zyc}, we conjecture the magnetic quiver of \eqref{D4_2} takes the form:
\begin{equation}
  \raisebox{-.5\height}{  \begin{tikzpicture}
	\begin{pgfonlayer}{nodelayer}
		\node [style=gauge1] (0) at (0, 0) {};
		\node [style=gauge1] (1) at (-1.5, 0) {};
		\node [style=gauge1] (3) at (1.5, 0) {};
		\node [style=none] (5) at (0.7, 0.5) {$\mathrm{USp}(4)$};
		\node [style=none] (6) at (-1.5, -0.5) {$\mathrm{SO}(2)$};
		\node [style=none] (7) at (1.5, -0.5) {$\mathrm{SO}(2)$};
		\node [style=gauge1] (8) at (0, 1.5) {};
		\node [style=none] (9) at (0, 2) {$\mathrm{SO}(2)$};
		\node [style=gauge1] (10) at (-1, -1.5) {};
		\node [style=gauge1] (11) at (1, -1.5) {};
		\node [style=none] (12) at (-1, -2) {$\mathrm{SO}(2)$};
		\node [style=none] (13) at (1, -2) {$\mathrm{SO}(2)$};
	\end{pgfonlayer}
	\begin{pgfonlayer}{edgelayer}
		\draw (8) to (0);
		\draw (0) to (3);
		\draw (0) to (1);
		\draw (10) to (0);
		\draw (0) to (11);
	\end{pgfonlayer}
\end{tikzpicture}}
\label{E6mirror}
\end{equation}
The unrefined Coulomb branch Hilbert series of (\ref{E6mirror}), computed using the prescription of \cite{Bourget:2020xdz}, matches that of the Higgs branch of (\ref{D4_2}). Both are equal to 
\begin{equation}
    \frac{\left( \begin{array}{c} 1+47 t^4+72 t^6+526 t^8+792 t^{10}+2065 t^{12}\\+2304 t^{14}+3218 t^{16}+\dots\mathrm{palindromic}\dots+t^{32} \end{array} \right)}{(1-t^2)^6(1-t^4)^7(1-t^6)}\end{equation} 
Conversely, the unrefined Higgs branch Hilbert series of (\ref{E6mirror}) matches with the unrefined Coulomb branch Hilbert series of (\ref{D4_2}), both being equal to 
\begin{equation}
    \frac{ \left(\begin{array}{c}
        1-4 t^2+16 t^4-11 t^6+29 t^8+31 t^{10}+12 t^{12}  \\
          +31 t^{14}+29 t^{16}-11 t^{18}+16 t^{20}-4  t^{22}+t^{24}
    \end{array} \right)}{\left(1-t^2\right)^4 \left(1-t^4\right)^5 \left(1-t^6\right)} \, .  
   \end{equation} 
   
This suggests, that (\ref{D4_2}) and (\ref{E6mirror}) form a mirror pair. We do not know a unitary quiver with the same moduli space as \eqref{E6mirror}, suggesting that in general for a non-linear quiver with mixed U and SU nodes it is unlikely that unitary, simply laced magnetic quivers can be found. It would be interesting to understand under which circumstances orthosymplectic magnetic quivers are available; and whether other types of quivers (e.g. non simply laced, wreathed) can make an appearance in that context. Note that both \eqref{D4_1} and \eqref{D4_2} could also be obtained from the the $E_6$ quiver
\begin{equation}
   \raisebox{-.5\height}{ \begin{tikzpicture}
        \node[gaugeJ,label=below:{U$(1)$}] (0) at (0,0) {};
        \node[gaugeJ,label=below:{U$(2)$}] (1) at (1,0) {};
        \node[gaugeJ,label=below:{U$(3)$}] (2) at (2,0) {};
        \node[gaugeJ,label=below:{U$(2)$}] (3) at (3,0) {};
        \node[gaugeJ,label=below:{U$(1)$}] (4) at (4,0) {};
        \node[gaugeJ,label=left:{U$(2)$}] (2u) at (2,1) {};
        \node[flavourJ,label=left:{$1$}] (2uu) at (2,2) {};
        \draw (0)--(1)--(2)--(3)--(4) (2)--(2u)--(2uu);
    \end{tikzpicture}} \,,
\end{equation}
whose Coulomb branch is the closure of the minimal nilpotent orbit of $E_6$, and Higgs branch is $\mathbb{C}^2 / \Gamma_{E_6}$, by trading every unitary node to a special unitary node. The 3d mirror of this $E_6$ quiver is the 3d reduction of the $E_6$ rank 1
theory \cite{Dasgupta:1996ij,Minahan:1996fg}. This theory has no representation as a quiver. This strongly suggests, that not every theory has a magnetic quiver (orthosymplectic, non-simply laced or otherwise) and that one has to turn to other objects, such as the quiverines of \cite{Closset:2020scj,Closset:2020afy}. We leave these questions for future exploration. 
 
\section*{Acknowledgements}
It is a pleasure to thank Santiago Cabrera for many discussions and his insights on brane lockings when we started this project. We also like to thank Stefano Cremonesi, Simone Giacomelli and Noppadol Mekareeya for discussions. Furthermore AB and JFG thank the team of the Carcans Workshop on 6d SCFTs and Geometry for hospitality. This work is supported by STFC grants ST/P000762/1 and ST/T000791/1. AB is supported by the ERC Consolidator Grant 772408-Stringlandscape, and by the LabEx ENS-ICFP: ANR-10-LABX-0010/ANR-10-IDEX-0001-02 PSL*. 

\clearpage

\appendix

\section{\texorpdfstring{The embedding $\mathcal{H}_{\mathrm{SU}} \supset \mathcal{H}_{\mathrm{U}/\mathrm{SU}}$}{The embedding HSU and HUSU}}
\label{app:quotient}

\subsection{The subvariety point of view}

In this work we use the fact that
\begin{equation}
    \mathcal{H}_{\mathrm{SU}} \supset \mathcal{H}_{\mathrm{U}/\mathrm{SU}}  \,
    \label{eq:UinsideSU}
\end{equation}
in order to motivate the web locking. This equation deserves some comments. For simplicity we will restrict ourselves to studying SQCD theories, i.e.\ linear quivers with a single gauge node. Let us start with
\begin{equation}
    \begin{tikzpicture}
        \node[gaugeJ,label=right:{SU$(k)$}] (0) at (0,0) {};
        \node[flavourJ,label=right:{$N\geq 2k$}] (1) at (0,1) {};
        \draw (0)--(1);
    \end{tikzpicture}
\end{equation}
which has the magnetic quiver
\begin{equation}
    \begin{tikzpicture}
        \node[gaugeJ,label=below:{$1$}] (0) at (0,0) {};
        \node (1) at (1,0) {$\cdots$};
        \node[gaugeJ,label=below:{$k$}] (2) at (2,0) {};
        \node[gaugeJ,label=left:{$1$}] (2u) at (2,1) {};
        \node (3) at (3,0) {$\cdots$};
        \node[gaugeJ,label=below:{$k$}] (4) at (4,0) {};
        \node[gaugeJ,label=right:{$1$}] (4u) at (4,1) {};
        \node (5) at (5,0) {$\cdots$};
        \node[gaugeJ,label=below:{$1$}] (6) at (6,0) {};
        \draw (0)--(1)--(2)--(3)--(4)--(5)--(6) (2)--(2u) (4)--(4u);
        \draw [decorate,decoration={brace,amplitude=2.5pt}] (6.2,-0.6)--(-0.2,-0.6);
        \node at (3,-1) {$N-1$};
    \end{tikzpicture}
    \label{eq:magQuivSQCD_SU_high}
\end{equation}
In this case the Higgs branch is the Higgs variety, as there are no nilpotent operators in the chiral ring. The global symmetry of the Higgs branch is $\mathrm{SU}(N)\times \mathrm{U}(1)_B$, the $\mathrm{U}(1)_B$ factor being the baryonic symmetry.

Following the conventions of \cite{Argyres:1996eh,Bourget:2019rtl} the generators of the chiral ring are the $N\times N$ meson matrix $M$ of degree $2$, and the baryonic generators $B^{i_1,\dots,i_k}$ and $\tilde{B}_{i_1,\dots,i_k}$ of degree $k$. The relations are
\begin{equation}
    \begin{split}
        0=&(\star B) \tilde{B}-\star (M^{k})\\
        0=&M \cdot (\star B)  \\
        0=&(\star \tilde{B}) \cdot M \\
        0=&(M - \frac{1}{k} \mathrm{Tr}(M) \mathbf{1}_{N}) \cdot B = \tilde{B} \cdot (M - \frac{1}{k} \mathrm{Tr}(M) \mathbf{1}_{N}) \\
        0=&M \cdot (M - \frac{1}{k} \mathrm{Tr}(M) \mathbf{1}_{N}) \\
        0=&B^{[i_1 i_2\dots  i_{k}} B^{j_1] j_2 \dots  j_{k}} \\
        0=&\tilde{B}_{[i_1 i_2\dots  i_{k}}  \tilde{B}_{j_1] j_2 \dots  j_{k}} \;.
    \end{split}
    \label{eq:SQCD_SU_rels}
\end{equation}
We can compare this to the theory
\begin{equation}
    \begin{tikzpicture}
        \node[gaugeJ,label=right:{U$(k)$}] (0) at (0,0) {};
        \node[flavourJ,label=right:{$N\geq 2k$}] (1) at (0,1) {};
        \draw (0)--(1);
    \end{tikzpicture}
\end{equation}
with the same $k$ and $N$ as before. The equations to describe its Higgs branch are much simpler. We only have the mesonic generators $M$, and the relations:
\begin{equation}
    M^2 = 0 \, , \qquad \mathrm{Tr}(M) = 0 \, ,  \qquad \mathrm{rank}(M) \leq k \, .
\end{equation}
The $U$ theory is obtained from the $SU$ theory by gauging the U$(1)_B$ symmetry. This means that the two Higgs branches are related by a hyper-K\"ahler quotient:
\begin{equation}
    \mathcal{H}_{\mathrm{U}} =\mathcal{H}_{\mathrm{SU}}///\mathrm{U}(1)_B\,.
    \label{eq:UinsideSUasHKquotient}
\end{equation}
This is visible on the magnetic quiver side as follows: the two U$(1)$ gauge nodes at the top of \eqref{eq:magQuivSQCD_SU_high} are combined into a single U$(1)$ by the quotient. We obtain the magnetic quiver of the $U$ theory:
\begin{equation}
    \begin{tikzpicture}
        \node[gaugeJ,label=below:{$1$}] (0) at (0,0) {};
        \node (1) at (1,0) {$\cdots$};
        \node[gaugeJ,label=below:{$k$}] (2) at (2,0) {};
        \node (3) at (3,0) {$\cdots$};
        \node[gaugeJ,label=below:{$k$}] (4) at (4,0) {};
        \node (5) at (5,0) {$\cdots$};
        \node[gaugeJ,label=below:{$1$}] (6) at (6,0) {};
        \node[gaugeJ,label=left:{$1$}] (u) at (3,1) {};
        \draw (0)--(1)--(2)--(3)--(4)--(5)--(6) (2)--(u)--(4);
        \draw [decorate,decoration={brace,amplitude=2.5pt}] (6.2,-0.6)--(-0.2,-0.6);
        \node at (3,-1) {$N-1$};
    \end{tikzpicture}\,.
\end{equation}
Equation \eqref{eq:UinsideSUasHKquotient} however does not immediately imply equation \eqref{eq:UinsideSU}. We can see that $\mathcal{H}_{\mathrm{U}}$ is a subvariety of $\mathcal{H}_{\mathrm{SU}}$ by adding the relations
\begin{equation}
    \mathrm{Tr}(M) = 0\textnormal{, }B=0\textnormal{ and }\tilde{B}=0
\end{equation}
to those in \eqref{eq:SQCD_SU_rels}. This is in agreement with \eqref{eq:UinsideSUasHKquotient} as $B$ and $\tilde{B}$ are exactly the generators of $\mathcal{H}_{\mathrm{SU}}$ which are charged under $\mathrm{U}(1)_B$, and the first equation $\mathrm{Tr}(M) = 0$ simply prevents the appearance of nilpotent operators. 

The case $k\leq N<2k$ works essentially along the same lines. In this case there are two magnetic quivers for the $SU$ theory, additionally there are nilpotent operators in the chiral ring. However nothing changes in the procedure of going from $\mathcal{H}_{\mathrm{SU}}$ to $\mathcal{H}_{\mathrm{U}}$.

For $N<k$ there are no baryonic generators for $\mathcal{H}_{\mathrm{SU}}$ and the U$(1)_B$ symmetry is carried by nilpotent operators alone. In this case one only needs to add the condition that the meson matrix is traceless in order to obtain $\mathcal{H}_{\mathrm{U}}$.

\subsection{A comment on Hasse diagrams}

To each Higgs branch discussed in this paper, one can associate a Hasse diagram, depicting the partial order of symplectic leaves and elementary transitions between them. 
It should be noted that the embedding (\ref{eq:UinsideSU}) may or may not translate into an inclusion of the respective Hasse diagrams. 

\paragraph{Example} 
Consider the two theories depicted below, with the Hasse diagram of their Higgs branches:\footnote{The FI parameter for the U(3) theory is taken to be vanishing, see Appendix \ref{bad} for more on this point.  } 
\begin{equation}
    \raisebox{-.5\height}{\begin{tikzpicture}
        \node[gauge,label=right:{SU(3)}] (0) at (0,0) {};
        \node[flavor,label=right:{4}] (1) at (0,1) {};
        \draw (0)--(1);
    \end{tikzpicture}}
    \raisebox{-.5\height}{\begin{tikzpicture}
        \node[hasse] (0) at (0,0) {};
        \node[hasse] (1) at (0,1) {};
        \node[hasse] (2) at (-1,2) {};
        \node[hasse] (3) at (1,2) {};
\draw (0) edge [] node[label=left:$a_3$] {} (1);
\draw (1) edge [] node[label=left:$a_1$] {} (2);
\draw (1) edge [] node[label=right:$a_1$] {} (3);
    \end{tikzpicture}} \qquad \qquad  \qquad \qquad  \raisebox{-.5\height}{\begin{tikzpicture}
        \node[gauge,label=right:{U(3)}] (0) at (0,0) {};
        \node[flavor,label=right:{4}] (1) at (0,1) {};
        \draw (0)--(1);
    \end{tikzpicture}}
    \raisebox{-.5\height}{\begin{tikzpicture}
        \node[hasse] (0) at (0,0) {};
        \node[hasse] (1) at (0,1) {};
        \node[hasse] (2) at (-1,2) {};
\draw (0) edge [] node[label=left:$a_3$] {} (1);
\draw (1) edge [] node[label=left:$a_1$] {} (2);
    \end{tikzpicture}}
\end{equation}
The theory on the left has a mesonic branch and a baryonic branch, intersecting along a quaternionic dimension 3 locus. When gauging the baryonic U(1), the baryonic branch is removed, and the mesonic branch is unaffected, as can be seen in the Hasse diagram. 

Consider now instead the two theories along with their Higgs branch Hasse diagrams
\begin{equation}
    \raisebox{-.5\height}{\begin{tikzpicture}
        \node[gauge,label=right:{SU(2)}] (0) at (0,0) {};
        \node[flavor,label=right:{4}] (1) at (0,1) {};
        \draw (0)--(1);
    \end{tikzpicture}}\quad
    \raisebox{-.5\height}{\begin{tikzpicture}
        \node[hasse] (0) at (0,0) {};
        \node[hasse] (1) at (0,2.5) {};
\draw (0) edge [] node[label=left:$d_4$] {} (1);
    \end{tikzpicture}} \qquad \qquad  \qquad \qquad  \raisebox{-.5\height}{\begin{tikzpicture}
        \node[gauge,label=right:{U(2)}] (0) at (0,0) {};
        \node[flavor,label=right:{4}] (1) at (0,1) {};
        \draw (0)--(1);
    \end{tikzpicture}}\quad
    \raisebox{-.5\height}{\begin{tikzpicture}
        \node[hasse] (0) at (0,0) {};
        \node[hasse] (1) at (0,1.5) {};
        \node[hasse] (2) at (0,2) {};
        \node at (0,2.5) {};
\draw (0) edge [] node[label=left:$a_3$] {} (1);
\draw (1) edge [] node[label=left:$a_1$] {} (2);
    \end{tikzpicture}}
\end{equation}
In this case, gauging the baryonic U(1) symmetry completely changes the structure of the Hasse diagram. Note in particular that the Higgs branch of the SU(2) theory has only a singularity at the origin, while the U(2) theory develops a non-trivial singular locus (corresponding to the fact that U(2) can be Higgsed to U(1)). 

\vspace{.5cm}

Hence we see that in general the Hasse diagram for $\mathcal{H}_{\mathrm{U}/\mathrm{SU}}$ does not have to be a subdiagram of the Hasse diagram of $\mathcal{H}_{\mathrm{SU}}$, although it can be in some specific cases. 

\section{A comment on bad theories}\label{bad}

In this paper we compute magnetic quivers of 3d $\mathcal{N}=4$ good, ugly and bad theories given by linear quivers with unitary or special unitary gauge nodes. The good theories enjoy the following properties:
\begin{enumerate}
    \item Every leaf closure in the moduli space is conical if masses and FI parameters are turned off, and there is a unique point in the moduli space where the theory flows to a fully interacting SCFT in the IR.
    \item The theory has a 3d mirror with the same moduli space, albeit the names Coulomb and Higgs are exchanged as well as mass and FI.
\end{enumerate}

\begin{figure}[ht]
\makebox[\textwidth][c]{
\scalebox{.85}{\begin{tikzpicture}
    \node at (0,0) {$\begin{tabular}{c|c|c|c|c|c|c|c|c|c|c}
& $x^0$ & $x^1$ & $x^2$ & $x^3$ & $x^4$ & $x^5$ & $x^6$ & $x^7$ & $x^8$ & $x^9$\\
\hline
{\color{red}NS5} & x & x & x & x & x & x & & & &  \\
\hline
D3 & x & x & x & & & & x & & & \\
\hline
{\color{blue}D5} & x & x & x & & & & & x & x & x
\end{tabular}$};
\node at (0,-5) {$\begin{tikzpicture}
        \begin{scope}[shift={(-2,1)}]
        \draw[->] (-3,1)--(-2,1);
        \node at (-1.5,1) {$(x^6)$};
        \draw[->] (-3,1)--(-3,2);
        \node at (-3,2.3) {$(x^3,x^4,x^5)$};
        \draw[->] (-3,1)--(-4,0);
        \node at (-5,0) {$(x^7,x^8,x^9)$};
        \end{scope}
        \draw[red] (0,0)--(0,4) (5,0)--(5,4);
        \node at (0,-0.5) {NS5};
        \node at (5,-0.5) {NS5};
        \draw[blue] (0.5,1)--(2.5,3) (2.5,1)--(4.5,3);
        \draw[dotted] (-1,2)--(6,2);
        \node at (-1.5,2) {origin};
        \draw[thick] (0,2)--(5,2);
        \node at (2.7,2.3) {$k$ D3};
        \draw [decorate,decoration={brace,amplitude=5pt}] (2.3,3.1)--(4.7,3.1);
        \node at (3.5,3.5) {$N$ D5};
        \node at (2,1.4) {\dots};
        \begin{scope}[shift={(7,1.5)}]
        \node[gaugeJ,label=below:{$k$}] (1) at (0,0) {};
        \node[flavourJ,label=above:{$N$}] (2) at (0,1) {};
        \draw (1)--(2);
        \end{scope}
    \end{tikzpicture}$};
    \node at (-8,0) {\Large(1)};
    \node at (-8,-5) {\Large(2)};
\end{tikzpicture}}}
\caption{(1) The Type IIB set-up: the 'x' mark the spacetime directions spanned by the various branes. (2) Depiction of the brane system for $\mathrm{U}(k)$ with $N$ flavors.}
\label{fig:setup}
\end{figure}
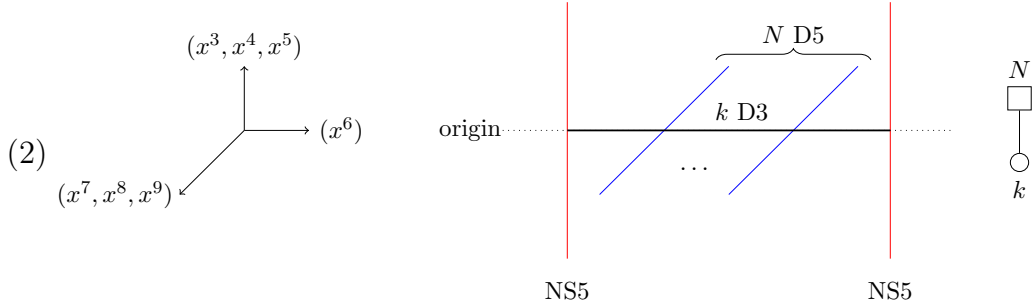

If the theory is bad, however, both of these statements are untrue in general. In the following we summarize the complications of bad theories and their solution following \cite{Yaakov:2013fza,Assel:2017jgo}, using the brane systems of \cite{Hanany:1996ie}. Our conventions for depicting brane systems are summarized in Figure \ref{fig:setup}. 

For simplicity we only consider $\mathrm{U}(k)$ SQCD, and illustrate through examples.

\paragraph{U(3) with 6 flavors.} As a warm-up we start with a good theory. The brane system for U(3) with 6 flavors at the origin of the moduli space is:
\begin{equation}
   \raisebox{-.5\height}{ \begin{tikzpicture}
        \draw[dotted] (-2,0)--(11,0);
        \draw[red] (0,-2)--(0,2) (9,-2)--(9,2);
        \draw[blue] (1,-1)--(3,1) (2,-1)--(4,1) (3,-1)--(5,1) (4,-1)--(6,1) (5,-1)--(7,1) (6,-1)--(8,1);
        \draw (0,0.1)--(9,0.1) (0,0)--(9,0) (0,-0.1)--(9,-0.1);
    \end{tikzpicture}}
\end{equation}

We can move to the Coulomb branch of the theory by moving all D3 branes along the NS5 branes:
\begin{equation}
    \raisebox{-.5\height}{   \begin{tikzpicture}
        \draw[dotted] (-2,0)--(11,0);
        \draw[red] (0,-2)--(0,2) (9,-2)--(9,2);
        \draw[blue] (1,-1)--(3,1) (2,-1)--(4,1) (3,-1)--(5,1) (4,-1)--(6,1) (5,-1)--(7,1) (6,-1)--(8,1);
        \draw (0,-0.6)--(9,-0.6) (0,-1.1)--(9,-1.1) (0,-1.6)--(9,-1.6);
    \end{tikzpicture}}
\end{equation}
At a generic point on the Coulomb branch the gauge group is broken to $\mathrm{U}(1)^3$ and all hypers are massive.

We can move to the Higgs branch of the theory by breaking the D3 branes into segments (respecting the S-rule) and moving the D3 segments along the D5 branes:
\begin{equation}
    \raisebox{-.5\height}{   \begin{tikzpicture}
        \draw[dotted] (-2,0)--(11,0);
        \draw[red] (0,-2)--(0,2) (9,-2)--(9,2);
        \draw[blue] (1,-1)--(3,1) (2,-1)--(4,1) (3,-1)--(5,1) (4,-1)--(6,1) (5,-1)--(7,1) (6,-1)--(8,1);
        \draw (0,0.1)--(2.1,0.1) (5.1,0.1)--(9,0.1);
        \draw (0,0)--(3,0) (6,0)--(9,0);
        \draw (0,-0.1)--(4-0.1,-0.1) (7-0.1,-0.1)--(9,-0.1);
        \begin{scope}[shift={(-0.7,-0.7)}]
        \draw (2,0)--(3,0) (6,0)--(7,0);
        \draw (3-0.1,-0.1)--(4-0.1,-0.1) (3+0.1,+0.1)--(4+0.1,+0.1) (5-0.1,-0.1)--(6-0.1,-0.1) (5+0.1,+0.1)--(6+0.1,+0.1);
        \draw (4-0.2,-0.2)--(5-0.2,-0.2) (4,0)--(5,0) (4+0.2,0.2)--(5+0.2,0.2);
        \end{scope}
    \end{tikzpicture}}
    \label{eq:U3-6F}
\end{equation}
At a generic point on the Higgs branch the gauge group is completely broken and we are left with nine massless free hypers.

Turning on all mass parameters (dashed lines), we can see that the Higgs branch is lifted, and the Coulomb branch is resolved:
\begin{equation}
    \raisebox{-.5\height}{   \begin{tikzpicture}
        \draw[dotted] (-2,0)--(11,0);
        \draw[red] (0,-2)--(0,2) (9,-2)--(9,2);
        \draw[blue] (1,-1)--(3,1) (2,-1+0.2)--(4,1+0.2) (3,-1+0.4)--(5,1+0.4) (4,-1+0.6)--(6,1+0.6) (5,-1+0.8)--(7,1+0.8) (6,-1+1)--(8,1+1);
        \draw[dashed] (3,0)--(3,0.2) (4,0)--(4,0.4) (5,0)--(5,0.6) (6,0)--(6,0.8) (7,0)--(7,1);
        \draw (0,-0.6)--(9,-0.6) (0,-1.1)--(9,-1.1) (0,-1.6)--(9,-1.6);
    \end{tikzpicture}}
\end{equation}

Similarly, turning on the FI parameter (dashed line), we can see that the Coulomb branch is lifted and the Higgs branch is resolved:
\begin{equation}
     \raisebox{-.5\height}{  \begin{tikzpicture}
        \draw[dotted] (-2,0)--(11,0);
        \draw[red] (0,-2)--(0,2) (9+0.5,-2+0.5)--(9+0.5,2+0.5);
        \draw[dashed] (9,0)--(9+0.5,0+0.5);
        \draw[blue] (1,-1)--(3,1) (2,-1)--(4,1) (3,-1)--(5,1) (4,-1)--(6,1) (5,-1)--(7,1) (6,-1)--(8,1);
        \draw (0,0.1)--(2.1,0.1);
        \draw (0,0)--(3,0);
        \draw (0,-0.1)--(4-0.1,-0.1);
        \begin{scope}[shift={(0.5,0.5)}]
        \draw (5.1,0.1)--(9,0.1) (6,0)--(9,0) (7-0.1,-0.1)--(9,-0.1);
        \end{scope}
        \begin{scope}[shift={(-0.7,-0.7)}]
        \draw (2,0)--(3,0) (6,0)--(7,0);
        \draw (3-0.1,-0.1)--(4-0.1,-0.1) (3+0.1,+0.1)--(4+0.1,+0.1) (5-0.1,-0.1)--(6-0.1,-0.1) (5+0.1,+0.1)--(6+0.1,+0.1);
        \draw (4-0.2,-0.2)--(5-0.2,-0.2) (4,0)--(5,0) (4+0.2,0.2)--(5+0.2,0.2);
        \end{scope}
    \end{tikzpicture}}
\end{equation}

Let us now focus on the Higgs phase when masses and FI are off. The magnetic quiver, which is the 3d mirror of the theory, can be read off from the system after doing a few HW transitions from \eqref{eq:U3-6F} annihilating all D3 branes stuck between a D5 and an NS5:
\begin{equation}
     \raisebox{-.5\height}{  \begin{tikzpicture}
        \draw[dotted] (0,0)--(9,0);
        \draw[red] (4.3,-2)--(4.3,2) (4.7,-2)--(4.7,2);
        \draw[blue] (1,-1)--(3,1) (2,-1)--(4,1) (3,-1)--(5,1) (4,-1)--(6,1) (5,-1)--(7,1) (6,-1)--(8,1);
        \begin{scope}[shift={(-0.7,-0.7)}]
        \draw (2,0)--(3,0) (6,0)--(7,0);
        \draw (3-0.1,-0.1)--(4-0.1,-0.1) (3+0.1,+0.1)--(4+0.1,+0.1) (5-0.1,-0.1)--(6-0.1,-0.1) (5+0.1,+0.1)--(6+0.1,+0.1);
        \draw (4-0.2,-0.2)--(5-0.2,-0.2) (4,0)--(5,0) (4+0.2,0.2)--(5+0.2,0.2);
        \end{scope}
        \begin{scope}[shift={(10,0)}]
        \node at (0,0) {$\mathsf{Q}_m=$};
        \node[gaugeJ,label=below:{$1$}] (1) at (1,0) {};
        \node[gaugeJ,label=below:{$2$}] (2) at (2,0) {};
        \node[gaugeJ,label=below:{$3$}] (3) at (3,0) {};
        \node[gaugeJ,label=below:{$2$}] (4) at (4,0) {};
        \node[gaugeJ,label=below:{$1$}] (5) at (5,0) {};
        \node[flavourJ,label=left:{$2$}] (3u) at (3,1) {};
        \draw (1)--(2)--(3)--(4)--(5) (3u)--(3);
        \end{scope}
    \end{tikzpicture}}
\end{equation}
One can straight forwardly convince oneself that the moduli space of the 3d mirror is indeed the same as that of U(3) with 6 flavors, and turning on masses on one side corresponds to turning on FI parameters on the other side and vice versa.

The Hasse diagram of the full moduli space of U(3) with 6 flavors can be obtained straightforwardly, e.g. from the brane system, or using different methods \cite{Cabrera:2016vvv,Cabrera:2017njm,Bourget:2019aer,Grimminger:2020dmg}. We use the conventions of \cite{Grimminger:2020dmg}:
\begin{equation}
    \raisebox{-.5\height}{   \begin{tikzpicture}[scale=1.5]
        \node[hasseJ,label=left:{\color{red}0},label=right:{\color{blue}0}] (1) at (0,0) {};
        \node[hasseJ,label=left:{\color{red}0},label=right:{\color{blue}5}] (2) at (1,1) {};
        \node[hasseJ,label=left:{\color{red}0},label=right:{\color{blue}8}] (3) at (2,2) {};
        \node[hasseJ,label=left:{\color{red}0},label=right:{\color{blue}9}] (4) at (3,3) {};
        \node[hasseJ,label=left:{\color{red}1},label=right:{\color{blue}0}] (1a) at (-1,1) {};
        \node[hasseJ,label=left:{\color{red}1},label=right:{\color{blue}5}] (2a) at (0,2) {};
        \node[hasseJ,label=left:{\color{red}1},label=right:{\color{blue}8}] (3a) at (1,3) {};
        \node[hasseJ,label=left:{\color{red}2},label=right:{\color{blue}0}] (1aa) at (-2,2) {};
        \node[hasseJ,label=left:{\color{red}2},label=right:{\color{blue}5}] (2aa) at (-1,3) {};
        \node[hasseJ,label=left:{\color{red}3},label=right:{\color{blue}0}] (1aaa) at (-3,3) {};
        \draw[blue] (1)--(2)--(3)--(4) (1a)--(2a)--(3a) (1aa)--(2aa);
        \draw[red] (1)--(1a)--(1aa)--(1aaa) (2)--(2a)--(2aa) (3)--(3a);
        \node at ($(1)!0.5!(2)$) {$a_5$};
        \node at ($(2)!0.5!(3)$) {$a_3$};
        \node at ($(3)!0.5!(4)$) {$a_1$};
        \node at ($(1a)!0.5!(2a)$) {$a_5$};
        \node at ($(2a)!0.5!(3a)$) {$a_3$};
        \node at ($(1aa)!0.5!(2aa)$) {$a_5$};
        \node at ($(1)!0.5!(1a)$) {$A_1$};
        \node at ($(1a)!0.5!(1aa)$) {$A_3$};
        \node at ($(1aa)!0.5!(1aaa)$) {$A_5$};
        \node at ($(2)!0.5!(2a)$) {$A_1$};
        \node at ($(2a)!0.5!(2aa)$) {$A_3$};
        \node at ($(3)!0.5!(3a)$) {$A_1$};
    \end{tikzpicture}}
\end{equation}

The lowest dimensional leaf is of dimension zero, the origin of the moduli space. The Hasse diagram of the 3d mirror is obtained by exchanging the red and blue colors.

\paragraph{U(3) with 4 flavors.} Let us now turn to a bad theory. We again pick the gauge group U(3), but this time we only have 4 flavors. Before turning to branes let us make a few observations:

\begin{enumerate}
    \item The theory suffers from so called incomplete Higgsing, which means that even on a generic point on the Higgs branch the gauge group is not completely broken. The gauge group can only be broken to a U(1) theory. This means that on a generic point on the Higgs branch there are still Coulomb directions in the full moduli space.
    \item The Coulomb branch is not a cone. It can be viewed as the Coulomb branch of pure U(1) where at each point there is a cone which is the Coulomb branch of U(2) with 4 flavors. In more mathematical terms, the Coulomb branch of U(3) with 4 flavors is the \emph{generalized affine Grassmannian slice} $\overline{\mathcal{W}}^{[4]}_{[-2]}$ of SL$(2,\mathbb{C})$ \cite{Braverman:2016pwk}.
\end{enumerate}

The brane system for U(3) with 4 flavors is:
\begin{equation}
    \raisebox{-.5\height}{   \begin{tikzpicture}
        \node at (0,-2.2) {A};
        \node at (7,-2.2) {B};
        \draw[dotted] (-2,0)--(9,0);
        \draw[red] (0,-2)--(0,2) (7,-2)--(7,2);
        \draw[blue] (1,-1)--(3,1) (2,-1)--(4,1) (3,-1)--(5,1) (4,-1)--(6,1);
        \draw (0,0.1)--(7,0.1) (0,0)--(7,0) (0,-0.1)--(7,-0.1);
    \end{tikzpicture}}
\end{equation}
where we labelled the two NS5 branes for reasons which will become apparent later.
Just like before we can move to the Coulomb branch of the theory by moving all D3 branes along the NS5 branes:
\begin{equation}
     \raisebox{-.5\height}{  \begin{tikzpicture}
        \node at (0,-2.2) {A};
        \node at (7,-2.2) {B};
        \draw[dotted] (-2,0)--(9,0);
        \draw[red] (0,-2)--(0,2) (7,-2)--(7,2);
        \draw[blue] (1,-1)--(3,1) (2,-1)--(4,1) (3,-1)--(5,1) (4,-1)--(6,1);
        \draw (0,-0.6)--(7,-0.6) (0,-1.1)--(7,-1.1) (0,-1.6)--(7,-1.6);
    \end{tikzpicture}}
\end{equation}
At a generic point on the Coulomb branch the gauge group is broken to U(1)${}^3$ and all hypers are massive.

However when attempting to move to the Higgs branch, we can only break two of the D3 branes into segments which we can move along the D5 branes. The remaining D3 brane can move along a Coulomb direction:
\begin{equation}
   \raisebox{-.5\height}{    \begin{tikzpicture}
        \node at (0,-2.2) {A};
        \node at (7,-2.2) {B};
        \draw[dotted] (-2,0)--(9,0);
        \draw[red] (0,-2)--(0,2) (7,-2)--(7,2);
        \draw[blue] (1,-1)--(3,1) (2,-1)--(4,1) (3,-1)--(5,1) (4,-1)--(6,1);
        \draw (0,0.1)--(2.1,0.1) (4.1,0.1)--(7,0.1);
        \draw (0,0)--(3,0) (5,0)--(7,0);
        \begin{scope}[shift={(-0.5,-0.5)}]
        \draw (2,0)--(3,0) (4,0)--(5,0);
        \draw (3-0.1,-0.1)--(4-0.1,-0.1) (3+0.1,+0.1)--(4+0.1,+0.1);
        \end{scope}
        \draw (0,-1.6)--(7,-1.6);
    \end{tikzpicture}}
\end{equation}
We can read a magnetic quiver for the subsystem of branes expressing the Higgs branch after doing a HW transition:
\begin{equation}
    \raisebox{-.5\height}{   \begin{tikzpicture}
        \node at (3.3,-2.2) {A};
        \node at (3.7,-2.2) {B};
        \draw[dotted] (0,0)--(7,0);
        \draw[red] (3.3,-2)--(3.3,2) (3.7,-2)--(3.7,2);
        \draw[blue] (1,-1)--(3,1) (2,-1)--(4,1) (3,-1)--(5,1) (4,-1)--(6,1);
        
        \begin{scope}[shift={(-0.5,-0.5)}]
        \draw (2,0)--(3,0) (4,0)--(5,0);
        \draw (3-0.1,-0.1)--(4-0.1,-0.1) (3+0.1,+0.1)--(4+0.1,+0.1);
        \end{scope}
        
        \draw (3.3,-1.6)--(3.7,-1.6);
        
        \begin{scope}[shift={(8,0)}]
        \node at (0,0) {$\mathsf{Q}_m=$};
        \node[gaugeJ,label=below:{$1$}] (1) at (1,0) {};
        \node[gaugeJ,label=below:{$2$}] (2) at (2,0) {};
        \node[gaugeJ,label=below:{$1$}] (3) at (3,0) {};
        \node[flavourJ,label=left:{$2$}] (2u) at (2,1) {};
        \draw (1)--(2)--(3) (2u)--(2);
        \end{scope}
    \end{tikzpicture}}
\end{equation}
This is the magnetic quiver (also 3d mirror) of U(2) with 4 flavors.

The Hasse diagram for the full moduli space of U(3) with 4 flavors is:
\begin{equation}
     \raisebox{-.5\height}{  \begin{tikzpicture}[scale=1.5]
        \node[hasseJ,label=left:{\color{red}1},label=right:{\color{blue}0}] (1) at (0,0) {};
        \node[hasseJ,label=left:{\color{red}1},label=right:{\color{blue}3}] (2) at (1,1) {};
        \node[hasseJ,label=left:{\color{red}1},label=right:{\color{blue}4}] (3) at (2,2) {};
        \node[hasseJ,label=left:{\color{red}2},label=right:{\color{blue}0}] (1a) at (-1,1) {};
        \node[hasseJ,label=left:{\color{red}2},label=right:{\color{blue}3}] (2a) at (0,2) {};
        \node[hasseJ,label=left:{\color{red}3},label=right:{\color{blue}0}] (1aa) at (-2,2) {};
        \draw[blue] (1)--(2)--(3) (1a)--(2a) ;
        \draw[red] (1)--(1a)--(1aa) (2)--(2a);
        
        \node at ($(1)!0.5!(2)$) {$a_3$};
        \node at ($(2)!0.5!(3)$) {$a_1$};
        \node at ($(1a)!0.5!(2a)$) {$a_3$};
        
        \node at ($(1)!0.5!(1a)$) {$A_1$};
        \node at ($(1a)!0.5!(1aa)$) {$A_3$};
        \node at ($(2)!0.5!(2a)$) {$A_1$};
    \end{tikzpicture} }
    \label{HasseDiagramU(3)}
\end{equation}
where the bottom leaf is of dimension 1, it is the Coulomb branch of pure U(1).

In comparison, the Hasse diagram of U(2) with 4 flavors is:
\begin{equation}
      \raisebox{-.5\height}{ \begin{tikzpicture}[scale=1.5]
        \node[hasseJ,label=left:{\color{red}0},label=right:{\color{blue}0}] (1) at (0,0) {};
        \node[hasseJ,label=left:{\color{red}0},label=right:{\color{blue}3}] (2) at (1,1) {};
        \node[hasseJ,label=left:{\color{red}0},label=right:{\color{blue}4}] (3) at (2,2) {};
        \node[hasseJ,label=left:{\color{red}1},label=right:{\color{blue}0}] (1a) at (-1,1) {};
        \node[hasseJ,label=left:{\color{red}1},label=right:{\color{blue}3}] (2a) at (0,2) {};
        \node[hasseJ,label=left:{\color{red}2},label=right:{\color{blue}0}] (1aa) at (-2,2) {};
        \draw[blue] (1)--(2)--(3) (1a)--(2a) ;
        \draw[red] (1)--(1a)--(1aa) (2)--(2a);
        
        \node at ($(1)!0.5!(2)$) {$a_3$};
        \node at ($(2)!0.5!(3)$) {$a_1$};
        \node at ($(1a)!0.5!(2a)$) {$a_3$};
        
        \node at ($(1)!0.5!(1a)$) {$A_1$};
        \node at ($(1a)!0.5!(1aa)$) {$A_3$};
        \node at ($(2)!0.5!(2a)$) {$A_1$};
    \end{tikzpicture}}
    \label{eq:HasseU(2)4}
\end{equation}
where the bottom leaf is zero dimensional. The Hasse diagram of the magnetic quiver of U(3) with 4 flavors is obtained from the Hasse diagram \eqref{eq:HasseU(2)4} by exchanging red and blue colors.

Now we can turn on the FI parameter and study the resolved Higgs branch. In order to do this we can leave no D3 brane suspended between the two NS5:
\begin{equation}
    \raisebox{-.5\height}{   \begin{tikzpicture}
        \node at (0,-2.2) {A};
        \node at (7+0.5,-2.2+0.5) {B};
        \draw[dotted] (-2,0)--(11,0);
        \draw[red] (0,-2)--(0,2) (7+0.5,-2+0.5)--(7+0.5,2+0.5);
        \draw[dashed] (7,0)--(7+0.5,0+0.5);
        \draw[blue] (1,-1)--(3,1) (2,-1)--(4,1) (3,-1)--(5,1) (4,-1)--(6,1);

        \draw (0,0.1)--(2.1,0.1);
        \draw (0,0)--(3,0);
        \draw (0,-0.1)--(4-0.1,-0.1);

        \begin{scope}[shift={(0.5,0.5)}]
        \draw (3.1,0.1)--(7,0.1);
        \draw (4,0)--(7,0);
        \draw (5-0.1,-0.1)--(7,-0.1);
        \end{scope}
        
        \begin{scope}[shift={(-0.5,-0.5)}]
        \draw (2,0)--(3,0) (4,0)--(5,0);
        \draw (3-0.1,-0.1)--(4-0.1,-0.1);
        \end{scope}
    \end{tikzpicture} }
\end{equation}

When we do the HW transitions in order to annihilate all D3 branes suspended between a D5 and an NS5 we have to change the order along the $x^6$ direction of the NS5 branes (A$\leftrightarrow$B). Now we can find the magnetic quiver for this resolved Higgs branch (keeping in mind that the mass parameter of the magnetic quiver is the FI parameter of the original quiver). 
\begin{equation}
    \raisebox{-.5\height}{   \begin{tikzpicture}
        \node at (4.5,-2.2) {A};
        \node at (2.5+0.5,-2.2+0.5) {B};
        \draw[dotted] (0,0)--(7,0);
        \draw[red] (4.5,-2)--(4.5,2) (2.5+0.5,-2+0.5)--(2.5+0.5,2+0.5);
        \draw[dashed] (2.5,0)--(2.5+0.5,0+0.5);
        \draw[blue] (1,-1)--(3,1) (2,-1)--(4,1) (3,-1)--(5,1) (4,-1)--(6,1);
        
        \begin{scope}[shift={(-0.5,-0.5)}]
        \draw (2,0)--(3,0) (4,0)--(5,0);
        \draw (3-0.1,-0.1)--(4-0.1,-0.1);
        \end{scope}
        
        \begin{scope}[shift={(8,0)}]
        \node at (0,0) {$\mathsf{Q}_m=$};
        \node[gaugeJ,label=below:{$1$}] (1) at (1,0) {};
        \node[gaugeJ,label=below:{$1$}] (2) at (2,0) {};
        \node[gaugeJ,label=below:{$1$}] (3) at (3,0) {};
        \node[flavourJ,label=left:{$1$}] (1u) at (1,1) {};
        \node[flavourJ,label=right:{$1$}] (3u) at (3,1) {};
        \draw (1u)--(1)--(2)--(3)--(3u);
        \end{scope}
    \end{tikzpicture}}
\end{equation}
This is the magnetic quiver / 3d mirror of U(1) with 4 flavors.

To summarize, we find the following magnetic quivers for U(3) with 4 flavors:

\begin{equation}
     \raisebox{-.5\height}{  \begin{tabular}{|c|c|c|}
        \hline
        \multirow{2}{*}{Theory} & \multicolumn{2}{c|}{Magnetic Quivers} \\\cline{2-3}
         & FI$=0$ & FI$\neq0$ \\\hline
         $\begin{tikzpicture}
             \node[gaugeJ,label=left:{U(3)}] (0) at (0,0) {};
             \node[flavourJ,label=left:{$4$}] (1) at (0,1) {};
             \draw (0)--(1);
             \node at (0,1.5) {};
             \node at (0,-0.5) {};
         \end{tikzpicture}$ & $\begin{tikzpicture}
                \node[gaugeJ,label=below:{$1$}] (1) at (1,0) {};
                \node[gaugeJ,label=below:{$2$}] (2) at (2,0) {};
                \node[gaugeJ,label=below:{$1$}] (3) at (3,0) {};
                \node[flavourJ,label=left:{$2$}] (2u) at (2,1) {};
                \draw (1)--(2)--(3) (2u)--(2);
                \node at (1,1.5) {};
                \node at (1,-0.5) {};
         \end{tikzpicture}$ & $\begin{tikzpicture}
                \node[gaugeJ,label=below:{$1$}] (1) at (1,0) {};
                \node[gaugeJ,label=below:{$1$}] (2) at (2,0) {};
                \node[gaugeJ,label=below:{$1$}] (3) at (3,0) {};
                \node[flavourJ,label=left:{$1$}] (1u) at (1,1) {};
                \node[flavourJ,label=right:{$1$}] (3u) at (3,1) {};
                \draw (1u)--(1)--(2)--(3)--(3u);
                \node at (1,1.5) {};
                \node at (1,-0.5) {};
         \end{tikzpicture}$ \\\hline
    \end{tabular} }
    \label{MQwithFI}
\end{equation}

Clearly the two magnetic quivers do not agree. Hence there is no clear way to define a 3d mirror of U$(3)$ with 4 flavors, even if the mirror theory would be supplemented by free hypers. In this paper we only provide the magnetic quivers in the case where FI parameters are turned off.

We can make the following statements about the magnetic quivers (Higgs branches) of U$(k)$ SQCD: Let $k\leq N<2k$,
\begin{itemize}
    \item if the FI parameter is turned off, then the magnetic quiver for U$(k)$ with N flavors is that of U($\left \lfloor{N/2}\right \rfloor$) with $N$ flavors. The smooth bottom leaf in the moduli space is the Coulomb branch of U$(k-\left \lfloor{N/2}\right \rfloor)$ with $0$ flavors if $N$ even and $1$ flavor if $N$ odd. This is the part of the gauge theory which remains massless on a generic point in the Higgs branch.
    \item if the FI parameter is turned on, then the magnetic quiver for U$(k)$ with $N$ flavors is that of U$(N-k)$ with N flavors. The Coulomb branch is lifted by the FI parameter.
\end{itemize}
If $N<k$, then no FI parameter can be turned on if supersymmetry is not to be broken. Other than that this case behaves just like before.

All of this is consistent with quiver subtraction which produces the Coulomb branch Hasse diagram. For $N$ odd and $N$ even we get respectively:
\begin{equation}
    \raisebox{-0.5\height}{\begin{tikzpicture}
        \node[hasseJ] (a) at (0,0) {};
        \node[hasseJ] (b) at (0,-2) {};
        \node[hasseJ] (c) at (0,-4) {};
        \node (cd) at (0,-5) {$\vdots$};
        \node[hasseJ] (d) at (0,-6) {};
        \node[hasseJ] (e) at (0,-8) {};
        \draw[red] (a)--(b)--(c)--(cd)--(d)--(e);
        
        \node at (-2,0) {$\begin{tikzpicture}
                \node[gaugeJ,label=left:{$k$}] (0) at (0,0) {};
                \node[flavourJ,label=left:{$N$}] (1) at (0,1) {};
                \draw (0)--(1);
                \node at (-3,0) {};
            \end{tikzpicture}$};
        \node at (-2,-2) {$\begin{tikzpicture}
                \node[gaugeJ,label=left:{$k-1$}] (0) at (0,0) {};
                \node[flavourJ,label=left:{$N-2$}] (1) at (0,1) {};
                \draw (0)--(1);
                \node at (-3,0) {};
            \end{tikzpicture}$};
        \node at (-2,-4) {$\begin{tikzpicture}
                \node[gaugeJ,label=left:{$k-2$}] (0) at (0,0) {};
                \node[flavourJ,label=left:{$N-4$}] (1) at (0,1) {};
                \draw (0)--(1);
                \node at (-3,0) {};
            \end{tikzpicture}$};
        \node at (-2,-6) {$\begin{tikzpicture}
                \node[gaugeJ,label=left:{$k-\left \lfloor{N/2}\right \rfloor+1$}] (0) at (0,0) {};
                \node[flavourJ,label=left:{$3$}] (1) at (0,1) {};
                \draw (0)--(1);
                \node at (-3,0) {};
            \end{tikzpicture}$};
        \node at (-2,-8) {$\begin{tikzpicture}
                \node[gaugeJ,label=left:{$k-\left \lfloor{N/2}\right \rfloor$}] (0) at (0,0) {};
                \node[flavourJ,label=left:{$1$}] (1) at (0,1) {};
                \draw (0)--(1);
                \node at (-3,0) {};
            \end{tikzpicture}$};
            
            \node at (0.7,-1) {$A_{N-1}$};
            \node at (0.7,-3) {$A_{N-3}$};
            \node at (0.7,-7) {$A_{2}$};
    \end{tikzpicture}}
    \qquad
    \raisebox{-0.5\height}{\begin{tikzpicture}
        \node[hasseJ] (a) at (0,0) {};
        \node[hasseJ] (b) at (0,-2) {};
        \node[hasseJ] (c) at (0,-4) {};
        \node (cd) at (0,-5) {$\vdots$};
        \node[hasseJ] (d) at (0,-6) {};
        \node[hasseJ] (e) at (0,-8) {};
        \draw[red] (a)--(b)--(c)--(cd)--(d)--(e);
        
        \node at (-2,0) {$\begin{tikzpicture}
                \node[gaugeJ,label=left:{$k$}] (0) at (0,0) {};
                \node[flavourJ,label=left:{$N$}] (1) at (0,1) {};
                \draw (0)--(1);
                \node at (-3,0) {};
            \end{tikzpicture}$};
        \node at (-2,-2) {$\begin{tikzpicture}
                \node[gaugeJ,label=left:{$k-1$}] (0) at (0,0) {};
                \node[flavourJ,label=left:{$N-2$}] (1) at (0,1) {};
                \draw (0)--(1);
                \node at (-3,0) {};
            \end{tikzpicture}$};
        \node at (-2,-4) {$\begin{tikzpicture}
                \node[gaugeJ,label=left:{$k-2$}] (0) at (0,0) {};
                \node[flavourJ,label=left:{$N-4$}] (1) at (0,1) {};
                \draw (0)--(1);
                \node at (-3,0) {};
            \end{tikzpicture}$};
        \node at (-2,-6) {$\begin{tikzpicture}
                \node[gaugeJ,label=left:{$k-N/2+1$}] (0) at (0,0) {};
                \node[flavourJ,label=left:{$2$}] (1) at (0,1) {};
                \draw (0)--(1);
                \node at (-3,0) {};
            \end{tikzpicture}$};
        \node at (-2,-8) {$\begin{tikzpicture}
                \node[gaugeJ,label=left:{$k-N/2$}] (0) at (0,0) {};
                \node at (-3,0) {};
            \end{tikzpicture}$};
            
            \node at (0.7,-1) {$A_{N-1}$};
            \node at (0.7,-3) {$A_{N-3}$};
            \node at (0.7,-7) {$A_{1}$};
    \end{tikzpicture}}
    \label{eq:generalizedHasse}
\end{equation}
This is in agreement with the Hasse diagram of generalized affine Grassmannian slices \cite[(1.1)]{Muthiah:2019jif}. The theory which is left after quiver subtraction in \eqref{eq:generalizedHasse} is precisely what remains on a generic point of the Higgs branch.

\providecommand{\href}[2]{#2}\begingroup\raggedright\endgroup


\begin{thebibliography}{10}

\bibitem{Intriligator:1996ex}
K.~A. Intriligator and N.~Seiberg, \emph{{Mirror symmetry in three-dimensional
  gauge theories}},
  \href{http://dx.doi.org/10.1016/0370-2693(96)01088-X}{\emph{Phys. Lett. B}
  {\bfseries 387} (1996) 513--519},
  [\href{https://arxiv.org/abs/hep-th/9607207}{{\ttfamily hep-th/9607207}}].

\bibitem{Hanany:1996ie}
A.~Hanany and E.~Witten, \emph{{Type IIB superstrings, BPS monopoles, and
  three-dimensional gauge dynamics}},
  \href{http://dx.doi.org/10.1016/S0550-3213(97)00157-0}{\emph{Nucl. Phys. B}
  {\bfseries 492} (1997) 152--190},
  [\href{https://arxiv.org/abs/hep-th/9611230}{{\ttfamily hep-th/9611230}}].

\bibitem{Aharony:1997ju}
O.~Aharony and A.~Hanany, \emph{{Branes, superpotentials and superconformal
  fixed points}},
  \href{http://dx.doi.org/10.1016/S0550-3213(97)00472-0}{\emph{Nucl. Phys. B}
  {\bfseries 504} (1997) 239--271},
  [\href{https://arxiv.org/abs/hep-th/9704170}{{\ttfamily hep-th/9704170}}].

\bibitem{Aharony:1997bh}
O.~Aharony, A.~Hanany and B.~Kol, \emph{{Webs of (p,q) five-branes,
  five-dimensional field theories and grid diagrams}},
  \href{http://dx.doi.org/10.1088/1126-6708/1998/01/002}{\emph{JHEP} {\bfseries
  01} (1998) 002}, [\href{https://arxiv.org/abs/hep-th/9710116}{{\ttfamily
  hep-th/9710116}}].

\bibitem{DeWolfe:1999hj}
O.~DeWolfe, A.~Hanany, A.~Iqbal and E.~Katz, \emph{{Five-branes, seven-branes
  and five-dimensional E(n) field theories}},
  \href{http://dx.doi.org/10.1088/1126-6708/1999/03/006}{\emph{JHEP} {\bfseries
  03} (1999) 006}, [\href{https://arxiv.org/abs/hep-th/9902179}{{\ttfamily
  hep-th/9902179}}].

\bibitem{Closset:2020afy}
C.~Closset, S.~Giacomelli, S.~Schafer-Nameki and Y.-N. Wang, \emph{{5d and 4d
  SCFTs: Canonical Singularities, Trinions and S-Dualities}},
  \href{http://dx.doi.org/10.1007/JHEP05(2021)274}{\emph{JHEP} {\bfseries 05}
  (2021) 274}, [\href{https://arxiv.org/abs/2012.12827}{{\ttfamily
  2012.12827}}].

\bibitem{Giacomelli:2020ryy}
S.~Giacomelli, N.~Mekareeya and M.~Sacchi, \emph{{New aspects of
  Argyres--Douglas theories and their dimensional reduction}},
  \href{http://dx.doi.org/10.1007/JHEP03(2021)242}{\emph{JHEP} {\bfseries 03}
  (2021) 242}, [\href{https://arxiv.org/abs/2012.12852}{{\ttfamily
  2012.12852}}].

\bibitem{Xie:2021ewm}
D.~Xie, \emph{{3d mirror for Argyres-Douglas theories}},
  \href{https://arxiv.org/abs/2107.05258}{{\ttfamily 2107.05258}}.

\bibitem{Dey:2021rxw}
A.~Dey, \emph{{Higgs Branches of Argyres-Douglas theories as Quiver
  Varieties}},  \href{https://arxiv.org/abs/2109.07493}{{\ttfamily
  2109.07493}}.

\bibitem{Bourget:2019rtl}
A.~Bourget, S.~Cabrera, J.~F. Grimminger, A.~Hanany and Z.~Zhong, \emph{{Brane
  Webs and Magnetic Quivers for SQCD}},
  \href{http://dx.doi.org/10.1007/JHEP03(2020)176}{\emph{JHEP} {\bfseries 03}
  (2020) 176}, [\href{https://arxiv.org/abs/1909.00667}{{\ttfamily
  1909.00667}}].

\bibitem{Cabrera:2018jxt}
S.~Cabrera, A.~Hanany and F.~Yagi, \emph{{Tropical Geometry and Five
  Dimensional Higgs Branches at Infinite Coupling}},
  \href{http://dx.doi.org/10.1007/JHEP01(2019)068}{\emph{JHEP} {\bfseries 01}
  (2019) 068}, [\href{https://arxiv.org/abs/1810.01379}{{\ttfamily
  1810.01379}}].

\bibitem{Gaiotto:2008ak}
D.~Gaiotto and E.~Witten, \emph{{S-Duality of Boundary Conditions In N=4 Super
  Yang-Mills Theory}},
  \href{http://dx.doi.org/10.4310/ATMP.2009.v13.n3.a5}{\emph{Adv. Theor. Math.
  Phys.} {\bfseries 13} (2009) 721--896},
  [\href{https://arxiv.org/abs/0807.3720}{{\ttfamily 0807.3720}}].

\bibitem{Dancer:2020wll}
A.~Dancer, A.~Hanany and F.~Kirwan, \emph{{Symplectic duality and implosions}},
   \href{https://arxiv.org/abs/2004.09620}{{\ttfamily 2004.09620}}.

\bibitem{Collinucci:2020kdm}
A.~Collinucci and R.~Valandro, \emph{{A string theory realization of special
  unitary quivers in 3 dimensions}},
  \href{http://dx.doi.org/10.1007/JHEP11(2020)157}{\emph{JHEP} {\bfseries 11}
  (2020) 157}, [\href{https://arxiv.org/abs/2008.10689}{{\ttfamily
  2008.10689}}].

\bibitem{Bourget:2019aer}
A.~Bourget, S.~Cabrera, J.~F. Grimminger, A.~Hanany, M.~Sperling, A.~Zajac
  et~al., \emph{{The Higgs mechanism \textemdash{} Hasse diagrams for
  symplectic singularities}},
  \href{http://dx.doi.org/10.1007/JHEP01(2020)157}{\emph{JHEP} {\bfseries 01}
  (2020) 157}, [\href{https://arxiv.org/abs/1908.04245}{{\ttfamily
  1908.04245}}].

\bibitem{Cremonesi:2014uva}
S.~Cremonesi, A.~Hanany, N.~Mekareeya and A.~Zaffaroni,
  \emph{{T$_{\rho}^{\sigma}$ (G) theories and their Hilbert series}},
  \href{http://dx.doi.org/10.1007/JHEP01(2015)150}{\emph{JHEP} {\bfseries 01}
  (2015) 150}, [\href{https://arxiv.org/abs/1410.1548}{{\ttfamily 1410.1548}}].

\bibitem{Argyres:1996eh}
P.~C. Argyres, M.~R. Plesser and N.~Seiberg, \emph{{The Moduli space of vacua
  of N=2 SUSY QCD and duality in N=1 SUSY QCD}},
  \href{http://dx.doi.org/10.1016/0550-3213(96)00210-6}{\emph{Nucl. Phys. B}
  {\bfseries 471} (1996) 159--194},
  [\href{https://arxiv.org/abs/hep-th/9603042}{{\ttfamily hep-th/9603042}}].

\bibitem{Assel:2017jgo}
B.~Assel and S.~Cremonesi, \emph{{The Infrared Physics of Bad Theories}},
  \href{http://dx.doi.org/10.21468/SciPostPhys.3.3.024}{\emph{SciPost Phys.}
  {\bfseries 3} (2017) 024},
  [\href{https://arxiv.org/abs/1707.03403}{{\ttfamily 1707.03403}}].

\bibitem{Hanany:2010qu}
A.~Hanany and N.~Mekareeya, \emph{{Tri-vertices and SU(2)'s}},
  \href{http://dx.doi.org/10.1007/JHEP02(2011)069}{\emph{JHEP} {\bfseries 02}
  (2011) 069}, [\href{https://arxiv.org/abs/1012.2119}{{\ttfamily 1012.2119}}].

\bibitem{Beem:2021jnm}
C.~Beem and C.~Meneghelli, \emph{{Geometric free field realization for the
  genus-two class S theory of type a1}},
  \href{http://dx.doi.org/10.1103/PhysRevD.104.065015}{\emph{Phys. Rev. D}
  {\bfseries 104} (2021) 065015},
  [\href{https://arxiv.org/abs/2104.11668}{{\ttfamily 2104.11668}}].

\bibitem{Argyres:2016xmc}
P.~Argyres, M.~Lotito, Y.~L\"u and M.~Martone, \emph{{Geometric constraints on
  the space of $ \mathcal{N}$ = 2 SCFTs. Part III: enhanced Coulomb branches
  and central charges}},
  \href{http://dx.doi.org/10.1007/JHEP02(2018)003}{\emph{JHEP} {\bfseries 02}
  (2018) 003}, [\href{https://arxiv.org/abs/1609.04404}{{\ttfamily
  1609.04404}}].

\bibitem{Closset:2020scj}
C.~Closset, S.~Schafer-Nameki and Y.-N. Wang, \emph{{Coulomb and Higgs Branches
  from Canonical Singularities: Part 0}},
  \href{http://dx.doi.org/10.1007/JHEP02(2021)003}{\emph{JHEP} {\bfseries 02}
  (2021) 003}, [\href{https://arxiv.org/abs/2007.15600}{{\ttfamily
  2007.15600}}].

\bibitem{Bourget:2021siw}
A.~Bourget, J.~F. Grimminger, A.~Hanany, M.~Sperling and Z.~Zhong,
  \emph{{Branes, Quivers, and the Affine Grassmannian}},
  \href{https://arxiv.org/abs/2102.06190}{{\ttfamily 2102.06190}}.

\bibitem{Bourget:2021zyc}
A.~Bourget, A.~Dancer, J.~F. Grimminger, A.~Hanany, F.~Kirwan and Z.~Zhong,
  \emph{{Orthosymplectic implosions}},
  \href{http://dx.doi.org/10.1007/JHEP08(2021)012}{\emph{JHEP} {\bfseries 08}
  (2021) 012}, [\href{https://arxiv.org/abs/2103.05458}{{\ttfamily
  2103.05458}}].

\bibitem{Bourget:2020xdz}
A.~Bourget, J.~F. Grimminger, A.~Hanany, R.~Kalveks, M.~Sperling and Z.~Zhong,
  \emph{{Magnetic Lattices for Orthosymplectic Quivers}},
  \href{http://dx.doi.org/10.1007/JHEP12(2020)092}{\emph{JHEP} {\bfseries 12}
  (2020) 092}, [\href{https://arxiv.org/abs/2007.04667}{{\ttfamily
  2007.04667}}].

\bibitem{Dasgupta:1996ij}
K.~Dasgupta and S.~Mukhi, \emph{{F theory at constant coupling}},
  \href{http://dx.doi.org/10.1016/0370-2693(96)00875-1}{\emph{Phys. Lett. B}
  {\bfseries 385} (1996) 125--131},
  [\href{https://arxiv.org/abs/hep-th/9606044}{{\ttfamily hep-th/9606044}}].

\bibitem{Minahan:1996fg}
J.~A. Minahan and D.~Nemeschansky, \emph{{An N=2 superconformal fixed point
  with E(6) global symmetry}},
  \href{http://dx.doi.org/10.1016/S0550-3213(96)00552-4}{\emph{Nucl. Phys.}
  {\bfseries B482} (1996) 142--152},
  [\href{https://arxiv.org/abs/hep-th/9608047}{{\ttfamily hep-th/9608047}}].

\bibitem{Yaakov:2013fza}
I.~Yaakov, \emph{{Redeeming Bad Theories}},
  \href{http://dx.doi.org/10.1007/JHEP11(2013)189}{\emph{JHEP} {\bfseries 11}
  (2013) 189}, [\href{https://arxiv.org/abs/1303.2769}{{\ttfamily 1303.2769}}].

\bibitem{Cabrera:2016vvv}
S.~Cabrera and A.~Hanany, \emph{{Branes and the Kraft-Procesi Transition}},
  \href{http://dx.doi.org/10.1007/JHEP11(2016)175}{\emph{JHEP} {\bfseries 11}
  (2016) 175}, [\href{https://arxiv.org/abs/1609.07798}{{\ttfamily
  1609.07798}}].

\bibitem{Cabrera:2017njm}
S.~Cabrera and A.~Hanany, \emph{{Branes and the Kraft-Procesi transition:
  classical case}},
  \href{http://dx.doi.org/10.1007/JHEP04(2018)127}{\emph{JHEP} {\bfseries 04}
  (2018) 127}, [\href{https://arxiv.org/abs/1711.02378}{{\ttfamily
  1711.02378}}].

\bibitem{Grimminger:2020dmg}
J.~F. Grimminger and A.~Hanany, \emph{{Hasse diagrams for 3d $ \mathcal{N} $ =
  4 quiver gauge theories \textemdash{} Inversion and the full moduli space}},
  \href{http://dx.doi.org/10.1007/JHEP09(2020)159}{\emph{JHEP} {\bfseries 09}
  (2020) 159}, [\href{https://arxiv.org/abs/2004.01675}{{\ttfamily
  2004.01675}}].

\bibitem{Braverman:2016pwk}
A.~Braverman, M.~Finkelberg and H.~Nakajima, \emph{{Coulomb branches of $3d$
  $\mathcal{N}=4$ quiver gauge theories and slices in the affine
  Grassmannian}},
  \href{http://dx.doi.org/10.4310/ATMP.2019.v23.n1.a3}{\emph{Adv. Theor. Math.
  Phys.} {\bfseries 23} (2019) 75--166},
  [\href{https://arxiv.org/abs/1604.03625}{{\ttfamily 1604.03625}}].

\bibitem{Muthiah:2019jif}
D.~Muthiah and A.~Weekes, \emph{{Symplectic leaves for generalized affine
  Grassmannian slices}},  \href{https://arxiv.org/abs/1902.09771}{{\ttfamily
  1902.09771}}.

\end{thebibliography}
\end{document}